\documentclass[12pt]{article}
\usepackage{lscape}
\usepackage{geometry}
\usepackage[onehalfspacing]{setspace}
\usepackage{amssymb}
\usepackage{amsfonts}
\usepackage{graphicx}
\usepackage{amsmath}
\usepackage{epstopdf}
\usepackage{natbib}
\usepackage{morefloats}
\usepackage{float}
\usepackage{comment}
\usepackage{pdflscape}
\usepackage{adjustbox}
\usepackage{subcaption}
\usepackage{longtable}
\usepackage{apalike}
\usepackage{xfrac}
\usepackage[dvipsnames,table]{xcolor}
\usepackage{etoolbox,siunitx}
\robustify\bfseries
\usepackage{xstring}

\usepackage{qtree}
\usepackage{multicol}
\usepackage{booktabs}
\usepackage{threeparttable}
\usepackage{tabularx}
\usepackage{dcolumn}
\newcolumntype{d}[1]{D..{#1}} 
 
\usepackage{siunitx}
 \usepackage{mathpazo} 
\usepackage[scaled]{helvet} 
\usepackage{courier} 
\normalfont
\usepackage[T1]{fontenc}
\usepackage{listings}
\usepackage{epigraph}
\def\sym#1{\ifmmode^{#1}\else\(^{#1}\)\fi}

\usepackage{xcolor} 
\definecolor{DarkerPineGreen}{RGB}{0, 90, 80} 
\definecolor{dukeblue}{rgb}{0.0, 0.0, 0.61}
\definecolor{darkblue}{RGB}{10, 10, 100}
 \definecolor{siena}{rgb}{0.91,0.45,0.32} 
\definecolor{darkred}{rgb}{0.8,0,0}
\definecolor{darkpowderblue}{rgb}{0.0, 0.05, 0.5}
 \definecolor{dpd2}{rgb}{0.0, 0.043, 0.43}
 \definecolor{dpd}{rgb}{0.0, 0.05, 0.5}
 
\definecolor{darkblue2}{HTML}{1e3986}
\definecolor{darkred2}{HTML}{B22222}
\definecolor{darkgreen2}{HTML}{1B9E77}

\usepackage[pdftex,bookmarks,colorlinks]{hyperref}
\hypersetup{
  colorlinks,
  citecolor=darkpowderblue,
  linkcolor=red,
  urlcolor=blue}
\usepackage{cleveref}

\usepackage{etoolbox}
 
\makeatletter
\patchcmd{\epigraph}{\@epitext{#1}}{\itshape\@epitext{#1}}{}{}
\makeatother

\makeatletter
\newcommand\halftiny{\@setfontsize\halftiny\@vipt\@viipt}
\newcommand\notsotiny{\@setfontsize\notsotiny{6.99}{9.2828}}
\newcommand\footscript{\@setfontsize\footscript{8.5}{10}}
\newcommand\notsolarge{\@setfontsize\notsolarge{12}{14}}
\newcommand\superlarge{\@setfontsize\superlarge{21.5}{24}}
\makeatother
 
\renewenvironment{abstract}
 {\small
  \begin{center}
  \bfseries \abstractname\vspace{-.5em}\vspace{0pt}
  \end{center}
  \list{}{
    \setlength{\leftmargin}{1.78cm}    \setlength{\rightmargin}{\leftmargin}  }  \item\relax}
 {\endlist}
\usepackage{graphicx}
\usepackage{wrapfig}
 
\allowdisplaybreaks
 
\usepackage{enumitem}

\usepackage{algorithm,algcompatible}
\usepackage{tikz}
\usetikzlibrary{positioning}

\begin{document}
\sloppy

    \title{\vspace*{-0.09cm} \fontsize{24.95}{22}  \textbf{\color{dpd} Opening the Black Box of Local Projections \\ \phantom{.}} \vspace*{0.4cm}}
\author{\hspace*{-0.3cm} Philippe Goulet Coulombe\thanks{%
Département des Sciences Économiques,  \href{mailto:p.gouletcoulombe@gmail.com}{\color{dpd}\texttt{goulet\_coulombe.philippe@uqam.ca}}.  For helpful discussions and  comments,  we thank  Maximilian G\"obel, Alain Guay, Julien Martin, Massimiliano Marcellino, Mikkel Plagborg-M{\o}ller, Gabriel Rodriguez Rondon, Francesco Simone Lucidi,  and {Boyuan Zhang}.  For research assistance, we thank Hugo Couture. The views expressed in this paper do not necessarily reflect those of the European Central Bank, Oesterreichische Nationalbank or the Eurosystem. \texttt{R} codes and slide deck are available \href{https://github.com/philgoucou/LPdecomposition}{\color{dpd}here} and \href{https://philippegouletcoulombe.com/research}{\color{dpd}here}, respectively. This draft: \today.}\\[-0.2cm] \hspace*{-0.3cm} \textbf{\texttt{\fontfamily{phv}\selectfont \notsolarge  Université du Québec à Montréal}} 
   \and 
\hspace*{1.3cm} Karin Klieber \\[-0.2cm] 
\hspace*{1.5cm} \textbf{\texttt{\fontfamily{phv}\selectfont \notsolarge European Central Bank, OeNB}}  
\vspace*{0.15cm}
}

\date{\vspace{0.7cm}
\small
\small 
\vspace{-0.25cm}
\Large
  }
\newgeometry{left=1.9 cm, right= 1.9 cm, top=2.3 cm, bottom=1.5 cm}
\maketitle

\date{\vspace{0.2cm}
\small
\small
\vspace{-0.4cm}
\large
  }
\maketitle
\begin{abstract}

\noindent Local projections (LPs) are widely used in empirical macroeconomics to estimate impulse responses to policy interventions.  Yet, in many ways, they are black boxes.  It is often unclear what mechanism or historical episodes drive a particular estimate.  We introduce a new decomposition of LP estimates into the sum of contributions of historical events, which is the product, for each time stamp, of a weight and the realization of the response variable.  
In the least squares case, we show that these weights admit two interpretations. First, they represent purified and standardized shocks. Second, they serve as proximity scores between the projected policy intervention and past interventions in the sample. Notably, this second interpretation extends naturally to machine learning methods, many of which yield impulse responses that, while nonlinear in predictors, still aggregate past outcomes \textit{linearly} via proximity-based weights. Applying this framework to shocks in monetary and fiscal policy, global temperature, and the excess bond premium, we find that easily identifiable events—such as Nixon’s interference with the Fed, stagflation, World War II, and the Mount Agung volcanic eruption—emerge as dominant drivers of often heavily concentrated impulse response estimates.

\end{abstract}
 
\thispagestyle{empty}
 

\clearpage
 
 
 \clearpage 
\setcounter{page}{1}
 \restoregeometry

\section{Introduction}

Local projections-based estimates of impulse response functions (IRFs), now ubiquitous in empirical macroeconomic analysis, are not regarded as black boxes. Yet, to an appreciable extent, they are. It is often unclear what transmission mechanism lies behind the curve, or how the  arbitrary inclusion/exclusion of control variables shapes the retrieved causal effects.  It is also difficult to know whether local projections (LP) estimates used to tell a cohesive story about certain economic events are actually sourced from those events. 

To elucidate this and other questions, we introduce a {new decomposition} of local projection estimates into a sum of contributions from historical events. Each contribution is the product of a \textit{weight} and the corresponding realization of the response variable for every time point in the sample. First, this decomposition serves as a powerful diagnostic tool, distinguishing whether an estimate is broadly supported by a wide range of historical events or dominated by a narrow set of episodes. This provides an off-the-shelf external validity assessment before using estimates to guide policy decisions or complement economic analysis with theoretical models explaining the empirical facts.  Second, analyzing individual historical weights and contributions reveals whether the evidence as perceived by the model conforms with external narrative knowledge: do the episodes we think are driving the causal effect estimates actually \textit{do so}? As such, it can prove useful to the ``empirical debugging'' process. For instance, when economists are faced with puzzles or other varieties of counterintuitive empirical findings, the decomposition can identify problematic periods or events that disproportionately contribute to the unexpected outcome.

In the case of LP estimated through least squares, {the weights have two interpretations}. The first, established through the well-known Frisch-Waugh-Lovell theorem, shows that the weights represent standardized, purified shocks that account for the influence of control variables. The second interpretation, inspired by the dual solution to  least squares problems, reveals that the weights correspond to pairwise proximity scores between the projected policy intervention and the historical interventions in the sample. As put forward in \cite{OLSA}, ordinary least squares (OLS) can also be framed as a similarity-based estimator in a transformed regressor space.  We extend these insights from predictions to coefficients, by showing that the weights underlying our decomposition correspond to Euclidean inner products between the representation of the intended policy on an orthogonal basis and the representations of past interventions in the same space. Therefore, OLS-based LPs implicitly construct an embedding of past interventions and predict the effect of a new one as a weighted average of past outcomes, giving greater weight to those associated with interventions that lie closer to the projected intervention in the  ``intervention space''.

The proximity interpretation of weights also enables us to extend the applicability range beyond linear models. Many {machine learning} (ML) algorithms produce local projections that, while nonlinear in regressors, are still \textit{linear combinations} of the response variable with weights corresponding to proximity scores. This means that, through our framework, nonlinear ML-based impulse responses can be decomposed and interpreted in \textit{the same way} as those from traditional linear methods. Weights differences between linear and nonlinear methods are attributable to different notions of proximity as constructed by models, and given the typically short length of macroeconomic time series, are  easy to compare graphically. Overall, our decomposition offers a pathway to integrating these more flexible methods into IRF analysis while maintaining a satisfactory level of transparency.

Lastly, we propose LP {concentration statistics}. Inspired by  prototypical measures of wealth inequality, we monitor the share of total weights (contributions) accounted for by the top 10\% of weights (contributions). While some degree of concentration is inevitable, our applications reveal that certain IRFs exhibit an alarmingly high level of it. 	Inescapably, elevated concentration threatens external validity. 

 \vskip 0.25cm  
{\noindent \sc \textbf{Applicability.}} In the empirical analysis, we focus on OLS and Random Forest, the latter being a powerful, nonlinear, off-the-shelf machine learning algorithm. While not explicitly used here, the decomposition framework is fully compatible with any estimator that is, or can be rewritten as, a linear combination of outcomes. This includes kernel methods, boosting, and neural networks, drawing on tools developed in \citet{dual}. The same applies to two-stage least squares, frequently used alongside local projections. In that context, our tools could be applied separately to both the first and second stage regressions, with the second stage being conceptually closest to our empirical application. {\color{black} The tool could also be applied in a panel data setting to decompose how both time periods and individual units contribute to the estimated effects.} Finally, further thinking is required to extend the framework to impulse response estimators that are not linear aggregations of a single outcome variable—such as certain Bayesian methods or VARs, for which we sketch a possible route using Shapley values in the conclusion.

 \vskip 0.25cm  
{\noindent \sc \textbf{Four Applications.}} We apply this decomposition to LPs obtained using various types of shocks, including monetary policy, fiscal policy, global temperature, and financial shocks.

For \textbf{monetary policy}, we find that impulse responses derived from vector autoregressions (VAR) identified via the Cholesky decomposition—often displaying the familiar price puzzle, where inflation rises in response to a contractionary monetary policy shock—are confounded by misinterpretations of stagflation episodes in the 1970s. \textcolor{black}{Expanding the information set in the LP, in the spirit of \cite{bernanke2005measuring},  shifts the inflation response in the expected direction by nullifying or inverting problematic contributions from these key 1970s episodes.} Shocks from \cite{romerromer2004} produce correctly signed impulse responses without requiring a plethora of controls. However, this is achieved not by canceling counterintuitive episodes found in the Cholesky VAR, but by forcefully offsetting them with stronger mid-1970s negative contributions. We find that these responses are almost entirely driven by politically motivated monetary \textit{loosening} episodes of the 1970s.

Further clarity emerges from Random Forest estimates, which allow for nonlinearities including shock-sign dependence. Consistent with the results of the linear model, we find little to no evidence from contractionary \cite{romerromer2004} shocks.  In contrast, the nonlinear LP for unexpected loosening shocks eliminates the early horizon price appreciation seen in the linear model, then realigns with the linear estimate at mid-range horizons and beyond. We find the underlying proximity weights to be heavily concentrated. Most of the evidence is attributable to two well-documented “Nixon’s calling” episodes. While this concentration supports the instrument’s validity—these appear to be genuine exogenous interventions to an otherwise systematic policy—it raises questions about the external validity of the estimates, particularly if used to anticipate the effects of \textit{contractionary} monetary policy nearly 50 years later.

For \textbf{fiscal policy}, we show that the state-dependent fiscal multipliers estimated by \cite{RZ2018} are almost entirely driven by a single historical event—World War II—in the recession scenario. The top 10\% of weights account for 66\% of the total weighting, and the top 10\% of contributions represent 90\% of the total absolute contributions. In the expansion scenario, the estimate is only slightly less concentrated, combining contributions from the later stages of World War II and the Korean War. Therefore, using estimates of fiscal policy identified by military spending shocks can be a delicate enterprise, as the effective number of observations supporting them is limited and those are concentrated in the distant past.

We also examine the effects of \textbf{global temperature shocks} on world GDP, based on \cite{bilal2024}. The original study finds that a 1°C increase in global temperature leads to a peak GDP decline of 12\% after six years, with negative effects still observable a decade later. \textcolor{black}{These estimates appear strikingly large, particularly in comparison to prior studies documenting modest effects of \textit{local} temperature increases on economic growth \citep{nordhaus1992optimal,dell2012temperature}.} Our analysis highlights key climatological events—such as the 1964 Mount Agung eruption and the late-1990s El Niño and La Niña cycles—as particularly influential on the estimated local projection coefficients, as they coincide (a decade later) with the peak of post-war economic growth in the former case and the height of the China boom in the latter. These observations motivate additional robustness checks, all of which reduce the intensity of the IRF. While medium-term effects remain sizable, our findings indicate that the effects of global temperature shocks on world GDP may be more transitory than depicted in the original study. 

Finally, for \textbf{financial shocks}, we compare linear and nonlinear responses, emphasizing how specific historical events are upweighted to generate the size- and sign-dependent impulse responses commonly observed in the literature. The findings confirm that large negative financial shocks have disproportionate effects on the economy, with the nonlinear model detecting a faster and more pronounced response than the linear counterpart. Perhaps surprisingly, the nonlinear model offers comparatively greater interpretability through its proximity weights. While the linear model aggregates a dense combination of many events, the nonlinear approach isolates a sparse historical structure. Large negative shock episodes form a distinct cluster, while other events are grouped as a set of small positive shocks of similar magnitude.  As a result, the average nonlinear effect is approximately the difference between the means of these two groups, making its historical foundations easier to communicate than those of the linear model.

 \vskip 0.25cm
{\noindent \sc \textbf{Literature $\bullet$ The Econometrics of Local Projections.}} Most existing econometric research has focused on the statistical properties of local projection \textit{estimators}, particularly their bias-variance trade-off relative to VARs \citep{kilian2011reliable,herbst2024bias,li2024local}, their large-sample behavior \citep{plagborg2021local}, and inference \citep{montiel_olea_local_2021}. In this paper, we propose tools to analyze a local projection \textit{estimate}. 

For this different objective, the literature is fairly scant. \cite{dufour_short_1998} and \cite{cloyne2023state} explore the decomposition of estimated impulse responses into direct and indirect effects, offering insights into the plausible transmission mechanisms underlying the responses—an approach reminiscent of VAR studies investigating the nature of such mechanisms \citep{bernanke1997systematic,bachmann2012confidence}.  Additionally, other standard interpretive tools from VAR analysis, such as forecast error variance decompositions, have been adapted for local projections \citep{gorodnichenko_forecast_2020}.  Our framework approaches the problem from another angle.  It does not focus on the mechanism or the economic structure behind the causal effect, but rather on how the estimates aggregates microscopic pieces of narrative evidence into a single number.  Those distinct interpretative goals are not irreconcilable: an ideal estimate should be (i)  backed by an economically plausible mechanism, and (ii) statistically supported by a diverse set of (preferably intuitive) data points. 

Despite their semantic similarity, our approach also differs from historical decompositions in vector autoregressions. The latter attribute realized values to specific shocks, offering a \textit{feature-based} decomposition of the observed dependent variable at a given point in time. In contrast, we explain how \textit{different realizations} of a single shock shape the full-sample \textit{coefficient estimate}, shifting the focus from explaining outcomes to understanding parameter formation.

Recent papers have explored themes related to our investigation, particularly around instrument relevance and robustness concerns when utilizing certain shock series in LPs. \cite{huber2024general} examine the time-varying volatility of instruments, introducing the notion of time-varying relevance. Similarly, in our weight series, the concept of heterogeneous instrument relevance emerges naturally, as some data points contribute far more to the final causal estimate than others. \cite{Barnichon2025} also highlight that the rarity of truly exogenous shocks creates an inevitable trade-off between credibility and efficiency. They show that fiscal multipliers derived from military spending shocks and the effects of monetary policy identified through narrative shocks can be fragile. To address this, they propose a technique to reinforce valid but weak instruments with potentially less valid ones.

\cite{kolesar2024dynamic} provide conditions under which linear LP estimates remain reliable even when the data-generating process (DGP) is nonlinear. They use an alternative weighting scheme to illustrate how different IRF estimates depend on positive vs. negative shocks and on distinct regions of the (non-Gaussian) shock distribution. These methods are inspired by \cite{yitzhaki1996using}’s work (later generalized in \cite{angrist1999empirical}), which showed that linear least squares coefficients can be expressed as weighted sums of slopes between adjacent observations, with weights determined by the empirical distribution of the independent variable. In this paper, we propose a different weighting scheme where the coefficients of interest are weighted sums of the target variables, with weights having the aforementioned intervention-proximity score meaning. This approach applies directly to both linear methods and nonparametric, nonlinear methods, such as those from machine learning. As such, our tools allow for a more proactive foray into the nonlinear world, as they help in understanding nonlinearities and time-variation captured flexibly by models featuring a limited set of functional form assumptions. Additionally, we emphasize time series visualization through cumulative contributions, offering a narrative understanding of the estimate by highlighting key events that drive results.

 \vskip 0.25cm
{\noindent \sc \textbf{Literature $\bullet$ Interpretable Machine Learning.}} The rapidly expanding field of machine learning interpretability seeks to understand what drives outputs from opaque statistical models. Much of this literature emphasizes post-hoc methods for explaining how predictors influence predictions in off-the-shelf ML tools, with notable applications in macroeconomic forecasting by \cite{buckmann2022interpretable} and \cite{anatomy}. We share a conceptual connection with this literature—namely, that a prediction or coefficient estimate is not the end of the story and that  it is equally important to understand how algorithms, whether simple or complex, arrive at their results.

Our approach differs from this set of techniques in one key aspect. Rather than focusing on how features drive predictions, we center on how historical episodes influence a causal estimate. This perspective aligns with two recent  contributions in the field. The first is the work of \cite{dual}, which represents macroeconomic forecasts from various ML algorithms as a portfolio of training target realizations, weighted by proximity scores reflecting similarities between current and past economic conditions. Our paper extends this dual interpretability framework—initially developed for machine learning forecasts (conditional expectations)—to coefficient estimates (differences of conditional expectations). 

The second recent advance considers the decomposition of metrics beyond raw predictions. For example, \cite{anatomy} and \cite{anatomy2} decompose \textit{by features} metrics such as root mean-squared errors, Sharpe ratios, and t-statistics using a fairly involved Shapley values-based scheme. Our contribution aligns with these approaches in that we seek to explain not just predictions, but also other consequential model outputs, such as coefficients. Yet everything we propose comes at little to no additional computational cost beyond estimating the model itself—a stark contrast with notoriously demanding Shapley value-based approaches, such as those in \citet{anatomy2}. 

 \vskip 0.25cm
{\noindent \sc \textbf{Literature $\bullet$ Robust Statistics, Representer Theorems, and Time-Variation.}} There are two distinct strands of mostly theoretical literature that decompose predictions or estimates using observation-specific weights. In both cases, the primary goal is not interpretability. 

The first is influence functions,  a concept from robust statistics  \citep{hampel1974influence,cook1980characterizations} with a certain number of applications in  econometrics, often for inference purposes \citep{chernozhukov2018double,farrell2020deep}. Influence functions assess the fragility of estimates to marginal deviations in the estimation sample and, in this regard, share a common objective with our approach, particularly with cumulative contributions from linear models. However, our reported metrics differ from the traditional influence function paradigm, as we do not estimate the influence of an observation as a pair of target and regressors---a fairly complicated objective in a time series context---but rather the impact of each purified shock. Additionally, we emphasize visualization through time series plots that accumulate into the final estimate, which greatly facilitates the extraction of relevant information based on historical analysis.

The second is the representer theorem from machine learning theory. It states that any minimizer of a regularized empirical risk functional over a reproducing kernel Hilbert space (RKHS) can be written as a finite linear combination of kernel functions evaluated at the training points \citep{kimeldorf1971some, scholkopf2001generalized}. With interpretability rather than reduced computation in mind, this paper extends these insights to the decomposition of coefficients rather than predictions and provides a more explicit economic interpretation of the “finite linear combination.” It does so by integrating the usual dual solution parameters into weights applied to outcomes and proposing a cohesive interpretation of least squares-based coefficients as intervention-proximity estimators in an orthonormal regressor space. 

Although some goals overlap, cumulative contributions also differ from methods that assess robustness to sample choices, such as expanding/rolling window approaches and time-varying parameter models. Contributions mechanically sum to the estimate of interest, and weights have the advantage of being derived from a single consistent model utilizing the entire sample. The visual analysis we employ may also evoke the CUSUM test, where the cumulative sum of residuals in a well-specified model remains within confidence bands derived from Brownian motion. Similarly, we expect cumulative contributions to progress more or less linearly from zero toward the estimate's value, unless the IRF is highly concentrated.

 \vskip 0.25cm
{\noindent \sc \textbf{Literature $\bullet$ Nonlinear and Nonparametric Local Projections.}} There is growing interest in exploring more sophisticated conditional expectations beyond those offered by OLS. For example, tree-based ensembles have been employed in \cite{mumtazimpulse2022}, \cite{paranhos_how_2024}, and \cite{hauzenberger_machine_2024} to obtain nonparametric LP estimates, moving past the traditionally dominant focus on manually defined nonlinearities \citep{RZ2018,paul_time-varying_2020,gonccalves2024state,nonparametricLP_2024}. However, as is often the case with off-the-shelf machine learning algorithms, it remains unclear what these methods uncover that cannot be captured by more explicitly specified models. Our contribution lies in clarifying which events are emphasized or de-emphasized in fancier estimates, offering a way to both probe and communicate ML-based results.

\vskip 0.25cm
{\noindent \sc \textbf{Outline.}} The paper is organized as follows. Section \ref{sec:dual} introduces {the historical foundation of LPs}, offers two interpretations of the weights, and extends these insights to ML-based LPs. Section \ref{sec:emp} presents empirical results across four applications. Section \ref{sec:con} concludes.

\section{Historical Foundations of Local Projections}\label{sec:dual}

We briefly review LPs here and refer the reader to \cite{jorda2024local} for a more detailed discussion. First, define \( y_t \) as the outcome variable of interest at time $t$. The controls, denoted by \( Z_t \), form a vector comprising exogenous or predetermined variables, including lagged values of both the outcome \( y_t \) and the policy intervention \( s_t \). Local projections provide a method to estimate the following population object. Formally, the impulse response is defined as
\begin{align}\label{diffrepre}
\text{IRF}_{s \to y}(h, \delta) \equiv \mathbb{E}[y_{t+h} \mid s_t = s^0 + \delta; Z_t] - \mathbb{E}[y_{t+h} \mid s_t = s^0; Z_t],
\end{align}
where \( h = 0, 1, \dots, H \), and \( \delta \) represents the size of the intervention.  This expression captures how an intervention at time \( t \) influences the average outcome \( y_{t+h} \) at a future horizon \( h \), relative to a baseline scenario with no intervention. A common convention is to normalize \( \delta = 1 \) or a one standard deviation of $s_t$,  with a reference scenario of \( s^0 = 0 \). Although shock size and signs do not matter in linear local projections, they do in a more general nonlinear nonparametric context, which we consider in Section \ref{sec:ml}. 


Provided \( s_t \) is exogenous, as in the case of preidentified shock series like monetary or fiscal policy shocks, the \textit{linear} LP can be easily estimated through a loop of ordinary least squares (OLS) regressions. The LP of \( y_{t+h} \) on \( s_t \) can be estimated with the regression
\begin{align}\label{linearlp} 
y_{t+h} =\xi_h + \beta_h s_t + \gamma_h' Z_t + v_{t+h}, \quad h = 0, 1, \ldots, H,
\end{align}
where \textcolor{black}{$\xi_h$ defines the intercept, $\gamma_h$ the coefficients on  controls, and $v_{t+h}$ the idiosyncratic error term. It follows that}  \( \text{IRF}_{s \to y}(h,1) = \beta_h \) by definition. By virtues of OLS being a linear estimator in $\boldsymbol{y}_{h}$, we have that
\begin{align}\label{linearlp2} 
\widehat{\text{IRF}}_{s \to y}(h, 1) =  \hat{\beta}_h  = \underbrace{[(\boldsymbol{X}' \boldsymbol{X})^{-1} \boldsymbol{X}']_{\{2, : \}}}_{\boldsymbol{w}} \boldsymbol{y}_{h} 
\end{align}
where $\boldsymbol{X} \in \mathbb{R}^{T \times K}$ and $\boldsymbol{y}_{h} \in \mathbb{R}^{T}$ is the training data.\footnote{\color{black} As a notational convention, bold symbols denote quantities spanning multiple observations, whether vectors or matrices. Vectors associated with a single observation (e.g., \( Z_t \)) are left unbolded.} It is assumed that \( s_t \) is placed in the second column of $\boldsymbol{X} = [\boldsymbol{1} \, \boldsymbol{s} \, \boldsymbol{Z}]$, {\color{black} so the notation $\{2,:\}$ refers to the second row of the projection matrix—i.e., the weights associated with \( s_t \)}.  By properties of projection and annihilator matrices, $\boldsymbol{1}' \boldsymbol{w} =0$ because $ \boldsymbol{X}$ includes an intercept.   The expression of $ \hat{\beta}_h $ as a weighted average of outcomes follows directly from the least squares formula and is noted, for instance, in \cite{davidson2004econometric}.\footnote{ Throughout the paper, all standard linear regression assumptions required for the consistency of point estimates are assumed to hold, as is customary in local projection applications. This includes coefficient stability, shock exogeneity, and inclusion of all relevant regressors. Importantly, these assumptions are \textit{not} required for the decomposition itself to hold: the decomposition is purely mechanical and applies to any OLS estimate, regardless of its causal interpretation. What these assumptions ensure, rather, is that the object being decomposed is worth decomposing at all.   Moreover, issues such as serial correlation and heteroskedasticity are typically addressed only at the inference stage, as they do not compromise consistency. Our decomposition of \textit{point estimates} follows suit, reflecting how evidence is aggregated across the sample under OLS  assumptions.} Our key contribution is the introduction of $\boldsymbol{w}$ for interpretation purposes, along with its (eventual) alternative interpretation as proximity scores, which establishes a connection to machine learning methods that are linear in $\boldsymbol{y}_{h} $.

In macroeconomic and financial forecasting studies, it is common to analyze forecasting performance (often measured by MSE) using time series plots of cumulative sums of squared errors \citep{WelchGoyal2008}. These plots help identify which historical episodes contribute most to a model's under- or out-performance. We can adopt a similar approach here, but for our estimated dynamic causal effects, as given by: 
\begin{align}\label{linearlp3} 
\widehat{\text{IRF}}_{s \to y}(h, 1) = \sum_{t=1}^T \underbrace{w_{t}y_{t+h}}_{c_{th}} = \sum_{t=1}^T c_{th} \, . 
\end{align}
Therefore, beyond the visualization of $w_t$ as a time series, we can examine \textit{contributions} \( c_{th} \) as a cumulative time series, \( C_{\tilde{T}h}  = \sum_{t=1}^{\tilde{T}} c_{th} \), which converges to \( \hat{\beta}_h \) when \( \tilde{T} = T \). As demonstrated in the empirical section, this straightforward plotting method proves highly effective in diagnosing empirical issues, such as the price puzzle \textcolor{black}{when estimating the effects of monetary policy shocks}, and in developing appropriate solutions. 

While much attention has been paid to the (small- or large-sample) statistical properties of this \textit{estimator} (e.g., \citealt{kilian2011reliable,plagborg2021local,gonccalves2024state,li2024local,herbst2024bias}), comparatively little effort has been devoted to anatomizing \textit{estimates} obtained from a fixed sample. The vector $\boldsymbol{w}$ and its byproducts are especially informative in this regard, arguably more so than $\gamma_h$ in \eqref{linearlp}, whose meaning becomes increasingly viscous beyond $h=1$. Moreover, the latter quantity changes with $h$ whereas the former is fixed across horizons  for a linear LP.\footnote{Note that $\boldsymbol{w}_{h} =\boldsymbol{w}$ for all $h$ is a specificity of OLS because $\boldsymbol{X}$  does not change. However, in a more general context of a machine learning model, where variables are selected and nonlinear transformations are created separately for each $h$, this will not be the case.}  Lastly, $\boldsymbol{w}$ has a more universal appeal than regression coefficients as it applies just as much to models with no coefficients. Indeed, many modern nonlinear conditional mean functions estimators are nonlinear in $\boldsymbol{X}$ but linear in $\boldsymbol{y}_{h}$, meaning they can also be represented as the product of a $\boldsymbol{w}$ and $\boldsymbol{y}_{h}$. The question is, how should we think about $w_t$ as a time series?

\subsection{$w_t$ as a Standardized and Purified Shock Series}\label{sec:purify}

In many linear applications, $w_t$ must, by construction, closely resemble $s_t$. This is particularly true if $s_t$ is a well-defined shock series, i.e., one that is unpredictable by its own lags or those of other variables, and orthogonal to other variables at time $t$. If $s_t$ satisfies these conditions exactly and is normalized to have variance 1 (as is often the case), we can write:
\begin{align}\label{eq_sum_simple}
\widehat{\text{IRF}}_{s \to y}(h, 1) = \frac{1}{T} \sum_{t=1}^T s_t y_{t+h}.
\end{align}
By construction, $\widehat{\text{IRF}}_{s \to y}(h, 1)$ is an unbiased estimator for ${\text{IRF}}_{s \to y}(h, 1)$, and so is each term $c_{th} = s_t y_{t+h}$. In this special case, the OLS estimator effectively reports an equally-weighted ensemble of unbiased but high-variance estimators of the causal effect. In this sense, it is directly analogous to cumulative squared error plots that sum to the mean-squared error.

That being said, in practice, $s_t$ often requires further purification, which justifies the inclusion of controls in the regression. Furthermore, incorporating relevant predictors—most notably, lags of the outcome variable—into the LP can lead to substantial efficiency gains by reducing the variance of the residuals. It is still possible  in this environment to interpret $ \boldsymbol{w} $ in the spirit of \eqref{eq_sum_simple}, leveraging a variation of the Frisch-Waugh-Lovell (FWL) theorem to derive a simple summation formula. 

A numerically equivalent $\hat{\beta}_{h}$ can be obtained by running a regression where $y_{t+h}$ remains the target, and the sole regressor is $\tilde{s}_t$, the residual from regressing $s_t$ on all other variables in the original regression, ${Z}_t$. Since this second regression is univariate, we have:  
\begin{align}\label{eq_sum_fwl0}
\boldsymbol{w} = \frac{\boldsymbol{s}' \boldsymbol{M}_{\boldsymbol{Z}} }{\boldsymbol{s}' \boldsymbol{M}_{\boldsymbol{Z}} \boldsymbol{s}} = \frac{\tilde{\boldsymbol{s}}}{\tilde{\boldsymbol{s}}'\tilde{\boldsymbol{s}}}
\end{align}
where $\boldsymbol{M}_{\boldsymbol{Z}}$ is the annihilator matrix projecting onto the space orthogonal to that spanned by $\boldsymbol{Z}$. 
We can define
\begin{align}\label{eq_sum_fwl}
{s}_t^* \equiv \frac{\tilde{s}_t}{{\text{Var}}(\tilde{s}_t)}
\end{align}
and recover a summation formula analogous to \eqref{eq_sum_simple}:
\begin{align}\label{eq_sum_fwl3}
\hat{\beta}_{h} = \frac{1}{T} \sum_{t=1}^T {s}_t^* y_{t+h},
\end{align}
where ${s}_t^*$ is the standardized, purged (from $\boldsymbol{Z}$) version of the shock $s_t$. Thus, with ${s}_t^*$ as the new shock series, all the insights related to \eqref{eq_sum_simple} remain valid. Specifically, $\hat{\beta}_{h}$ continues to represent an equally weighted ensemble of unbiased, high-variance estimates of the causal effects, and these individual contributions can be analyzed as they cumulatively form $\hat{\beta}_{h}$.

Note that, unlike the standard statement of the FWL theorem, the formulation in  \eqref{eq_sum_fwl0}--\eqref{eq_sum_fwl3} does not residualize \( y_{t+h} \) with respect to \( \boldsymbol{Z} \), preserving a more direct connection to the simpler case in \eqref{eq_sum_simple}, which also applies upweighting and downweighting to raw \( y_{t+h} \). Due to the idempotency of \( \boldsymbol{M}_{\boldsymbol{Z}} \), this representation is numerically equivalent to the classical double residualization approach.

In this paper's framework, shutting down a particular $c_{th}$ corresponds to turning off \textit{the purified shock} ${s}_t^*$ at a specific time. This is a reasonable experiment because purified shocks are uncorrelated by construction. Thus, the approach does not evaluate the influence of a specific time point, but rather the influence of \textit{each realized intervention}. In this sense, it is more palatable than the alternative perspective, where ${s}_t^*$ captures the influence of the heavily autocorrelated $y_{t+h}$ on $\hat{\beta}_h$. In Section \ref{sec:links}, we provide a more detailed discussion on the connection between our approach and influence functions, leverage, as well as expanding and rolling window estimation.

The formulation of $\hat{\beta}_{h}$ as a weighted sum of outcomes using purified shocks, as in \eqref{eq_sum_fwl3}, opens a window onto broader methodological debates surrounding the identification of macroeconomic shocks.  When $s_t$ is truly exogenous, we expect $s_t^* \approx s_t$, so the weights $w_t$ closely track the original shock series. This alignment is especially plausible in applications where shocks are themselves constructed via regressions on control variables \citep{romerromer2004,aruoba2024identifying}. However, the literature has questioned the exogeneity of narrative shocks, with \citet{nakamura2018high} emphasizing their potential predictability. Others argue that in-sample shock predictability may arise mechanically due to overfitting. These concerns, originally raised in the context of auxiliary shock regressions \citep{nakamura2018identification}, extend naturally to LPs  with an expansive set of controls leading to an overly aggressive orthogonal projection matrix $\boldsymbol{M}_{\boldsymbol{Z}}$ in \eqref{eq_sum_fwl0}. What \eqref{eq_sum_fwl3} offers in this context is a diagnostic tool: by comparing the resulting $w_t$ and $c_{th}$ series, researchers can evaluate how different shock-cleaning strategies reweight the underlying data in constructing dynamic causal effect estimates. 

\subsection{$w_t$ as a Proximity Weights Series}\label{sec:prox}

It seems generally intuitive to estimate the effects of a given policy by assigning greater weight to realized outcomes of time points (or individuals) that have experienced similar interventions. In this section, we show that the weights $w_t$ estimated by OLS follow this principle. Beyond its conceptual appeal, this alternative proximity interpretation has the additional advantage of relying not on coefficients, but on conditional forecasts that are linear in $\boldsymbol{y}_h$. Indeed, revisiting the \textit{difference-of-expectations representation} in \eqref{diffrepre} allows us to reinterpret \( \boldsymbol{w}_t \) and \( c_{th} \) in a way that extends beyond the least-squares framework. While most machine learning methods lack bona fide coefficients, many still yield predictions that are linear in the target, much like OLS.

Specifically, for OLS, we can compute $\widehat{\text{IRF}}_{s \to y}(h, \delta)$ in two equivalent ways:
\begin{align}\label{forecast_diff}
\widehat{\text{IRF}}_{s \to y}(h, \delta) = \delta \hat{\beta}_h = \hat{y}_{\tau + h}^\delta - \hat{y}_{\tau +h}^0 = \underbrace{(\boldsymbol{w}^\delta - \boldsymbol{w}^0)}_{\boldsymbol{w}} \boldsymbol{y}_{h} \, ,
\end{align}
where \( \tau \) is a hypothetical new observation, and \( \boldsymbol{w}^\mathcal{S} = X_\tau^\mathcal{S} (\boldsymbol{X}' \boldsymbol{X})^{-1} \boldsymbol{X}' \), with \( X_\tau^\mathcal{S} \equiv [ {1} \,  \, \mathcal{S} \, {Z}_\tau  ] \in \mathbb{R}^{K} \) for \( \mathcal{S} \in \{0, \delta\} \), representing the two dosage scenarios. In \cite{OLSA}, OLS is reframed as a similarity-based estimator, where out-of-sample \textit{predictions} are expressed as weighted linear combinations of the training values of the target variable, with weights interpreted as proximity scores. As shown in \eqref{forecast_diff}, local projection estimates can be computed—numerically equivalently—as the difference between two conditional forecasts at time $\tau$ for horizon $h$, given economic conditions ${Z}_\tau$.\footnote{In linear models, $\widehat{\text{IRF}}_{s \to y}(h, \delta)$ is invariant to the context ${Z}_\tau$, rendering the IRF itself time-invariant. Thus, we can freely assign values to ${Z}_\tau$ in linear models, as these contributions are canceled out in \eqref{forecast_diff}. However, retaining ${Z}_\tau$ anyhow allows for an alternative interpretation of $\boldsymbol{w}$ as a measure of proximity, which becomes particularly meaningful when conditional expectations are nonlinear and the invariance to ${Z}_\tau$ no longer holds.} If the weight vectors $\boldsymbol{w}^\delta$ and $\boldsymbol{w}^0$ each carry a proximity interpretation, then so should their difference, $\boldsymbol{w}$.

Building on this intuition and substituting in the OLS solution for ``out-of-sample'' predictions, the weight vector $\boldsymbol{w}$ admits the following equivalent representation:
\begin{align}\label{forecast_diff_alt}
\boldsymbol{w} = X_\tau^\delta (\boldsymbol{X}' \boldsymbol{X})^{-1} \boldsymbol{X}' - X_\tau^0 (\boldsymbol{X}' \boldsymbol{X})^{-1} \boldsymbol{X}'.
\end{align}
This expression, representing the difference between out-of-sample projection matrices based on the \textit{primal} solution to least squares, gains additional interpretive depth through the numerically equivalent \textit{dual} solution. Although rarely utilized in the standard \( K < T \) setting, the OLS estimator can be equivalently expressed as:
\[
\hat{\boldsymbol{b}}_{\text{OLS}} = (\boldsymbol{X}'\boldsymbol{X})^{-1} \boldsymbol{X}'\boldsymbol{y}_h = \boldsymbol{X}' (\boldsymbol{X} \boldsymbol{X}')^{+} \boldsymbol{y}_h \, ,
\]
leveraging the properties of the Moore-Penrose pseudoinverse, denoted by \( (\boldsymbol{X}\boldsymbol{X}')^{+} \). The generalized inverse satisfies algebraic conditions that guarantee a solution even when the matrix \( \boldsymbol{X} \) has fewer predictors than observations (\( K < T \)).

In the context of standard local projections estimated via OLS, this reformulation enables a \text{proximity-based interpretation} of \( \hat{\beta}_h \), where:
\[
\boldsymbol{w} = \underbrace{(X_\tau^\delta \boldsymbol{X}' - X_\tau^0 \boldsymbol{X}')}_{\text{\scriptsize Proximity Differential}} \quad \times  \underbrace{(\boldsymbol{X}\boldsymbol{X}')^{+}}_{\text{\scriptsize Proximity Denominator}}.
\]
This highlights that \( \boldsymbol{w} \) is the product of two components, both quantifying similarity between data points. Specifically, \( X_\tau^\mathcal{S} \boldsymbol{X}' \) is a vector stacking inner products \( \langle X_\tau^\mathcal{S}, X_t \rangle \), which measure the \textit{alignment} (or proximity) in \( \mathbb{R}^K \) between the hypothetical scenario at time \( \tau \) and each observation \( t \) in the estimation sample. \( \boldsymbol{X}\boldsymbol{X}' \) is the \textit{Gram matrix}, encoding pairwise proximities among all observations in the estimation sample within \( \mathbb{R}^K \).

While this formula offers no advantage for estimating local projections---being more computationally demanding than standard least squares while yielding identical results---it provides an insightful interpretation of \( \boldsymbol{w} \). At this point, the question is how should we think of the effect of \( (\boldsymbol{X} \boldsymbol{X}')^{+} \), the ``Proximity Denominator''.

\subsubsection{A Proximity-Based Representation of $\hat{\beta}_h$: the Uncorrelated $X_t$ Case}\label{sec:uncorr}

Let us first abstract from the Proximity Denominator and assume that \( \frac{1}{T}\boldsymbol{X}'\boldsymbol{X} = I_P \), which also implies \( (\boldsymbol{X}\boldsymbol{X}')^{+} = T \). In this case, the primal and dual solutions for OLS collapse to the same formula:
\begin{align}\label{eq:this}
\hat{\beta}_{h} &= \frac{1}{\delta T} \sum_{t=1}^{T} \left[  \left\langle X_{\tau}^\delta  , X_t  \right\rangle -  \left\langle X_{\tau}^0  , X_t  \right\rangle   \right] y_{t+h} \, .
\end{align}
Here, \( w_t = \frac{\left\langle X_{\tau}^\delta , X_t  \right\rangle -  \left\langle X_{\tau}^0 , X_t  \right\rangle}{\delta T} \), reflecting the proximity differential between conditions at time \( t \) and the projected conditions for a dose of \( \delta \) versus 0.  In other words, \( w_t \) increases if imposing \( s_\tau = \delta \) on \( X_{\tau}^{\mathcal{S}} \) brings economic conditions at \( \tau \) (represented as a point in \( \mathbb{R}^K \)) closer to those of \( t \) than when setting \( s_\tau = 0 \). \textcolor{black}{We reintroduce the more general dosage parameter $\delta$ in place of 1. Since the proximity interpretation extends to machine learning methods, where the magnitude and sign of $\delta$ affect impulse response estimates, retaining $\delta$—though redundant in linear models—is preferable.}  

We can further explore the geometry by rewriting the expression in terms of cosine similarity:
\begin{align}\label{eq:cos}
\hat{\beta}_{h} &= \frac{1}{\delta T} \sum_{t=1}^{T} \left[ \| {X}_t\|_2 \left( \| X_{\tau}^\delta \|_2 \cos(\theta^\delta_{\tau t}) - \| X_{\tau}^0 \|_2 \cos(\theta^0_{\tau t}) \right) \right] y_{t+h}.
\end{align}
Here, $\theta^{\delta}_{\tau t}$ denotes the angle (in degrees) between the vectors $X_{\tau}^{\delta}$ and $X_t$, so that $\cos(\theta^{\delta}_{\tau t}) = \frac{\left\langle X_{\tau}^\delta  , X_t  \right\rangle}{\| X_{\tau}^\delta \|_2 \| X_t \|_2}$, with an analogous definition for $\theta^{0}_{\tau t}$. The notation $\| \cdot \|_2$ refers to the standard Euclidean norm, that is, the square root of the sum of squared components. From \eqref{eq:cos}, we see that both alignment (through the cosine terms) and vector magnitudes matter. Intuitively, if \( s_t \approx \delta \), the vector \( X_{\tau}^\delta \) should be closer to \( X_t \) in \( \mathbb{R}^K \) than \( X_{\tau}^0 \). In a linear model with uncorrelated predictors, this geometric perspective may appear unnecessarily complex, as the expression simplifies to \eqref{eq_sum_simple}, where the focus reduces to one-dimensional similarity. Still, it provides a useful foundation for correlated predictors.

\subsubsection{A Proximity-Based Representation of $\hat{\beta}_h$: the Correlated $X_t$ Case}

Incorporating the ``Proximity Denominator'' is relatively straightforward, though it makes the interpretation of proximity differentials slightly more abstract. The primal formulation,
\[
\hat{\beta}_{h} = \frac{1}{\delta} \left[ X_\tau^\delta (\boldsymbol{X}' \boldsymbol{X})^{-1} \boldsymbol{X}' - X_\tau^0 (\boldsymbol{X}' \boldsymbol{X})^{-1} \boldsymbol{X}' \right] \boldsymbol{y}_h,
\]
can be rewritten as
\[
\hat{\beta}_{h} = \frac{1}{\delta} \sum_{t=1}^{T} \left[ \langle F_{\tau}^\delta, F_t \rangle - \langle F_{\tau}^0, F_t \rangle \right] y_{t+h},
\]
where \( (\boldsymbol{X}' \boldsymbol{X})^{-1} = \boldsymbol{U} \Lambda^{-1} \boldsymbol{U}' \) is the eigen-decomposition of the precision matrix, \( F_t = X_t \boldsymbol{U} \Lambda^{-\frac{1}{2}} \), and \( F_\tau^{\mathcal{S}} = X_{\tau}^{\mathcal{S}} \boldsymbol{U} \Lambda^{-\frac{1}{2}} \) for \( \mathcal{S} \in \{0, \delta\} \). Here, \(\Lambda\) is the diagonal matrix of eigenvalues of $(\boldsymbol{X}' \boldsymbol{X})^{-1}  $, and \(\boldsymbol{U}\) is the matrix of corresponding eigenvectors. These matrices ensure that \( F_t \) and \( F_\tau^\mathcal{S} \) are orthogonal representations of \( X_t \) and \( X_{\tau}^{\mathcal{S}} \), respectively. Formally, since \( \boldsymbol{F}'\boldsymbol{F} = I_P \) holds, we are effectively brought back, within a new embedding, to the uncorrelated scenario described in Section \ref{sec:uncorr}. \textcolor{black}{This extends \cite{OLSA}'s insights on OLS predictions to the interpretation of OLS coefficients—that is, the primal solution is a dual solution in disguise.}

Thus, \( \hat{\beta}_{h} \) is an estimator that upweights outcomes \( y_{t+h} \) in periods where there is a large gap between the pairwise proximity scores of evaluation factors \( F_\tau^\delta \) and economic conditions at the time of \( F_t \), relative to those of \( F_\tau^0 \) and \( F_t \). The factors \( F_\tau^\delta \) and \( F_\tau^0 \) represent two scenarios that differ solely in their dosage choice \textit{in the original space}, meaning that multiple entries of \( F_\tau^\delta \) and \( F_\tau^0 \) differ, unlike in the case of uncorrelated \( X_t \)'s.

Importantly, transforming to \( F_t \) and \( F_\tau^\mathcal{S} \) entails no loss of information, as the number of factors equals the number of predictors. The role of the ``Proximity Denominator'' is thus to express the same information in a space where summing simple inner products serves as a proper scoring rule for quantifying proximity. In essence, the matrices \( (\boldsymbol{X}' \boldsymbol{X})^{-1} \) or \( (\boldsymbol{X} \boldsymbol{X}')^{+} \) can be viewed as pre-processing steps applied to the data \( X_t \), allowing the simplified formula for uncorrelated predictors in \eqref{eq:this} to hold.

\subsubsection{OLS as a Nearest-Interventions Estimator}

 First, note that \( X_{\tau}^\delta - X_\tau^0 = [ \, 0 \, \delta \,  \,  0 \,  \,  0 \, \dots \, 0  \,  \,  ]  \), which nullifies the other components of the inner product, regardless of the values of \( Z_t \). Returning to the \textit{uncorrelated} case for illustrative purposes, this implies that we can rewrite  \eqref{eq:this} more compactly as:
\begin{align}\label{eq:this2}
\hat{\beta}_{h} &= \frac{1}{\delta T} \sum_{t=1}^{T} \left\langle X_{\tau}^\delta - X_{\tau}^0, X_t - X_t^0 \right\rangle y_{t+h},
\end{align}
where \( X_t^0 \) represents any fictitious counterfactual scenario for time \( t \), where \( s_t \) is forced to 0 instead of its observed value. In this formulation, \( \hat{\beta}_{h} \) upweights outcomes \( y_{t+h} \) corresponding to episodes where the intervention and no-intervention differentials most closely resemble those being evaluated at time \( \tau \). This reveals that \( w_t \) is effectively a series of proximity scores between the intended intervention and historical interventions.

In the more realistic case with correlated \( X_t \), a similar representation emerges by defining \( F_t^0 = \boldsymbol{0} \) as the counterfactual scenario at time \( t \):
\begin{align}
\hat{\beta}_{h} &= \frac{1}{\delta} \sum_{t=1}^{T} \left\langle F_{\tau}^\delta - F_{\tau}^0, F_t - \boldsymbol{0} \right\rangle y_{t+h}.
\end{align}
This highlights that \( w_t \), obtained through OLS, assigns higher values to periods where the set of interventions (\( \mathcal{I}_t = \boldsymbol{0} \rightarrow F_t \)) most closely aligns with the intended evaluation (\( \mathcal{I}_{\tau} = F_{\tau}^0 \rightarrow F_{\tau}^\delta \)). This can be represented in two equivalent forms:
\begin{align}\label{eq:cosinepolicy}
\hat{\beta}_{h} &= \frac{1}{\delta} \sum_{t=1}^{T} \left\langle \mathcal{I}_{\tau}, \mathcal{I}_t \right\rangle y_{t+h} = \frac{\|\mathcal{I}_{\tau} \|_2 }{\delta} \sum_{t=1}^{T} \cos(\theta^{\mathcal{I}}_{\tau t}) \|  \mathcal{I}_t \|_2 \, y_{t+h}.
\end{align}
Here, $w_t$ corresponds to the inner product proximity scores between $\mathcal{I}_{\tau}$ and $\mathcal{I}_t$, which are not scale-invariant and embed information about both vector alignment and magnitudes. The second representation, based on cosine similarity, explicitly separates these components: $w_t$ is expressed as the product of the scale-invariant alignment in $\mathbb{R}^K$ (cosine similarity) and the magnitudes (measured by the $l_2$-norm) of the two interventions being compared. Intuitively, a large $\| \mathcal{I}_{\tau} \|_2$ (e.g., due to a high $\delta$) scales $\hat{\beta}_{h}$ linearly upward. A large $\| \mathcal{I}_t \|_2$ indicates that observation $t$ is particularly informative and carries leverage, being located farther from the origin $\boldsymbol{0}$ in the orthonormal feature space.\footnote{\color{black} In the uncorrelated case, \( \| \mathcal{I}_\tau \|_2 = |\delta| \), so the expression simplifies to \( \hat{\beta}_h = \text{sign}(\delta) \times \sum_{t=1}^{T} \cos(\theta^{\mathcal{I}}_{\tau t}) \| \mathcal{I}_t \|_2 \, y_{t+h} \).} In Section \ref{sec:emp}, we report the two components separately ($\cos(\theta^{\mathcal{I}}_{\tau t})$ and $| \mathcal{I}_t |_2$) to disentangle the respective roles of alignment and scale in the baseline monetary policy results.

{\color{black} All in all, this discussion highlights  that traditional regression-based estimators are not so different—at least at a conceptual level—from those explicitly formulated as matching or similarity-based estimators, as in, e.g.,  \citet{angrist_semiparametric_2018}.}

\subsection{Machine Learning Models}\label{sec:ml}

Local projections are typically framed as the difference between two conditional expectations. However, their application is almost exclusively linear, and when nonlinearities are explored, these are often rigidly specified, such as employing two regimes based on a predefined state variable. While such approaches can capture average nonlinear effects \citep{kolesar2024dynamic}, they may overlook finer nuances in the heterogeneity of treatment effects. These nuances could be crucial for policymakers implementing actual policy {interventions}. As emphasized in modern studies focusing on treatment effect heterogeneity \citep{athey2019ensemble}, causal inference is fundamentally a \textit{prediction} about the effects of policy. Hence, nonlinearity and time-variation are not just ornaments, they are about better predicting the effect of one's actions in varying economic environments.

However, one not-so-minor roadblock is the apparent lack of transparency of fully nonparametric methods, especially those pertaining to the class of machine learning algorithms.  The good news is that our interpretation tool applies to most ML algorithms the same way it applies to linear models. Indeed, nonlinearities formulated in the covariate space simply imply a different notion of proximity, and a different \textit{linear combination} of $\boldsymbol{y}_h$.  Hence,  the anatomies of linear and nonparametric  IRFs can be contrasted by comparing their weights and contributions. 

\subsubsection{Random Forest Local Projections}\label{sec:rflp}


Among modern ML techniques, Random Forest (RF) is highly popular because it allows for complex nonlinearities, handles high-dimensional data, is mostly exempt from overfitting, and requires very little tuning to do so. As such, it has proven effective in predicting many things, including macroeconomic aggregates \citep{medeiros2019,chen2019off,GCLSS2018}.  In the context of more structural macroeconomic analysis, Random Forests and related tree-based ensemble methods have been applied to generalized time-varying parameters \citep{MRFjae} and to local projection \citep{mumtazimpulse2022,paranhos_how_2024}. A review of the algorithm can be found in Appendix \ref{sec:RFrev}.

What is particularly relevant for our application is that proximity weights are straightforward to recover in Random Forest---which can be seen as an adaptive nearest neighbors algorithm---and are well-studied in predictive modeling (see \citealt{dual} and references therein). The ability to express RF predictions as a convex combination of \(\boldsymbol{y}\) was noted by \cite{lin2006random} and involves simple operations on estimation outputs. While our nonlinear exploration in this paper focuses on Random Forests for simplicity, the proposed decomposition is also applicable to any machine learning algorithm that generates predictions linear in the target variable, such as kernel methods, boosting, and neural networks, by utilizing tools introduced in \cite{dual}.

First, we define the nonparametric local projection as the difference between two RF projections:
\begin{align}\label{forecast_diff22}
\widehat{\text{IRF}}_{s \to y}(h, \delta, \tau) = \hat{y}_{\tau + h}^\delta - \hat{y}_{\tau + h}^0,
\end{align}
for a fictitious new observation \( \tau \). Unlike in linear models, \( \widehat{\text{IRF}}_{s \to y}(h, \delta, \tau) \) depends on the context vector \( Z_\tau \) and is not necessarily proportional to \( \delta \), i.e., \( \widehat{\text{IRF}}_{s \to y}(h, \delta, \tau) \neq \delta \times \widehat{\text{IRF}}_{s \to y}(h, 1, \tau) \). The resulting impulse response is time-varying, with the sign and size of the shock influencing the outcome in a non-trivial way. To obtain the average nonlinear effect, we can average over many \( Z_\tau \), with natural candidates for draws being those from the training sample (\( Z_{1:T} \)). This produces a ``dataset'' of IRFs, allowing us to report summary statistics such as the mean or explore heterogeneity through clustering analyses.

\subsubsection{Retrieving Weights in Random Forest}\label{sec:retreive}

As evident from the difference of conditional expectations formulation in \eqref{forecast_diff}, the weights for the two predictions can be recovered separately. This allows us to express \( \widehat{\text{IRF}}_{s \to y}(h, \delta, \tau) = \boldsymbol{w}_{\tau h} (\delta) \boldsymbol{y}_h \), where \( \boldsymbol{w}_{\tau h} (\delta) \) represents the difference between two sets of weights. 

Importantly, \( \boldsymbol{w}_{\tau h} (\delta) \) depends on the context (indexed by \( \tau \)), the horizon \( h \), and the dosage $\delta$.  Unlike linear models, which apply the same out-of-sample projection matrix to \( \boldsymbol{y}_h \) (e.g., \( X_\tau^\delta (\boldsymbol{X}' \boldsymbol{X})^{-1} \boldsymbol{X}' \)), the weights in RF are specialized for each horizon. This is because the algorithm performs feature selection and captures nonlinearity independently for each horizon. 

The Random Forest prediction for \(\tau\) at horizon $h$ for scenario $\mathcal{S} \in \{ 0, \delta\}$ is 
\[ \hat{y}_{\tau + h}^{\mathcal{S}} = \frac{1}{B} \sum_{b=1}^B \mathcal{T}_{bh}(X_\tau^{\mathcal{S}}) \]
where \(B\) is the number of trees in the RF. Each single tree \(\mathcal{T}_{b h}\) delivers a prediction according to the following rule:
\[
\mathcal{T}_{bh}(X_\tau^{\mathcal{S}}) = \frac{1}{\sum_{t=1}^T I\left(t \in \mathcal{P}_{bh}(X_\tau^{\mathcal{S}})\right)}\sum_{t=1}^T y_{t+h} I\left(t \in \mathcal{P}_{bh}(X_\tau^{\mathcal{S}})\right) = \sum_{t=1}^T w_{b\tau t h}^{\mathcal{S}} y_{t+h}
\]
where \(\mathcal{P}_{bh}\) is the partition implied by the tree and its conditioning information for observation \(\tau\), and \(w_{b\tau t h}^{\mathcal{S}} = \frac{I\left(t \in \mathcal{P}_{bh}(X_\tau^{\mathcal{S}})\right)}{\sum_{t'=1}^T I\left(t' \in \mathcal{P}_{bh}(X_\tau^{\mathcal{S}})\right)}\). Then, by reordering sums, we get the desired representation
\[
\hat{y}_{\tau+h}^\mathcal{S} = \frac{1}{B} \sum_{b=1}^B \mathcal{T}_{bh}(X_\tau) = \frac{1}{B} \sum_{b=1}^B \sum_{t=1}^T w_{b\tau t h}^{\mathcal{S}} y_{t+h} = \sum_{t=1}^T \underbrace{\frac{1}{B} \sum_{b=1}^B w_{b\tau t h}^{\mathcal{S}}}_{w_{\tau t h}^{\mathcal{S}}} y_{t+h} = \boldsymbol{w}_{\tau h}^\mathcal{S} \boldsymbol{y}_h
\]
for intervention scenario $\mathcal{S}$, context indexed by $\tau$, and the horizon $h$. In words, to generate $\boldsymbol{w}_{\tau h}^\mathcal{S} $ in the RF case, one can follow these steps: determine which leaf observation $\tau$ falls into for a given tree (based on its $X_\tau^\mathcal{S}$), identify the corresponding in-sample observations for the leaf and their weights (calculated as \(\sfrac{1}{\text{leaf size}}\)), assign these weights to the relevant in-sample observations (\(w_{b \tau t h}\)), and then aggregate these ``votes'' across all trees in the ensemble.

\vskip 0.15cm

{\noindent \sc \textbf{Context-Specific IRF.}} To obtain the weights for a \( \tau \)-specific RF-based local projection, we compute
\begin{align}\label{w_rf}
\boldsymbol{w}_{\tau h}(\delta) = \boldsymbol{w}_{\tau h}^\delta - \boldsymbol{w}_{\tau h}^0 \, ,
\end{align}
and cumulative contributions can then be calculated and visualized as for linear models. The notation emphasizes that these weights are more heterogeneous and adaptive compared to OLS, as the portfolio choice of \( y_{t+h} \)'s depends on the context, the sign and size of the intervention, and the horizon under study.

\vskip 0.15cm

{\noindent \sc \textbf{Unconditional IRF.}} To compute weights for average effects across multiple economic conditions (\( Z_{\tau}, \, \tau = 1, \dots, T \)), we rearrange the sums as follows
\begin{align*}
\widehat{\text{IRF}}_{s \to y}(h, \delta) &= \frac{1}{T} \sum_{\tau=1}^T \widehat{\text{IRF}}_{s \to y}(h, \delta, \tau) \\ &= \frac{1}{T} \sum_{\tau=1}^T \sum_{t=1}^T w_{\tau t h} (\delta) y_{t+h} \\
&= \sum_{t=1}^T y_{t+h} \underbrace{\sum_{\tau=1}^T \frac{w_{\tau t h}(\delta)}{T}}_{w_{th}(\delta)} \\ &= \boldsymbol{w}_h(\delta) \boldsymbol{y}_h.
\end{align*}
Even the average nonlinear effect can differ from OLS. First, nonlinear algorithms like RF perform internal feature engineering and variable selection, resulting in effective controls that may differ from the original inputs. Second, the effect depends on \( \delta \) in a non-trivial way \textcolor{black}{ with potentially disproportional responses}. Comparing weights for different choices of $\delta$ to that of OLS makes explicit from where more sophisticated (and less transparent) conditional mean algorithms derive their estimates.

\subsubsection{Summarizing and Understanding Estimated Nonlinearities}\label{sec:rfsum}

As we now know, the nonparametric approach produces an IRF that depends on the input \( Z_{\tau} \). Exploring this causal effect heterogeneity (or time-variation) is worthwhile, as in this flexible model, the intervention's impact depends on the sign, size, \textit{and} the economic conditions at the time of implementation. To investigate these variations, we can adopt an intermediate approach between examining all \( \tau \)-specific IRFs and the full-sample average: analyzing clusters or groups.  For a predefined cluster of observations \( \mathcal{C}_j \), the IRFs can be decomposed similarly to the average:
\begin{align}\label{forecast_diff33}
\widehat{\text{IRF}}_{s \to y}(h, \delta, \tau \in \mathcal{C}_j) 
&= \frac{1}{|\mathcal{C}_j|} \sum_{\tau \in \mathcal{C}_j} \widehat{\text{IRF}}_{s \to y}(h, \delta, \tau) \\
&= \frac{1}{|\mathcal{C}_j|} \sum_{\tau \in \mathcal{C}_j} \sum_{t=1}^T w_{\tau t h} (\delta) y_{t+h} \\
&= \sum_{t=1}^T y_{t+h} \underbrace{\sum_{\tau \in \mathcal{C}_j} \frac{w_{\tau t h} (\delta)}{|\mathcal{C}_j|}}_{w_{tj h} (\delta)} \\
&= \boldsymbol{w}_{j h} (\delta) \boldsymbol{y}_h \, .
\end{align}
In the empirical section, we apply a simple \( k \)-means algorithm to the ``IRF dataset'', a \( T \times H \) matrix, treating the time dimension as observations and \( H \) (horizons) as characteristics. This allows us to assess whether, for a fixed \( \delta \), there is substantial heterogeneity in policy effects or if a single homogeneous group dominates. This approach can be viewed as a data-driven method for defining regimes, contrasting with the manually defined regimes typically used in local projections \citep{auerbach2012measuring,RZ2018}.

\subsection{Two Impulse Response Concentration Statistics}\label{sec:derivatives}


To quantify the concentration of IRF estimates, we introduce two summary statistics. The first measures the share of total absolute weights concentrated in the top \( Q\% \), indicating how much the IRF relies on a narrow subset of observations. This statistic primarily reflects the distribution of purified shocks and how many observations the model interprets as proximate to the policy intervention. The second applies the same approach to contributions, incorporating both the weighting structure and extreme realizations of the dependent variable. Since IRFs are linear aggregates of contributions, this metric helps diagnose whether concentration arises from weights, extreme outcomes, or both.  

Concentration in distributions can be measured in various ways. We adopt a concentration ratio that captures the proportion of the total sum of absolute weights contributed by the top \( Q\% \) of observations:  
\[
\text{\textbf{WC}}(\hat{\beta}_h) = \frac{\sum_{q=1}^{\lfloor Q \times \sfrac{T}{100} \rfloor} |w_{hq}|}{\sum_{q=1}^T |w_{hq}|}
\]
where \( q \) indexes absolute weights, ordered from largest (\( q=1 \)) to smallest (\( q=T \)). This approach parallels well-known measures of income and wealth inequality, offering an intuitive interpretation—for instance, stating that 50\% of \( \hat{\beta}_h \) is driven by just 5\% of observations indicates substantial concentration. For linear models, this measure is independent of the horizon \( h \), as it depends solely on the regressors. However, in machine learning settings, where predictors and their nonlinear transformations are optimized for each horizon, weighting structures can vary, making \text{\textbf{WC}} horizon-dependent.  

The concentration of contributions is defined similarly:  
\[
\text{\textbf{CC}}(\hat{\beta}_h) = \frac{\sum_{q=1}^{\lfloor Q \times \sfrac{T}{100} \rfloor} |c_{hq}|}{\sum_{q=1}^T |c_{hq}|}
\]
where \( q \) indexes absolute contributions, ordered from largest (\( q=1 \)) to smallest (\( q=T \)). Unlike weights, contributions depend on the response variable \( y_{t+h} \), making \text{\textbf{CC}} inherently horizon-dependent, regardless of model linearity.  

These measures are reported throughout our empirical analysis (see Section~\ref{sec:emp}) and should be closely monitored by researchers assessing the robustness of IRF estimates.

\subsection{Relationship to Other Methods for Assessing Influence}\label{sec:links}


What thought experiment does the $c_{th}$ path represent, and how does it compare to other methods evaluating the influence of specific data points on an estimate?

\vskip 0.15cm
{\noindent \sc \textbf{Link to Influence Functions.}} Shutting down specific $c_{th}$ terms and comparing the resulting IRF to the original estimate naturally evokes a connection to influence function analysis—a key tool from robust statistics \citep{hampel1974influence,cook1980characterizations}. The classic treatment in econometrics is provided by \cite{newey1994large}. Influence functions also play a central role in the theoretical analysis of semiparametric estimators \citep{ichimura2022influence} and in causal inference with machine learning models \citep{chernozhukov2018double,farrell2020deep}. A popular application is found in \cite{Firpo2009}, which uses specific influence functions of individual observations as targets to estimate unconditional quantile regressions.

The relationship is formalized as:  
\begin{align}\label{IFlink} 
c_{th} &=  w_t y_{t+h} \\
 &=  w_t (\hat{y}_{t+h} + \hat{\nu}_{t+h}) \\
&=  w_t \hat{y}_{t+h} +  \underbrace{w_t \hat{\nu}_{t+h}}_{\approx \text{IF}_t} \, ,  
\end{align}
where ${\text{IF}_t}$ denotes the influence function of data point $t$. By properties of projection matrices, we have  
\begin{align}
\sum_{t=1}^T w_t \hat{y}_{t+h} &= [(\boldsymbol{X}' \boldsymbol{X})^{-1} \boldsymbol{X}']_{\{2, : \}} \boldsymbol{P}_{\boldsymbol{X}} \boldsymbol{y}_h=  \hat{\beta}_h, \quad \quad \text{and} \label{ooo}\\ \quad \sum_{t=1}^T w_t \hat{\nu}_{t+h} &= [(\boldsymbol{X}' \boldsymbol{X})^{-1} \boldsymbol{X}']_{\{2, : \}} \boldsymbol{M}_{\boldsymbol{X}} \boldsymbol{y}_h = 0.
\end{align}
The first term, shown in \eqref{ooo}, captures the systematic component—how the shock variable contributes to each in-sample prediction, given the model’s estimated structure. Under the purified and standardized shock interpretation of $w_t$ from Section \ref{sec:purify}, we obtain  
\[
w_t \hat{y}_{t+h} = \frac{(s_t^*)^2}{T} \hat{\beta}_h.
\]  
Thus, when $s_t^*$ is far from the origin—regardless of sign, given the model's linearity—it yields a substantial contribution in the direction of the aggregate effect $\hat{\beta}_h$.  

The second term corresponds to the thought experiment: ``What if this data point had been different?''  In a cross-sectional context with \textit{i.i.d.} data—thus excluding dynamic features such as lags of $s_t$ and $y_t$—this term is exactly the formula for the influence function of data point $t$ on $\hat{\beta}_h$. However, with dependent data featuring autocorrelated $\hat{\nu}_{t+h}$ and time-stamped information from $t$ being utilized in multiple roles through a distributed lag structure, the interpretation becomes murkier. 

Therefore, a key challenge in this setting is evaluating the influence of a single \textit{observation} (i.e., the $[y_{t+h} \enskip \boldsymbol{X}_t]$ pair). Several studies have addressed the limitations posed by time series data, extending influence function methodologies to account for the temporal structure of information—whether in influencing predictions \citep{ghosh2020approximate,zhang2024timeinf} or autoregressive parameters \citep{kunsch1984infinitesimal}. This is indeed a valuable endeavor, as the usual $\text{IF}_t$ calculations do not directly correspond to the original thought experiment of removing observation $t$ from the estimation.

However, we consider these concerns peripheral to our main line of investigation for two reasons. First, we interpret $c_{th}$ as representing a thought experiment focused on shutting down interventions. Specifically, turning off a particular $c_{th}$ corresponds to deactivating the \textit{purified shock} $s_t$ (or ${s}_t^*$) at a given point in time. This is a reasonable experiment because purified shocks are, by construction, uncorrelated. From an influence function perspective, this approach does not assess the influence of a specific time point or a particular realization of $y_{t+h}$, but rather the influence of \textit{each realized intervention}. In this sense, it is more conceptually sound than the alternative view, where ${s}_t^*$ represents the influence function of the highly autocorrelated $y_{t+h}$ on $\hat{\beta}_h$. The key distinction is that it is consistent with the data to momentarily switch on or off a non-serially correlated object—an assumption that does not hold in the opposite scenario. 

Second, our visualization emphasizes cumulative contributions, which significantly smooth the $c_{th}$ values through an integration-like filtering effect. Consequently, $\sum_{t=1}^{\tilde{T}} c_{th}$ inherently incorporates substantial neighboring information. Since $c_{th}$ can be fairly noisy due to the influence of ${s}_t^*$, examining raw $c_{th}$ values offers limited value. As a compromise, moving average filters could be applied, effectively incorporating neighboring data points and enhancing the interpretability of $c_{th}$, and we report one such example for monetary policy shocks in the appendix.

\vskip 0.15cm
{\noindent \sc \textbf{Link to Leverage.}}  \cite{davidson2004econometric} provide an extensive geometric treatment of leverage and influence in a regression context, emphasizing that, by construction, regression coefficients are weighted averages of the target variable. The relationship between the contribution term $c_{th}$ and classical leverage statistics is
\begin{align}\label{IFlink} 
    c_{th} =   w_t \hat{y}_{t+h}+  \underbrace{w_{t}\hat{\nu}_{t+h}}_{= (1 - \boldsymbol{H}_{tt}) \, \text{LOOI}_t} \, ,
\end{align}
where $\text{LOOI}_t$ represents the leave-one-out influence of data point $t$, and $\boldsymbol{H}_{tt}$ is the $t$-th diagonal element of the hat matrix (i.e., the leverage of observation $t$). The term $(1 - \boldsymbol{H}_{tt})$ serves as a \textit{deflator} that adjusts for the fact that observations with high leverage exert greater control over their own fitted values, thereby reducing their residual variance.

The term $(1 - \boldsymbol{H}_{tt})$ enters this formulation because, unlike the classical influence function---which measures the sensitivity of parameter estimates to {infinitesimal perturbations}---the discrete influence function captures the {finite effect} of specific observations on the estimator. In classical influence analysis, the infinitesimal nature of the perturbation implicitly accounts for leverage effects through linear approximation, obviating the need for an explicit $(1 - \boldsymbol{H}_{tt})$ adjustment.  However, unlike influence functions, which mechanically sum to 0, and our cumulative $c_{th}$ series, which converges to $\hat{\beta}_h$, $\text{LOOI}_t$  do not sum to 0. Therefore, they are less appealing for our objective of decomposing $\hat{\beta}_h$.

Lastly, it is important to note that the direct connection to influence functions and leverage statistics holds primarily in linear models. For various machine learning models, such as Random Forests, which we will employ in our empirical section, no closed-form influence function exists. This is due, in part, to the absence of a well-defined gradient—since Random Forests rely on a greedy algorithm—and the fact that in-sample residuals are misleading due to the well-documented benign overfitting phenomenon \citep{belkin2019reconciling,TBTP}.

\vskip 0.15cm

{\noindent \sc \textbf{Link to Expanding- and Rolling-Window Schemes.}}  The formulations below compare cumulative contributions with the coefficients estimated from expanding window (EW) and rolling window regressions with window size $\omega$ (RW). Each of these quantities provides distinct insights into how different periods drive estimation results:
\begin{align}
C_{\tilde{T}h}  &= \sum_{t=1}^{\tilde{T}} [(\boldsymbol{X}' \boldsymbol{X})^{-1} \boldsymbol{X}']_{\{2, \, t \}} y_{t+h}, \tag{Cumulative Contributions} \label{eq:CumulativeAttribution} \\
\hat{\beta}_{\tilde{T}h}^{\text{EW}} &= \sum_{t=1}^{\tilde{T}} [(\boldsymbol{X}_{\{1: \tilde{T}, \, : \}}' \boldsymbol{X}_{\{ 1: \tilde{T}, \, : \}})^{-1} \boldsymbol{X}']_{\{2, \, t \}} y_{t+h}, \tag{Expanding Window} \label{eq:ExpandingWindow} \\
\hat{\beta}_{\tilde{T}h}^{\text{RW}} &= \sum_{t=\tilde{T}-\omega}^{\tilde{T}} [(\boldsymbol{X}_{\{(t-\omega): t, \, : \}}' \boldsymbol{X}_{\{(t-\omega): t, \, :  \}})^{-1} \boldsymbol{X}']_{\{2, \, t \}} y_{t+h}. \tag{Rolling Window} \label{eq:RollingWindow}
\end{align}
The closest conceptual parallel is between cumulative contributions ($C_{\tilde{T}h}$) and expanding window estimates ($\hat{\beta}_{\tilde{T}h}^{\text{EW}}$). Both generate sequences that eventually converge to the full-sample coefficient. Indeed, the cumulative sum of $\Delta \hat{\beta}_{\tilde{T}h}^{\text{EW}}$'s eventually lands on $\hat{\beta}_{h}$. These differences, however, stem from a series of different models rather than a single consistent specification.  Therefore, they cannot be interpreted as the product of a coherent set of weights (seen either as a purified shock or proximity scores) and the target.

Mathematically, the distinction between these approaches lies in how the covariance matrix is handled. The cumulative contribution approach \textit{pre-multiplies} $\boldsymbol{X}'$ by a precision matrix $(\boldsymbol{X}' \boldsymbol{X})^{-1}$ estimated from the full sample. That is, it learns the multivariate distribution of $\boldsymbol{X}$ using the entire dataset. This allows contributions to be accumulated based on a single consistent proximity structure rather than an ever-changing one that depends only on the information available at time $\tilde{T}$. Additionally, cumulative contributions tend to be more stable, particularly in the earlier parts of the sample, where expanding windows use a possibly unreliable precision matrix $(\boldsymbol{X}_{\{1:\tilde{T}, \, : \}}' \boldsymbol{X}_{\{1:\tilde{T}, \, : \}})^{-1}$ due to limited sample size.

The one-sided rolling window estimator $\hat{\beta}_{\tilde{T}h}^{\text{RW}}$ does not produce a sequence that converges to a final coefficient using the full sample. Still, like many time-varying parameter methods (such as those employing random walks), it conveys information into how different periods influence the estimated coefficients. This is directly relevant to our analysis, as periods with significantly different $\hat{\beta}_{\tilde{T}h}^{\text{RW}}$ will exert influence on $\hat{\beta}_{\tilde{T}h}$, attracting it in their direction. However, achieving an assessment that is both highly localized (as $c_{th}$ is) and accurate presents a challenge due to the inherent bias-variance trade-off in time-varying parameter models. More often than not, acceptable variance levels come at the cost of overly smooth estimates, which ultimately reduce the degree of localization \citep{GC2019,MRFjae}.

In summary, the cumulative contribution approach differs fundamentally from both expanding and rolling window methods. Its attached weights are derived from a single consistent model and may be more efficient in handling the covariance structure of $\boldsymbol{X}_t$.

\section{Empirical Applications}\label{sec:emp}

We consider four applications in this section to demonstrate the usefulness of visualizing contributions and weights within our proposed framework. The first examines the effects of monetary policy shocks on the economy, with a particular focus on inflation, a variable often associated with puzzles in the literature. The second focuses on fiscal policy, analyzing its impact through estimates that, while consequential, exhibit considerable variation across studies. The third explores the effect of global temperature shocks on world GDP, a question which has naturally received a lot of attention lately. Finally, the fourth investigates the impact of financial shocks on the economy, emphasizing how incorporating nonlinearities in the conditional mean can alter results compared to linear specifications.

We present results through cumulative contributions and weights. As previously discussed, by construction, for a given horizon, we have $\hat{\beta}_h = \sum_{t=1}^{\tilde{T}} c_{th}$ when $\tilde{T}$ reaches $T$, the full length of the estimation sample. We refer to the resulting time series as the \textbf{evidence curve} of $\hat{\beta}_h$.  Weights are reported either in their raw form or smoothed using a 6-month moving average, which helps highlight specific events. While smoothing may seem unexpected for series typically considered white noise, it is important to note that when $y_{t+h}$ is particularly smooth---as with cumulative IRFs at longer horizons---a sequence of small, consecutive shocks of the same sign can exert an influence comparable to that of a single large shock. A moving average filter can elicit such pockets of instrument relevance.

\subsection{Monetary Policy}\label{sec:mp}

We decompose the response of the economy to monetary policy shocks, focusing on the price level and unemployment, which are central to the Fed's dual mandate. We compare two shock series with estimates dating back to at least 1970: the  \citet[][henceforth R\&R]{romerromer2004} monetary policy shocks and Cholesky-identified shocks from a VAR. The latter is well-known for frequently producing the so-called price puzzle, where monetary tightening counterintuitively leads to higher inflation. Using our framework, we investigate the underlying reasons for this puzzle (or its absence) and assess whether our explanation aligns with existing interpretations in the literature. 

The main specification includes inflation, unemployment, industrial production, federal funds rate, S\&P500 stock market index, and the respective shock series. The latter is obtained from i) R\&R and ii) a Cholesky VAR estimated with the listed variables (excluding the shock) in the same order. Data is taken from FRED-MD \citep{mccrackenng}, with all variables including 12 lags. We report cumulative IRFs throughout, including 84\% confidence bands. To account for serial correlation in the error term, we apply the Newey-West correction to the standard errors \citep{NeweyWest1987}.\footnote{\color{black} If a less passive approach to heteroskedasticity—such as stochastic volatility or generalized least squares (GLS)—were used, the contributions and weights would adjust accordingly. This raises an intriguing possibility: although feasible GLS is often set aside in favor of OLS with corrected inference, visualizing the implied data weighting may help users better understand and accept differences in point estimates.}

\vspace{0.25em}

\begin{figure}[t]
  \caption{\normalsize{Responses to Contractionary Monetary Policy Shocks}} \label{fig:mp_contrib_var_romer}
    \centering
    \vspace*{-0.2cm}
    \includegraphics[width=\textwidth, trim = 0mm 0mm 0mm 0mm, clip]{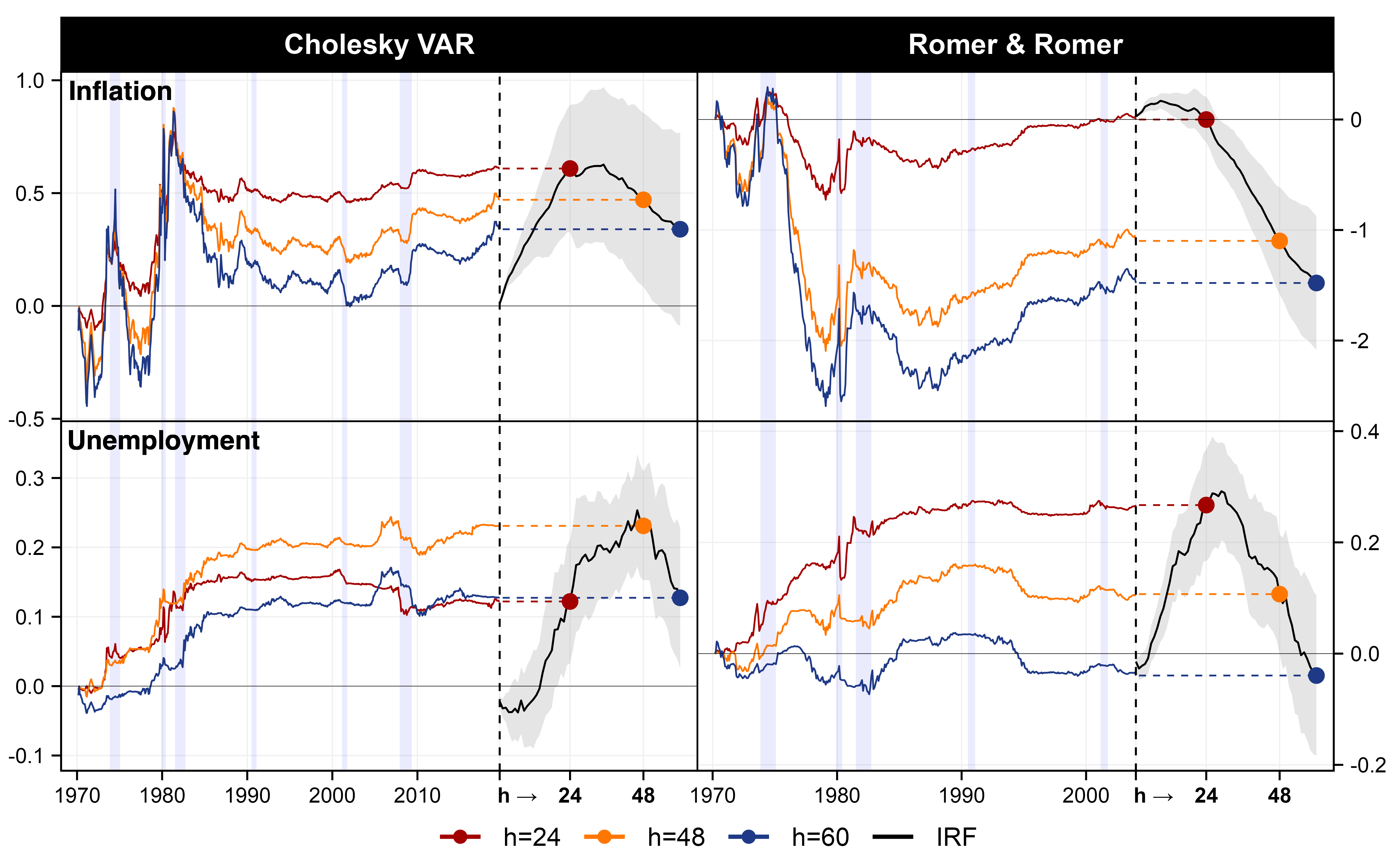}
     
    \begin{threeparttable}
    \centering
    \vspace*{-0.6cm}
    \begin{minipage}{\textwidth}
      \begin{tablenotes}[para,flushleft]
    \setlength{\lineskip}{0.2ex}
    \notsotiny 
  {\textit{Notes}: The plot shows responses of US inflation and unemployment to a one standard deviation contractionary monetary policy shock. The left panels are identified via a 4-variable VAR with short run restrictions. The estimation sample starts in 1970M3 and ends in 2019M9, marked by the vertical dashed line. The right panels use instrumental variable identification with the instrument from \cite{romerromer2004} with the sample starting in 1970M3 and ending in 2004M1. Solid colored lines present cumulative contributions $\sum_{t=1}^T c_{th}$, summing to the final predicted value shown as dots. Full impulse response functions are shown in black with Newey-West 84\% confidence bands. Lavender shading corresponds to NBER recessions.}
    \end{tablenotes}
  \end{minipage}
  \end{threeparttable}

\end{figure}

{\noindent \sc \textbf{General Observations.}}  In Figure \ref{fig:mp_contrib_var_romer}, we observe that both the Cholesky VAR and R\&R shocks produce broadly similar impulse response functions for unemployment. The evidence supporting the responses 24 months after the shock is concentrated between 1970 and the mid-1980s. Within this 15-year span, the evidence is relatively evenly distributed, indicating that the IRFs are not driven by any single event.

\begin{figure}[t]
  \caption{\normalsize{Proximity Scores for Monetary Policy Shocks}} \label{fig:mp_weights_var_romer}
    \centering
    \vspace*{-0.2cm}
    \includegraphics[width=\textwidth, trim = 0mm 0mm 0mm 0mm, clip]{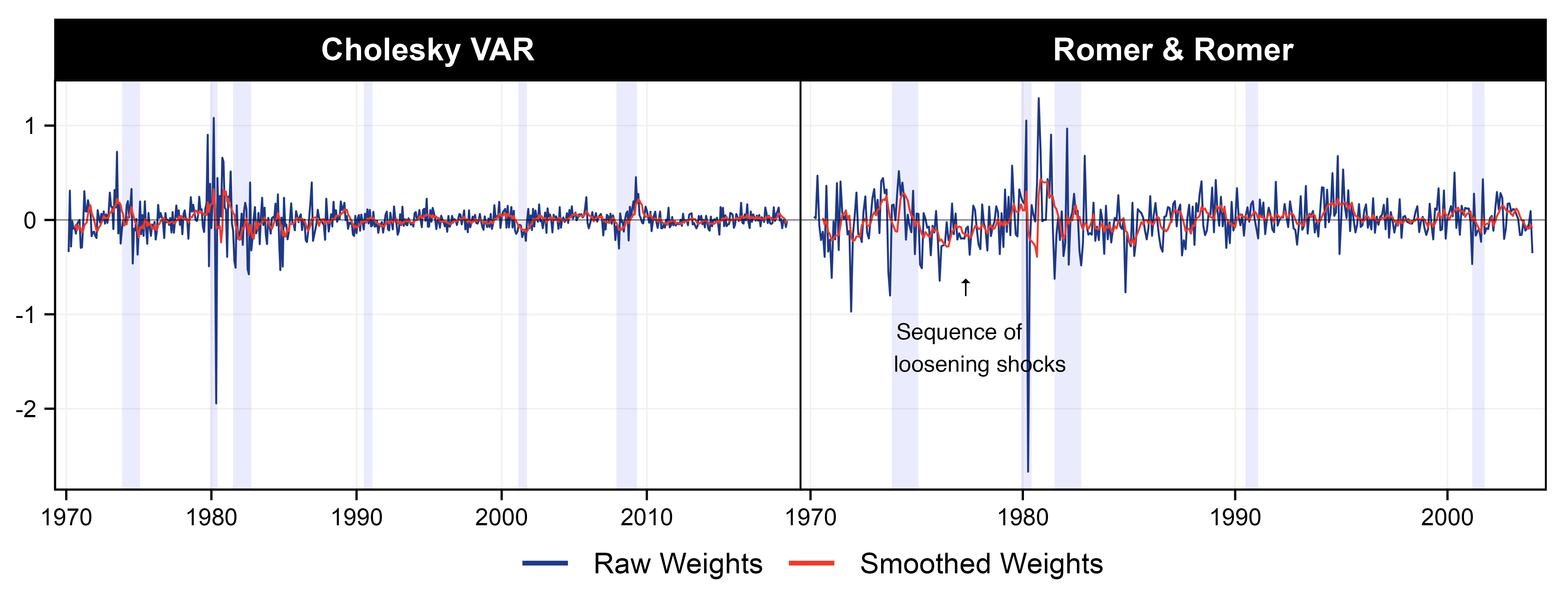}
     
    \begin{threeparttable}
    \centering
    \vspace*{-0.6cm}
    \begin{minipage}{\textwidth}
      \begin{tablenotes}[para,flushleft]
    \setlength{\lineskip}{0.2ex}
    \notsotiny 
  {\textit{Notes}: The plot displays proximity scores between the intended policy intervention and past interventions. Smoothed weights are averaged over six months. The left panel is estimated with a 4-variable VAR with short run restrictions. The estimation sample starts in 1970M3 and ends in 2019M9. The right panel use instrumental variable identification with the instrument from \cite{romerromer2004} with the sample starting in 1970M3 and ending in 2004M1. Lavender shading corresponds to NBER recessions.}
    \end{tablenotes}
  \end{minipage}
  \end{threeparttable}

\end{figure}

In Figure \ref{fig:mp_contrib_var_romer}, the Cholesky VAR and R\&R shocks yield markedly different impulse response functions for inflation, with the heart of the disagreement centered on the interpretation of the 1970s. The Cholesky VAR shocks appear to conflate the dynamics of stagflation, attributing episodes of both high unemployment and high inflation to monetary policy. As shown in Figure \ref{fig:mp_weights_var_romer}, the Cholesky VAR assigns a series of weights ($w_t$) in tightening territory during both the first and second inflation spirals of the 1970s. While there are sporadic mild negative contributions from 1976 to 1978, these are largely offset by the strong attribution of the inflation spirals to monetary tightening.

In contrast, the R\&R shocks exhibit a clear sequence of consistently negative weights preceding the second inflation spiral, as seen in Figure \ref{fig:mp_weights_var_romer}. Although R\&R shocks also reflect some counterintuitive evidence for the 1974–1975 period, this is substantially counterbalanced by the prolonged stretch of monetary loosening (purified) shocks from 1976 to 1978, occurring 48 to 60 months before inflation peaks at its post-war summit. \textcolor{black}{This period of monetary expansion marks the final years of the Burns era and the start of G. William Miller's tenure as Federal Reserve chairman, who was known for his dovish stance against inflation \citep{romerromer2004}.} These latter contributions completely offset the various temporal regions where evidence pushes in the direction of a puzzle. Finally, it is also worth noting that these "correct" results seem to be based almost entirely on evidence from \textit{ loosening of} shocks. We will investigate this claim further within our nonlinear specification, which can accommodate this and other subtleties absent from linear models.

As emphasized by \eqref{eq:cosinepolicy} and the more basic form in \eqref{eq:cos}, the weights—under their proximity-based interpretation—can be split into two components: the scale-invariant proximity term, captured by vector alignment through $\cos(\theta_{\tau t}^{\mathcal{I}})$, and the magnitude of the encoded economic conditions at time $t$, denoted by $F_t$. We report both quantities separately in Figure \ref{fig:mp_cosine}. The smoothed sequence of the cosine term sharpens the delineation of monetary policy regimes, revealing periods of exogenous expansion or contraction with greater clarity. A striking example is the prolonged expansionary stance of the mid-1970s, which emerges distinctly in the R\&R specification but is notably absent under the Cholesky VAR. This suggests that, when seeking historical interventions analogous to recent monetary loosening, the R\&R shocks anchor strongly to the mid-1970s, whereas the Cholesky-based view does not. As for the norm term, $||F_t||_2$, it remains relatively stable across the sample. Still, two episodes receive systematically elevated weights—sometimes nearly double—regardless of directional alignment: the 1973-1975 recession and the early 1980s twin recessions. This reflects the intensity of economic conditions during those periods, which leads to higher weights regardless of directional alignment. 

\vspace{0.25em}

{\noindent \sc \textbf{Addressing Puzzles.}} Both \cite{balke1994understanding} and \cite{hanson2004price} find evidence suggesting that the price puzzle, when identified using a recursive Cholesky ordering, is associated primarily with the 1959–1979 sample period. The evidence curves in Figure \ref{fig:mp_contrib_var_romer} concur, and provide a \textit{why}. Those closely track inflation and interest rate movements in the 1970s, suggesting that peak contributions are likely driven by reverse causality and omitted variables, particularly those related to inflation expectations. Incorporating additional information into the local projection using VAR-identified shocks could correct the “faulty” attribution of the two inflationary spirals of the 1970s to monetary tightening. Ultimately, such an adjustment could yield an IRF with the correct sign.

\begin{figure}[t!]
  \caption{\normalsize{Price Puzzle Resolutions for Monetary Policy Shocks}} \label{fig:mp_contrib_var_var}
    \centering
    \vspace*{-0.2cm}
    \includegraphics[width=\textwidth, trim = 0mm 0mm 0mm 0mm, clip]{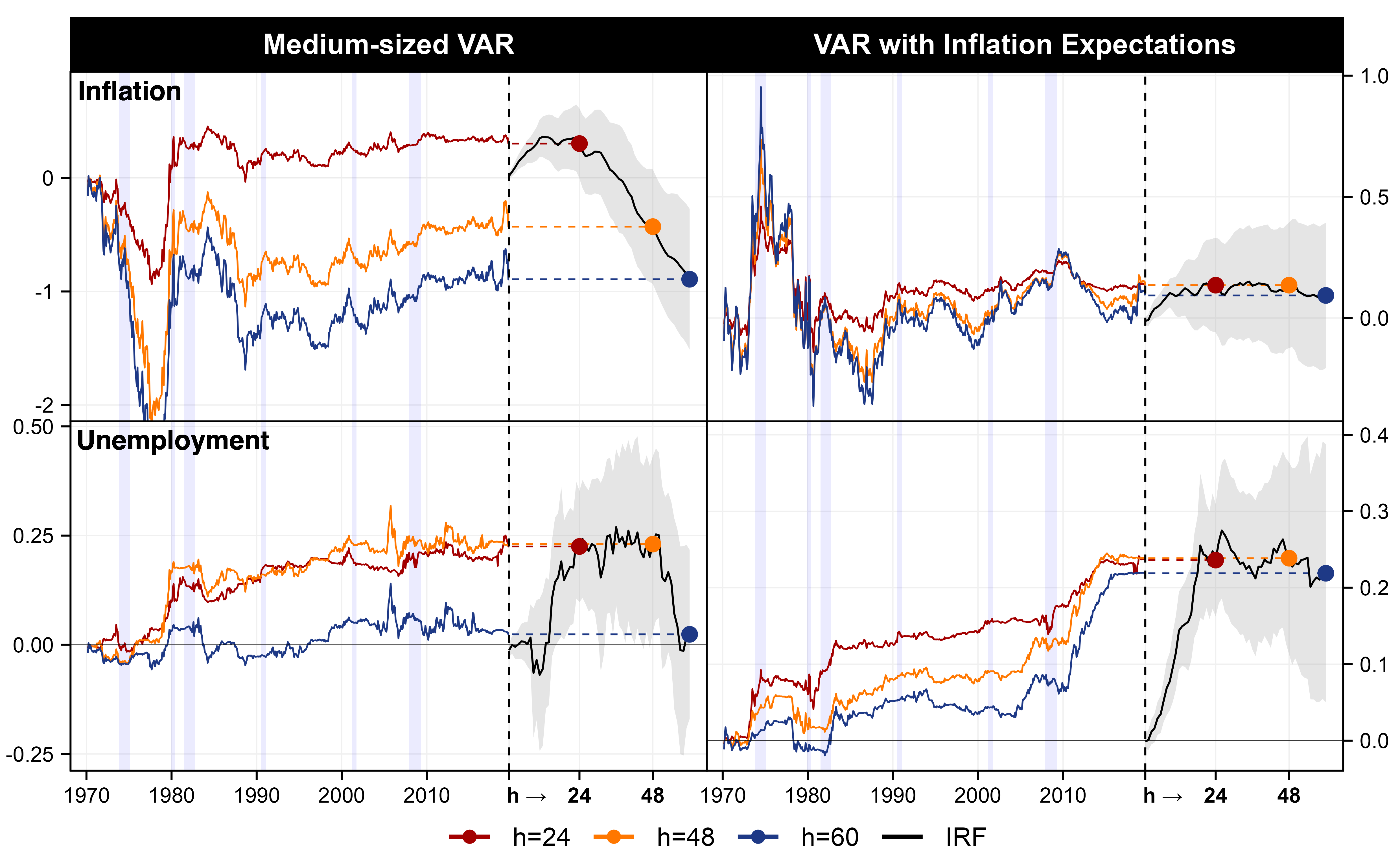}
     
    \begin{threeparttable}
    \centering
    \vspace*{-0.6cm}
    \begin{minipage}{\textwidth}
      \begin{tablenotes}[para,flushleft]
    \setlength{\lineskip}{0.2ex}
    \notsotiny 
  {\textit{Notes}: The plot shows responses of US inflation and unemployment to a one standard deviation contractionary monetary policy shock. The left panels are identified via a medium-sized VAR with short run restrictions. The right panels use inflation expectations in addition to the 4-variable VAR identified via short run restrictions. In both cases, the estimation sample starts in 1970M3 and ends in 2019M9, marked by the vertical dashed line. Solid colored lines present cumulative contributions $\sum_{t=1}^T c_{th}$, summing to the final predicted value shown as dots. Full impulse response functions are shown in black with Newey-West 84\% confidence bands. Lavender shading corresponds to NBER recessions.}
    \end{tablenotes}
  \end{minipage}
  \end{threeparttable}

\end{figure}

In Figure \ref{fig:mp_contrib_var_var} and \ref{fig:mp_weights_var_var} (Appendix), we evaluate two approaches: (1)  enriching the local projection with a medium-sized VAR information set, and (2) directly proxying for inflation expectations.\footnote{The medium-sized VAR includes 21 variables with 18 lags, covering various sectors of the economy (FRED-MD mnemonics: RPI, DPCERA3M086SBEA, INDPRO, CE16OV, CUMFNS, UNRATE, PAYEMS, HOUST, S.P.500, FEDFUNDS, T10YFFM, AAAFFM, BAAFFM, WPSFD49207, WPSID62, OILPRICEx, CPIAUCSL, CPIULFSL, PCEPI, CES0600000008, M2SL). For the specification with inflation expectations, we include the median expected price change during the next 12 months from the Michigan Survey of consumers and the median forecast for the following quarter and the current year from the Survey of Professional Forecasters.}  Accounting for inflation expectations, incorporating forward-looking variables, or including additional variables to address omitted variable bias are established solutions to the price puzzle, as discussed in \cite{christiano1996effects, christiano1999monetary}, \cite{bernanke2005measuring}, and \cite{brissimis2006forward}.  

We find that the former is sufficient to ``correct'' the IRF, while the latter is not.  For the ``Medium-sized VAR'' information set case, the now-correctly signed IRF closely resembles that of the R\&R shocks and highlights the same dominant negative contributors---namely, the sequence of monetary loosening shocks from 1976 to 1978. However, the period leading up to the 1980 recession continues to exert a strong contribution in the opposite direction, dampening the overall magnitude of the estimated response.

In the case of the expectations proxies, the second cluster of counterintuitively signed contributions is nullified; however, the first cluster persists, continuing to push the IRF upward. Combined with the weak contributions from loosening shocks in 1977, this adjustment proves insufficient to tilt the IRF into negative territory. This aligns with the findings of \cite{hanson2004price}---where many forward-looking variables are considered and do not prove sufficient to correct for the price puzzle, especially for the period before the 1980s.

Regardless of the specific econometric specification, an important pattern emerges. A correctly signed and significant IRF of the effect of monetary policy shocks on inflation, using data extending back to the 1970s, consistently requires weights that capture the clear sequence of loosening shocks from 1976 to 1978. Without these, neither the augmented VAR specifications nor the simpler R\&R shock-based models yield intuitive IRFs.


Figure \ref{fig:mp_contrib_ma} (Appendix) illustrates an alternative perspective on contributions. Rather than evaluating evidence curves through the cumulative contributions of training observations, this approach uses moving averages of contributions, introducing less smoothing. The most influential contributions originate from the 1970s and 1980s. While the Cholesky VAR fails to capture the negative contributions of the mid-1970s loosening shocks, both the R\&R specification and the medium-sized VAR clearly identify them. These negative contributions are substantial enough to offset other influences from the same period that exhibit counterintuitive signs.

\vspace{0.25em}

\begin{figure}[t!]
  \caption{\normalsize{Nonlinear Responses of Inflation to Monetary Policy Shocks}} \label{fig:mp_contrib_rf}
    \centering
    \vspace*{-0.2cm}
    \includegraphics[width=\textwidth, trim = 0mm 0mm 0mm 0mm, clip]{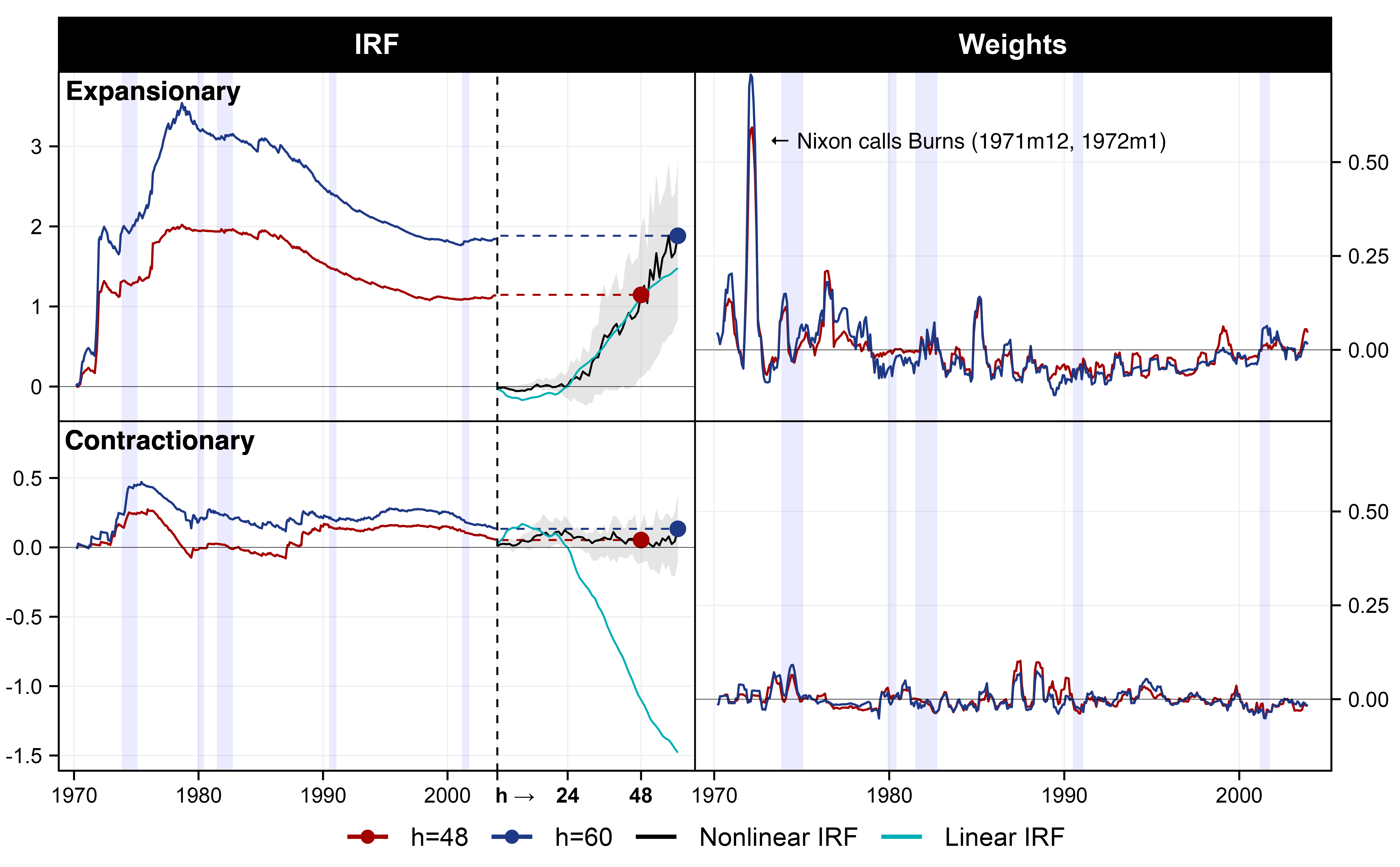}
     
    \begin{threeparttable}
    \centering
    \vspace*{-0.6cm}
    \begin{minipage}{\textwidth}
      \begin{tablenotes}[para,flushleft]
    \setlength{\lineskip}{0.2ex}
    \notsotiny 
  {\textit{Notes}: The plot shows responses of US inflation to a one standard deviation for a expansionary (sign flipped) and contractionary monetary policy shock. The shock is identified via the \cite{romerromer2004} instrumental variable. The left panels present cumulative contributions $\sum_{t=1}^T c_{th}$ in solid colored lines, summing to the final predicted value shown as dots. Full impulse response functions are shown in black, \textcolor{black}{defined by the mean and the 16$^{th}$ and 84$^{th}$ percentile of the tree distribution (see Section \ref{sec:RFrev}).} Right panels show the corresponding weights ($w_{th}$) over time in 6-month moving averages. Lavender shading corresponds to NBER recessions.}
    \end{tablenotes}
  \end{minipage}
  \end{threeparttable}

\end{figure}

{\noindent \sc \textbf{Nonlinear Enlightenment.}} We complete the analysis with our nonlinear IRFs for R\&R shocks (see Figure \ref{fig:mp_contrib_rf}). {\color{black} The nonlinear model is a Random Forest, following the methodology outlined in Section~\ref{sec:ml}, with implementation details provided in Appendix~\ref{sec:RFrev}.}

We address two questions: First, is there any evidence of effects from tightening shocks in an R\&R shock-based local projection? Second, does the early hump observed with R\&R shocks persist in a nonlinear model where shock size \textcolor{black}{and sign} matter? The answer to the first question is negative---IRFs from R\&R tightening shocks are effectively null.  R\&R acknowledge in their original paper that much of the exogenous variation in interest rates is attributable to political influence---and that the pressure is almost always unidirectional. Therefore, it is natural to ask whether there is any quantitative evidence from  R\&R we should use to forecast the effects of future monetary tightening decisions.

The answer to the second question is positive---separating positive and negative shocks yields more intuitive IRFs for positive shocks: flat for nearly two years before rising to align with the linear IRF (in turquoise in Figure \ref{fig:mp_contrib_rf}). Comparing weights across horizons highlights the key events driving these patterns. The first substantial weight stems from Nixon’s pressure on Arthur Burns to loosen monetary policy ahead of the 1972 election  \citep{drechsel2024estimating}, accounting for over half of the IRF at horizon $h=48$. Thus, when tasked with identifying an event resembling an exogenous, unsystematic policy intervention, the framework points to the economically unjustified monetary easing of late 1971 and early 1972. The IRF at $h=60$ rises further due to the sequence of loosening shocks from 1976--1978, also prominent in the linear model (\textcolor{black}{see Figure \ref{fig:mp_weights_rf} in the appendix for a comparison of linear and nonlinear weights.}). In both cases, R\&R shocks capture evidence primarily from expansionary episodes driven by political pressure on the Fed during the 1970s.

While this concentration supports the instrument’s validity—reflecting genuine exogenous shocks to an otherwise systematic policy—it raises concerns about the external validity of the estimates, particularly when applied to forecast the effects of contractionary monetary policy nearly 50 years later. Specifically, it is unclear whether the observed delays and magnitudes in the IRF  are representative of the current economic environment. Post-pandemic evidence suggests that inflation has not responded to monetary tightening faster than with a two-year lag, casting doubt on the relevance of 1970s dynamics for contemporary policy analysis.

\vspace{0.25em}

{\noindent \sc \textbf{Further Nonlinearities.}} Finally, we dig out additional nonlinearities by clustering the IRFs as described in Section \ref{sec:ml}. In Figure \ref{fig:mp_cluster_rf} (Appendix), we report two clusters for each shock sign.


For expansionary monetary policy shocks, the first cluster corresponds in great part to the episode of Nixon pressuring Arthur Burns \textcolor{black}{and covers 57\% of observations.} This cluster yields an IRF closely resembling the one obtained from the full nonlinear model. The second cluster, \textcolor{black}{grouping the remaining 43\% of observations,} appears to be primarily driven by the prolonged sequence of monetary loosening shocks from 1976 to 1978. While the overall shape of the IRF is similar to that of the first cluster, the magnitude is notably smaller and the inflation response is more sluggish, taking 36 months to become apparent. This suggests a potential difference in the economic response between a large, discrete shock and a series of smaller, gradual shocks. Overall, the clustering analysis highlights the significance of politically motivated shocks, which are grouped within one cluster, and episodes of sustained monetary loosening, clustered separately. 

For contractionary shocks, the clustering analysis confirms that the evidence is relatively weak. The dominant cluster, supported by 95\% of the data, shows an insignificant IRF consistent with the full-sample estimate (Cluster 1 in Figure \ref{fig:mp_cluster_rf}). However, a very small cluster, based on a limited number of data points from the late 1980s, exhibits an IRF featuring a price puzzle. While this result is derived from a narrow subset of observations (5\% of the sample), it is nonetheless notable that, in a general nonlinear model, any strong evidence stemming from contractionary R\&R shocks appears to point in the wrong direction.

\subsection{Government Spending}\label{sec:fiscal}

Building on the work of \cite{auerbach2012measuring,Auerbach2013}, and \cite{RZ2018}, a substantial literature has explored the heterogeneous effects of fiscal policy, particularly focusing on how fiscal multipliers vary with the state of the economy. A central challenge in this literature is identifying an exogenous fiscal policy impulse, as it is more often than not endogenously shaped by prevailing economic conditions. \cite{ramey2011} and  \citet[][henceforth RZ]{RZ2018} championed the use of military spending shocks as a solution, arguing that such shocks are driven by causes largely exogenous to the US economy and government spending decisions. 

 As in many macroeconomic studies of this type, truly exogenous impulses are rare, limiting the number of effective data points. While the weak IV literature addresses inferential challenges in such settings, we ask the related question: How many historical experiments contribute to this estimate? The sparse case is particularly consequential for both inference and external validity.

\begin{figure}[t!]
  \caption{\normalsize{Responses of Real GDP to Government Spending Shock}} \label{fig:fiscal_contrib_stat}
    \centering
    \vspace*{-0.3cm}
    \includegraphics[width=\textwidth, trim = 0mm 0mm 0mm 0mm, clip]{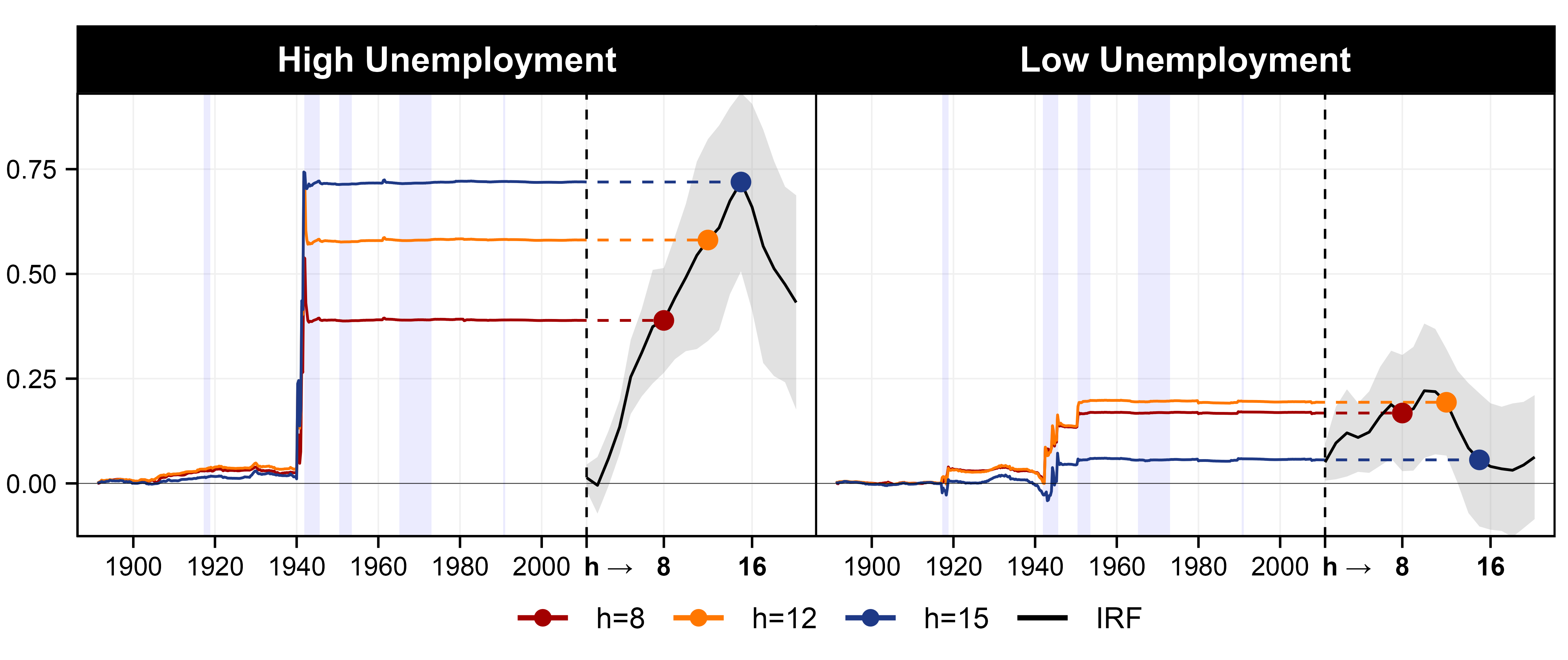}
     
    \begin{threeparttable}
    \centering
    \begin{minipage}{\textwidth}
    \vspace*{-0.2cm}
      \begin{tablenotes}[para,flushleft]
    \setlength{\lineskip}{0.2ex}
    \notsotiny 
    {\textit{Notes}: The plot shows cumulated responses of real GDP to a government spending shock as defined in \cite{RZ2018} in periods of high versus low unemployment (above/below 6.5\%). Data are transformed to stationarity. The estimation sample starts in 1891Q1 and ends in 2010Q4. Cumulative contributions $\sum_{t=1}^T c_{th}$ are presented in solid colored lines, summing to the final predicted value shown as dots. Full impulse response functions are shown in black with 95\% confidence bands applying Newey-West standard errors. Lavender shading corresponds to WWI, WWII, Korean war, Vietnam war, and Gulf war.}
    \end{tablenotes}
  \end{minipage}
  \end{threeparttable}

\end{figure}

As it turns out, for fiscal policy effects identified via government spending shocks, the answer could hardly by clearer in Figure \ref{fig:fiscal_contrib_stat}. First,  considering the causal effect during times of slack---defined by the unemployment rate exceeding 6.5\%---the estimate is  derived almost entirely from a single event: World War II. \textcolor{black}{This is also reflected in concentration statistics (Table \ref{tab:fiscal} in the appendix), where contributions reach a 90\% concentration level for all three horizons.} Second, the low-unemployment $\hat{\beta}_h$ is a composite of the late WWII period (when unemployment was declining) and the Korean War. \textcolor{black}{Again, concentration measures for contributions amount to 81\% for the shorter horizons and 77\% for $h=15$}.

Figure \ref{fig:fiscal_weights_stat} (Appendix) shows raw shocks and weights, either as such or smoothed using a 4-quarter moving average. While raw shocks indicate heightened military shocks volatility around the onset of WWII, the period’s significance is far more evident in the weights, particularly in the smoothed version for the high-unemployment regime. There, we observe a distinct window of instrument relevance opening around 1940Q2 and closing in 1942Q4, while the rest of the sample remains mostly flat at zero. 

\vspace{0.25em}

\begin{figure}[t]
  \caption{\normalsize{Responses of Real GDP to Government Spending Shocks for Different Subsamples}} \label{fig:fiscal_trim_stat}
    \centering
    \vspace*{-0.2cm}
    \includegraphics[width=\textwidth, trim = 0mm 0mm 0mm 0mm, clip]{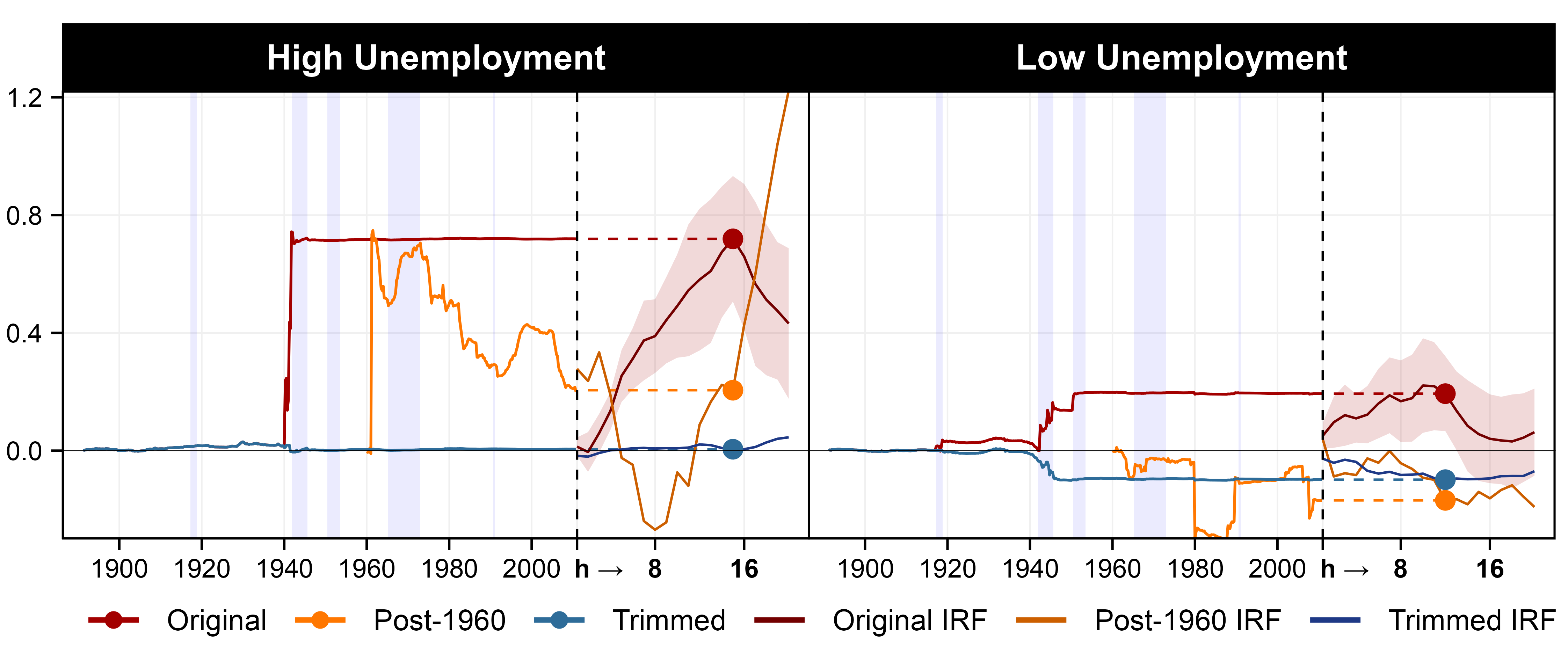}
     
    \begin{threeparttable}
    \centering
    \begin{minipage}{\textwidth}
    \vspace*{-0.2cm}
      \begin{tablenotes}[para,flushleft]
    \setlength{\lineskip}{0.2ex}
    \notsotiny 
    {\textit{Notes}: The plot shows responses of real GDP growth to a government spending shock, as identified in \cite{RZ2018}, for three different estimation samples. \textit{Post-1960 IRF} results from estimating the model starting in 1960Q1 and ending in 2010Q4. For \textit{Trimmed IRF}, we set weights at the 1st and 99th percentiles to zero for each horizon, and then re-cumulate the remaining contributions to obtain the IRF. \textit{Original IRF} refers to results obtained using the full sample, as done in the main analysis (Figure \ref{fig:fiscal_contrib_stat}). For the evidence curves we focus on the peak horizon of the original IRF. Cumulative contributions $\sum_{t=1}^T c_{th}$ are presented in solid colored lines, summing to the final predicted value shown as dots. Full impulse response functions are shown in black with 95\% confidence bands applying Newey-West standard errors. Lavender shading corresponds to WWI, WWII, Korean war, Vietnam war, and Gulf war.}
    \end{tablenotes}
  \end{minipage}
  \end{threeparttable}

\end{figure}

{\noindent \sc \textbf{Adversarial Robustness Checks.}} In Figure \ref{fig:fiscal_trim_stat}, we perform two robustness checks. First, we re-estimate the IRFs for both regimes after trimming the top and bottom 1\% of weights, in the spirit of a trimmed-mean estimator.\footnote{Since the usual IRFs are effectively weighted averages of \( T \times c_{th} \), another natural robustness-enhancing alternative is the trimmed-mean estimator based on contributions rather weights. We get even more negative results from this alternative.} The results are striking: neither regime's IRF survives this test, as both collapse to nearly zero. \textcolor{black}{Concentrations of weights collapse to 36\% and 49\% in the high- and low-unemployment regimes, respectively, and the corresponding measures for contributions show values around 55\% for the former and near 70\% for the latter (see Table \ref{tab:fiscal} in the Appendix).} Second, we re-estimate the models using data from 1960 onward, excluding all previously identified key contributors. This not only removes early influential episodes but also allows the model to reconstruct its embedding (via \( F_t \)), potentially bringing other episodes to shine. However, no such emergence occurs—both post-1960 IRFs exhibit either erratic shapes, incorrect signs, or both.

How should we interpret results with this degree of sparsity? While they clearly identify the historical precedents driving the estimates, they also raise concerns about external validity. The empirical recession regime estimates derived from military spending shocks, despite spanning over a century of data, can largely be reduced to a before-and-after analysis of WWII-era government spending. Can we confidently predict the effects of fiscal stimulus in the next US recession based on a single, highly unique event from more than 80 years ago? Such sparsity enhances transparency and interpretability, but comes with risky extrapolation for future policy analysis.

This paper is not the first to put dents in the armor of military spending shocks. \cite{kolesar2024dynamic} highlight that relying solely on military buildups, without corresponding retrenchments, introduces asymmetries that complicate interpretation. This aligns with \cite{Barnichon2025}, who emphasize the low power of the instrument as a key efficiency concern, leading to wide and potentially uninformative confidence intervals or estimates that are highly sensitive to specification choices.

\vspace{0.25em}

{\noindent \sc \textbf{Specification Choices: Growth Rates vs. Ratios.}}  In Figures \ref{fig:fiscal_contrib_levels} and \ref{fig:fiscal_weights_levels} (Appendix), we present results based on ratio-based estimates, following exactly the regression design in RZ. However, the stationarity of these ratios remains uncertain. Moreover, the impact on $\hat{\beta}_h$ of choosing stationarized data (e.g., growth rates) over levels is not always straightforward in applied settings.  Weights and cumulative contributions help illustrate how the use of potentially non-stationary data influences the estimated coefficients.

Whether the analysis uses levels/ratios or first differences, the estimates are strongly influenced by a narrow set of historical events from the distant past.  In the recession regime, WWII is the sole significant positive contributor, while the overall trending behavior negatively affects contributions across the 100-year sample. Figure \ref{fig:fiscal_weights_levels} (Appendix) also reveals another, albeit more modest, period of instrument relevance just before WWII. However, since these negative weights coincide with strong growth in the following years, they ultimately contribute to shrinking the size of the fiscal multiplier in the recession regime.

A similar pattern emerges in the expansion regime, where the Korean War drives the estimates, while downward trends reduce contributions from approximately 0.5 in the 1950s to below 0.2. This decline primarily reflects the combination of mostly positive military spending shocks with a  multi-decadal deceleration in GDP growth from 1950 to the early 2000s. It is rather unclear whether this association is desirable. 

Consequently, the IRF using stationary data in Figure \ref{fig:fiscal_contrib_stat} peaks about 0.2 higher at $h=15$  in the recession case than when using the partly trending ratio data. A similar but more moderate shrinkage also occurs in the expansion case. Therefore, we see that gazing at cumulative contributions and weights can not only be helpful at understanding the role of the exogenous variable of interest, but also the non-trivial effects of the design of the regression into which it is included. 

\vspace{0.25em}

{\noindent \sc \textbf{Other Fiscal Policy Effects Estimates.}} Figures \ref{fig:fiscal_contrib_benzeev} and \ref{fig:fiscal_weights_benzeev} (Appendix) present cumulative contributions and weights for alternative specifications of government spending shocks on real economic activity. Specifically, we focus on the \citet[][henceforth BZP]{benzeev2017} and \cite{ramey2011} specifications. The former identifies fiscal shocks by defining exogenous defense spending shocks as those orthogonal to current defense spending, while also best capturing the trajectory of defense spending shocks over a five-year horizon. The latter differs slightly from the specification used earlier, as it does not distinguish between high- and low-unemployment regimes and focuses on a later sample, starting in 1947—much later than RZ.

In Figure \ref{fig:fiscal_contrib_benzeev} (Appendix), we see that both shocks yield relatively similar estimates for \( h=5 \), with \( \hat{\beta}_5 \) heavily influenced by the Korean War, accounting for more than half of the reported value in both cases. However, results diverge at \( h=9 \) and \( h=12 \), where BZP produces more moderate estimates compared to \( h=5 \), while the IRF from \cite{ramey2011} remains elevated and even increases further. This discrepancy arises partly from how the years immediately preceding the Korean War contribute to later horizons. For \cite{ramey2011},  these years contribute in the same direction as earlier horizons, whereas for BZP, they push the IRF toward zero. There is some indication that these estimates might remain viable even without their most influential historical contribution, as the curves steadily climb in the expected direction from the 1950s onward. \textcolor{black}{Additionally, concentration measures suggest that the distribution of weights as well as concentrations are more dispersed (with a \textbf{WC}($\hat{\beta}_h$) of 40\% and \textbf{CC}($\hat{\beta}_h$) ranging from 0.52 and 0.58 across different horizons).}

Figure \ref{fig:fiscal_trim_benzeev} (Appendix) presents the trimmed and post-1960 estimates for the \cite{ramey2011} and BZP specifications. Both fail the post-1960 test, as their IRFs take the wrong sign. The trimming test, however, is less severe. The \cite{ramey2011} specification still produces a positive impulse response with a shape similar to the original, despite the Korean War being visibly trimmed out. This resilience is driven by numerous small shocks that continue to contribute in the expected direction. In contrast, the BZP IRF, which was already weaker in statistical significance than \cite{ramey2011} in its original form, turns negative more quickly and appears broadly insignificant.

\cite{barnichon2022} and BZ build on RZ by using military spending shocks to identify fiscal policy effects, specifically examining asymmetries between spending increases and decreases. Despite differing identification strategies, both rely on military spending as the primary instrument for exogenous variation. Their mixed findings—\cite{barnichon2022} detects asymmetries, while \cite{BenZeevRameyZubairy2023} does not—align with our results, suggesting that fiscal multipliers identified through military shocks are fragile due to their reliance on a highly limited set of events. Further narrowing the analysis to detect heterogeneity  amplifies this concentration issue.

\subsection{Global Temperature Shocks}\label{sec:climate}

In an influential recent paper, \citet[][henceforth BK]{bilal2024} employ local projections to analyze the dynamic causal impact of global temperature shocks on world GDP, starting from 1960. Their findings suggest that a 1°C increase in global temperature results in a gradual decline in world GDP, peaking at a 12\% reduction after six years. This effect is statistically significant at the 5\% level for the period spanning years 3 through 7. Furthermore, the temperature shocks exhibit a persistent effect on global temperature levels, which remain elevated by more than 0.5°C several years after the initial shock. Even a decade later, GDP does not fully revert to its pre-shock trajectory, implying a degree of lasting economic damage.

We reassess the robustness of the claim using our new set of tools. Given the limited number of data points—annual data from 1960 to 2019, yielding between 48 and 58 observations depending on the forecast horizon \textcolor{black}{with the specification incorporating two lags}—it should be relatively straightforward to identify which pairs of years and shocks are driving the show.

\begin{figure}[t]
  \caption{\normalsize{Responses to Global Temperature Shocks}} \label{fig:climate_full}

      \centering
      \vspace*{-0.3cm}
      \includegraphics[width=1\textwidth, trim = 0mm 0mm 0mm 0mm, clip]{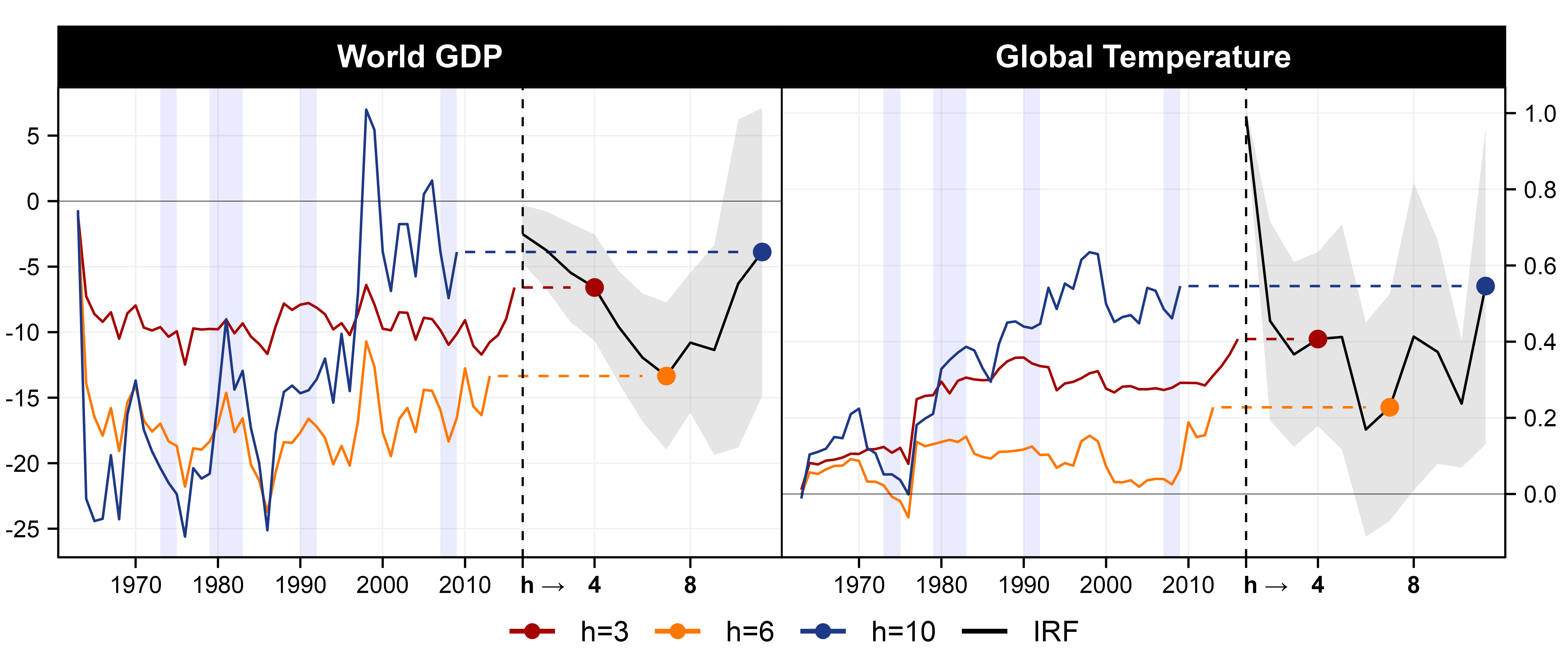}

    \begin{threeparttable}
    \centering
    \begin{minipage}{\textwidth}
    \vspace*{-0.2cm}
      \begin{tablenotes}[para,flushleft]
    \setlength{\lineskip}{0.2ex}
    \notsotiny 
    {\textit{Notes}: The plot shows responses of world real GDP (left panel) and global average temperature (right panel) to a global temperature shock as identified by \cite{bilal2024}.  The estimation sample starts in 1963 and ends in 2019, in yearly frequency. Cumulative contributions $\sum_{t=1}^T c_{th}$ are presented in solid colored lines, summing to the final predicted value shown as dots. Full impulse response functions are shown in black with 90\% confidence bands applying Newey-West standard errors. Lavender shading corresponds to NBER recessions.}
    \end{tablenotes}
  \end{minipage}
  \end{threeparttable}

\end{figure}

\begin{figure}[t]
  \caption{\normalsize{Proximity Scores and Realizations of $y_{t+h}$ for Global Temperature Shock}} \label{fig:climate_full2}
    \begin{center}
    \vspace*{-0.8cm}

      \hspace*{-0.35cm} \includegraphics[width=1.04\textwidth, trim = 0mm 0mm 0mm 0mm, clip]{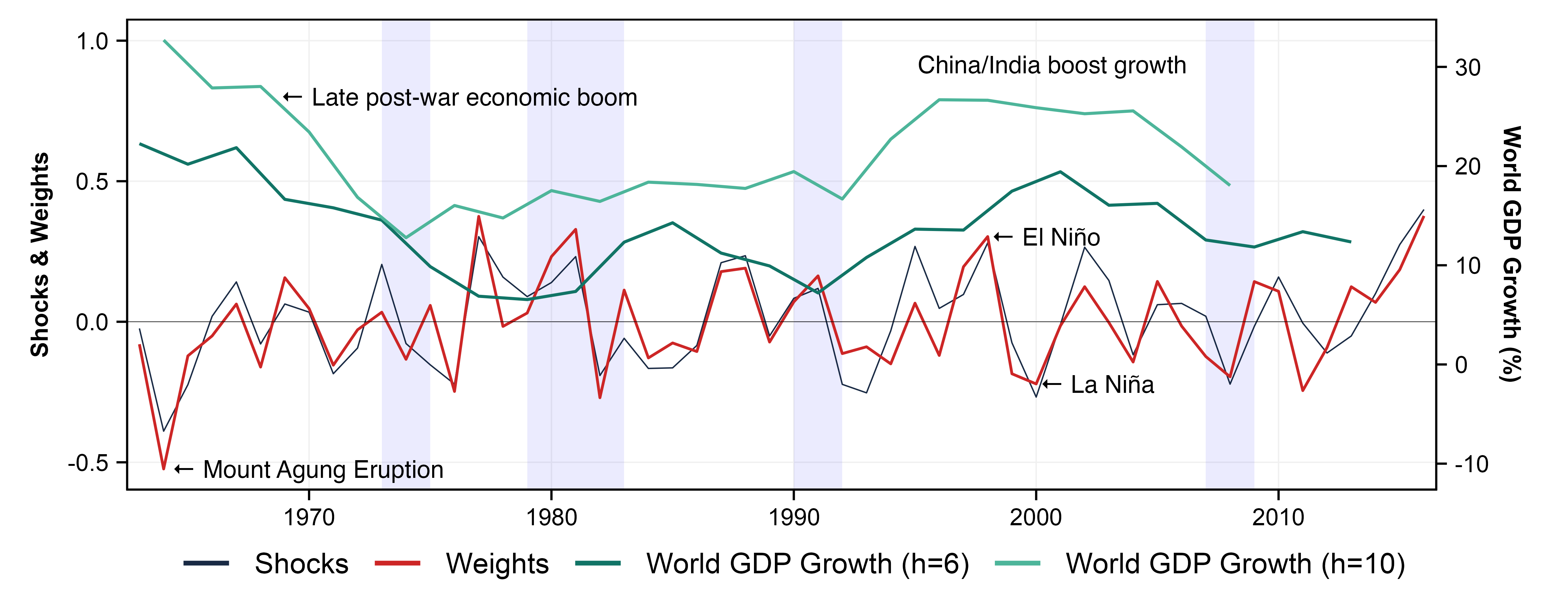}

    \end{center}
     
    \begin{threeparttable}
    \centering
    \begin{minipage}{\textwidth}
    \vspace*{-0.9cm}
      \begin{tablenotes}[para,flushleft]
    \setlength{\lineskip}{0.2ex}
    \notsotiny 
    {\textit{Notes}: The plot displays the global temperature shock series as identified by \cite{bilal2024}, the proximity scores between the anticipated shock and past shocks, and realizations of world real GDP growth for horizon $h \in \{6,10\}$ as used in the estimation. Shock series is scaled to match the mean absolute deviation of the weights. Note that we smooth the target to obtain cumulative LPs, and that longer horizons result in a shorter estimation sample. }
    \end{tablenotes}
  \end{minipage}
  \end{threeparttable}

\end{figure}

Results are presented in Figures \ref{fig:climate_full} and \ref{fig:climate_full2}, which display cumulative contributions and the corresponding weights alongside the values of the response variable, respectively. By examining the evidence curves derived from the key impulse response function for world GDP, we observe the clear influence of two significant climatological events on the estimated effects. These events are the 1964 cooling shock, the late-1990s El Niño, and the subsequent La Niña event. Despite these observations, the overall concentration levels are more moderate than in previous applications, with \textbf{WC}$(\hat{\beta}_h)=0.25$  and \textbf{CC}($\hat{\beta}_h$) ranging from 0.27 ($h=3$) to 0.33  ($h=10$). Therefore, some more investigation is needed. 

The evidence curve for the response of global temperature to a global temperature shock shows broad support from the data, with trajectories generally trending upward, resembling random walks with drift. This pattern indicates strong evidence that the effect of a temperature shock on global temperature at time $t$ is partially offset—by approximately 50\%—by the end of the following year, with the cumulative effect stabilizing in subsequent years.

\vspace{0.25em}

{\noindent \sc \textbf{Eruption of Mount Agung and the Post-war Economic Boom.}} The massive cooling shock of 1964 resulted from the eruption of Mount Agung in Indonesia in 1963. This volcanic eruption released large quantities of sulfur dioxide (SO$_2$) into the stratosphere, forming a reflective aerosol layer that reduced solar radiation and caused a “volcanic winter.” This shock coincides with a significant period of post-war economic growth for the following 10 years, which happens to be the largest in the sample for both the 6-years and 10-years moving average (see Figure  \ref{fig:climate_full2}). The post-war boom is well-documented as a time of substantial global GDP growth, which slowed markedly in the mid-1970s due to the oil price shocks and geopolitical turmoil, such as the Iranian Revolution.

This pairing of a negative cooling shock with a decade of extraordinary growth, clearly visible at the beginning of Figure  \ref{fig:climate_full2}, significantly affects the estimated long-run impacts of climate shocks on world GDP. We see in Figure \ref{fig:climate_full} that the early contribution of the 1964 shock is by far the most important contributor to the large negative effects, after 3, 6, and 10 years. Still, there is reason to believe that world GDP's rapid growth---driven by post-war reconstruction, productivity gains, booming trade, and global investment under the Bretton Woods system, with stable energy prices sustaining growth---may be unaccounted for.

\vspace{0.25em}

{\noindent \sc \textbf{El Niño and La Niña Events of the Late 1990s, and the China/India Boom.}} The late-1990s El Niño event represents another pivotal climatological shock, marked by a significant global temperature increase. The 1997–1998 El Niño is one of the strongest on record, characterized by warming of the Pacific Ocean and disrupted weather patterns worldwide. In terms of IRF analysis, this event is paired with elevated economic growth in the mid-2000s, right before the Great Financial Crisis (GFC). We see well the synchronization of such events in Figure \ref{fig:climate_full2}. This pairing leads to an upward temperature shock coinciding with a period of economic expansion, which shatters the longer-horizon IRF estimates due to the contrarian relationship between temperature shocks and economic conditions.

Following this, the late-1990s La Niña event—a period marked by cooler-than-average ocean temperatures in the equatorial Pacific—had the opposite climatological effect. La Niña tends to generate extreme weather events, including droughts and floods, which have varied economic impacts. Interestingly, this event coincided with an era of steady economic growth in many regions. The evidence curve reveals that this period of cooler temperatures mitigated some of the ``counterintuive'' effects of the El Niño event, as colder temperatures are now paired with sustained expansion.

A key question is whether these climatological events played a role, to some extent, in the sparkling global economic growth of the early 2000s, as there is no shortage of alternative explanations. The early 2000s were characterized by significant global GDP growth due to the rapid economic expansion of emerging markets, particularly China and India. China’s accession to the World Trade Organization (WTO) in 2001 catalyzed global trade, while the rise of the information technology sector contributed to productivity gains worldwide. Additionally, the global commodity boom fueled by growing demand in emerging economies supported higher income levels in resource-exporting countries, contributing to a synchronized global expansion. 

\vspace{0.25em}

\begin{figure}[t]
  \caption{\normalsize{Effects of Global Temperature Shocks on World GDP for Different Subsamples}} \label{fig:climate_subsample}

      \centering
      \vspace*{-0.3cm}
      \includegraphics[width=1\textwidth, trim = 0mm 0mm 0mm 0mm, clip]{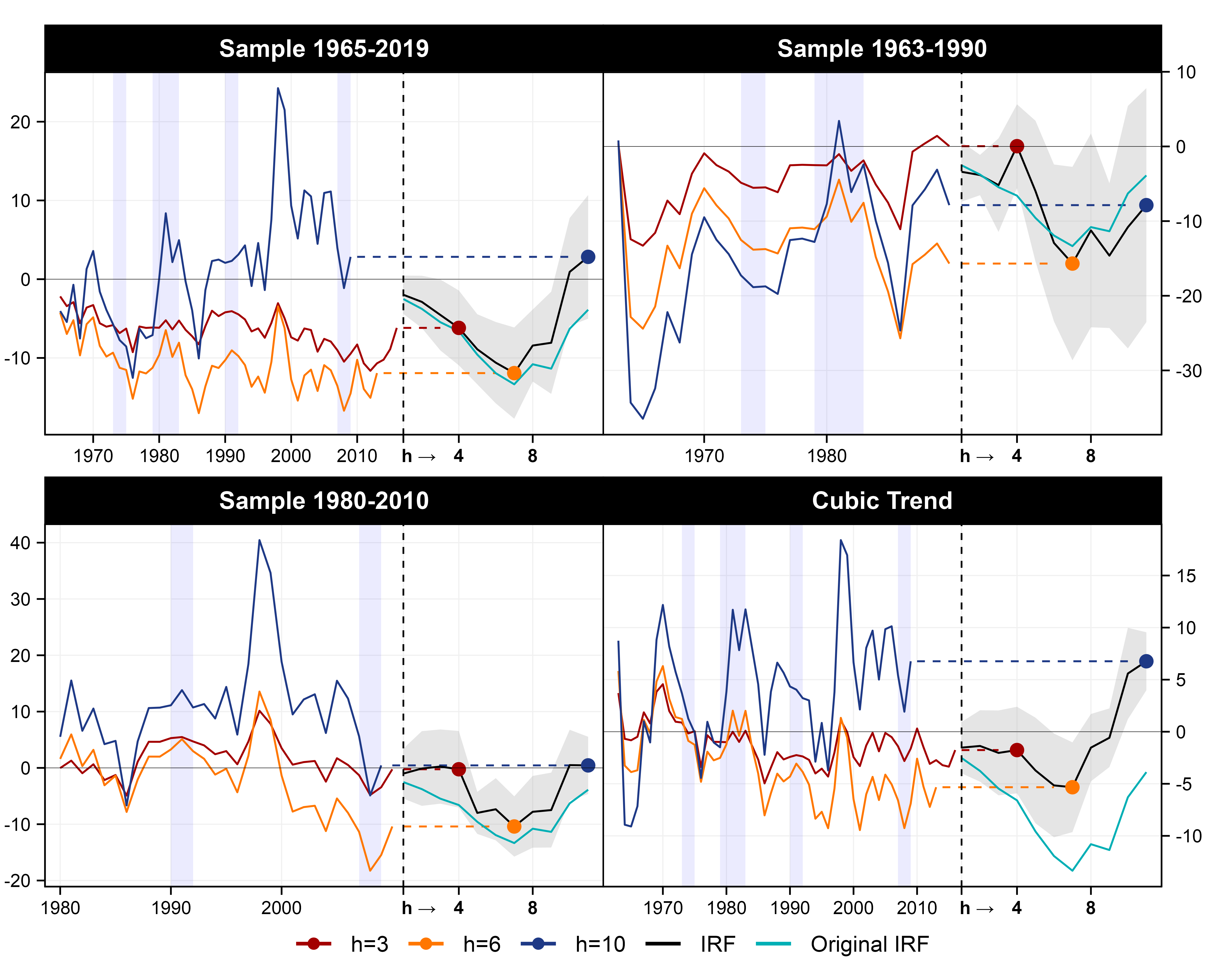}

    \begin{threeparttable}
    \centering
    \begin{minipage}{\textwidth}
    \vspace*{-0.2cm}
      \begin{tablenotes}[para,flushleft]
    \setlength{\lineskip}{0.2ex}
    \notsotiny 
    {\textit{Notes}: The plot shows responses of world real GDP to a global temperature shock, as identified in \cite{bilal2024}, for four different estimation samples. Cumulative contributions $\sum_{t=1}^T c_{th}$ are presented in solid colored lines, summing to the final predicted value shown as dots. Full impulse response functions are shown in black with 90\% confidence bands applying Newey-West standard errors. The dashed line marks the end of the estimation sample and changes for each scenario. \textit{Original IRF} refers to the results obtained using the full sample, as is done in the main analysis (Figures \ref{fig:climate_full} and \ref{fig:climate_full2}). Lavender shading corresponds to NBER recessions. }
    \end{tablenotes}
  \end{minipage}
  \end{threeparttable}

\end{figure}

{\noindent \sc \textbf{Adversarial Robustness Checks.}} The information gathered from Figures \ref{fig:climate_full} and \ref{fig:climate_full2} suggests four robustness checks on BK's original specification (see Figure \ref{fig:climate_subsample}).

\begin{enumerate}
    \item \textit{Excluding the 1964 Cooling Shock.}  The first robustness check examines the exclusion of the 1964 cooling shock by estimating from 1985 onward. Removing this shock reveals that the IRF remains robust for early and mid-range horizons (4 to 8 years), but the long-run effects (8 to 10 years) collapse to zero.   This suggests that conclusions about persistent economic effects of climate shocks are heavily reliant on the inclusion of this singular event, making the long-run impacts fragile.
    
  BK also performs a similar robustness check by excluding all volcanic eruptions, recognizing that temperature shocks from such events may have different implications than those from other sources. Their results also indicate that this modification attenuates the long-run effects. However, the eruption of Mount Agung stands out as the most influential, as it coincides with a period of exceptionally strong post-war economic growth. Consequently, the robustness check conducted here is inherently more adversarial to the baseline results than the one in BK.

    \item \textit{Excluding the El Niño and La Niña Events.}  The second robustness check focuses on the exclusion of the late-1990s El Niño and subsequent La Niña events. While one might expect these significant climatological events to influence the results substantially, their opposing dynamics largely cancel each other out in the IRF analysis. The 1997–1998 El Niño, characterized by a global temperature surge, is followed by a cooling period associated with La Niña. This cancellation effect means that excluding these events does not materially change the results. The transitory effects of climate shocks remain significant and consistent with the main findings, but no evidence emerges to suggest a stronger or more persistent long-run impact when these events are excluded. BK perform a similar robustness check and reach comparable findings. The reason is clear: although El Niño and La Niña are linked to periods of very strong economic growth years later, their opposing effects largely offset each other.

    \item \textit{Estimating from the 1980s Onward.} The third robustness check estimates the IRFs using data from the 1980s onward, effectively excluding the 1960s and 1970s. This period avoids the influence of the 1964 cooling shock, the post-war boom, and the oil crises. The results confirm that the mid-range IRF (horizons 4 to 8 years) remains broadly consistent with the original findings, although it becomes marginally more muted. Importantly, the long-run effects (horizons beyond 8 years) clearly converge to zero, further supporting the conclusion that the persistent impacts of global temperature shocks on GDP are not robust to changes in the sample period. 

    \item \textit{Including a Flexible Time-Trend.}  The fourth robustness check introduces a cubic time trend to capture long-run structural changes in world GDP growth. The goal is to approximate the effect of a random walk parameter for the intercept while accounting for long-run economic dynamics excluded from the baseline regression. This flexible trend can accommodate periods of elevated growth at the beginning of the sample, prolonged slower growth phases, and subsequent recoveries. By allowing for smooth, data-driven shifts in GDP growth, the cubic trend provides a nonparametric yet economically meaningful way to control for long-term factors influencing global output beyond the modeled climate shocks.  

    The inclusion of this flexible trend is supported by the adjusted $R^2$, which steadily increases up to the inclusion of a fourth polynomial term. Empirically, the effect of this trend is to dampen the original impact of the Mount Agung volcanic eruption shock and reduce the influence of exceptional economic growth events. Notably, it shifts the long-run IRF into counterintuitive positive territory due to the role of the El Niño event. More importantly, it reduces the peak effect after six years from 12\%—as reported in the original paper—to just 5\%. This highlights that specification choices regarding long-run world GDP growth not only shape persistent effects but also influence transitory ones, at least quantitatively.

\end{enumerate}

\noindent These findings suggest that the effects of global temperature on GDP are likely less persistent than originally stated. In this regard, they align more closely with earlier studies, such as \cite{nordhaus1992optimal} and \cite{dell2012temperature}, which indicate relatively modest long-run impacts. Moreover, even the observed transitory impacts appear to align with periods of robust global economic growth---namely 1965–1974 and the early 2000s---usually assumed to have been driven by structural factors unrelated to climate shocks.

\subsection{Financial Shocks}\label{sec:ebp}

In our final application, we decompose the responses of real activity and the interest rate to an adverse financial shock. Following the recent literature \citep{mumtazimpulse2022,barnichon2022,hauzenberger_machine_2024}, we examine potential nonlinearities in shock transmission and explain them by comparing evidence curves from a linear, OLS-based model and a nonlinear specification based on a Random Forest.

Data is taken from FRED-MD, except for the excess bond premium (EBP), which serves as our shock series and is taken from \cite{gilchrist2012}. It spans the period from 1974M1 to 2024M5. We adopt a standard setup that includes industrial production, CPI inflation, unemployment rate, EBP, national financial conditions index (NFCI), S\&P500 stock market index, and federal funds rate \citep[as in, e.g., ][]{forni2024}. We include 12 lags for all variables and a linear trend, \textcolor{black}{and account for serial correlation in the error term by applying the Newey-West standard errors \citep{NeweyWest1987}.} To assess potential asymmetries between the linear and nonlinear models, we estimate the response to a large shock of two standard deviations. 

\begin{figure}[t]
  \caption{\normalsize{Responses to an Adverse Financial Shock}} \label{fig:fin_contrib}
    \centering
    \vspace*{-0.2cm}
    \includegraphics[width=\textwidth, trim = 0mm 0mm 0mm 0mm, clip]{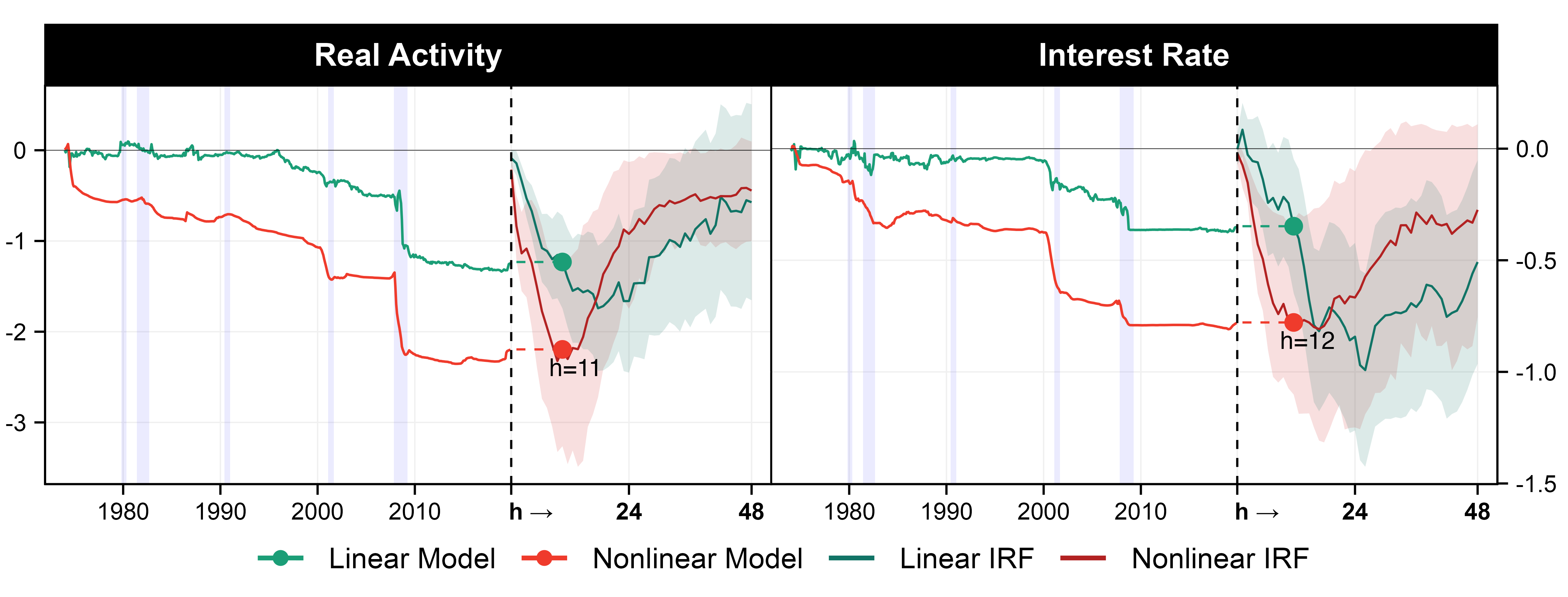}
     
    \begin{threeparttable}
    \centering
    \begin{minipage}{\textwidth}
    \vspace*{-0.2cm}
      \begin{tablenotes}[para,flushleft]
    \setlength{\lineskip}{0.2ex}
    \notsotiny 
    {\textit{Notes}: The plot presents responses of  industrial production and the federal funds rate to an adverse financial shock, measured by the excess bond premium \cite{gilchrist2012}. The sample starts in 1974M1 and ends in 2024M5. The linear model refers to OLS-based estimation, the nonlinear model uses a Random Forest. Evidence curves for both models are shown for the peak horizon from estimating the Random Forest model. Cumulative contributions $\sum_{t=1}^T c_{th}$ are displayed before the dashed line, which marks the end of the estimation sample. Full impulse response functions are shown after the dashed line. In the linear model, we report 84\% confidence bands applying Newey-West standard errors. For the nonlinear model, we present the mean and the 16$^{th}$ and 84$^{th}$ percentile of the tree distribution (see Section \ref{sec:RFrev}). }
    \end{tablenotes}
  \end{minipage}
  \end{threeparttable}

\end{figure}

\vspace{0.25em}
{\noindent \sc \textbf{Main Findings.}} 
Aligning with the theoretical literature \citep{brunnermeier2014}, we find that large adverse financial shocks trigger far-reaching effects in the real economy, with responses disproportionately larger than those predicted by the linear model (see left panel in Figure \ref{fig:fin_contrib}). The nonlinear model detects a faster and more pronounced reaction, with real activity peaking at $h=11$, whereas the linear model shows a slower adjustment, reaching its maximum effect only after 18 horizons. A similar pattern emerges for interest rate responses (see right panel in Figure \ref{fig:fin_contrib}). The nonlinear model adjusts considerably faster---peaking after one year compared to 26 months in the linear model. Both models yield responses of similar magnitude. 

While the evidence curve of the linear model suggests that financial shocks became more relevant after 2000, the nonlinear evidence curve already begins trending downward in the 1970s. This finding indicates that financial shock transmission is not merely a recent phenomenon. As discussed in the vast literature on financialization, the 1970s mark the beginning of a more finance-driven economy, with deregulation measures spurring the transition. The process then gained momentum in the 1990s and 2000s, as capital markets expanded and financial innovation accelerated \citep{Epstein2005,krippner2005}.

\begin{figure}[t]
  \caption{\normalsize{Proximity Scores for Financial Shock}} \label{fig:fin_weights}
  \begin{center}
        \vspace*{-0.7cm}
    \begin{subfigure}[t]{0.5\textwidth}
      \centering
      \includegraphics[width=\textwidth, trim = 0mm 0mm 0mm 0mm, clip]{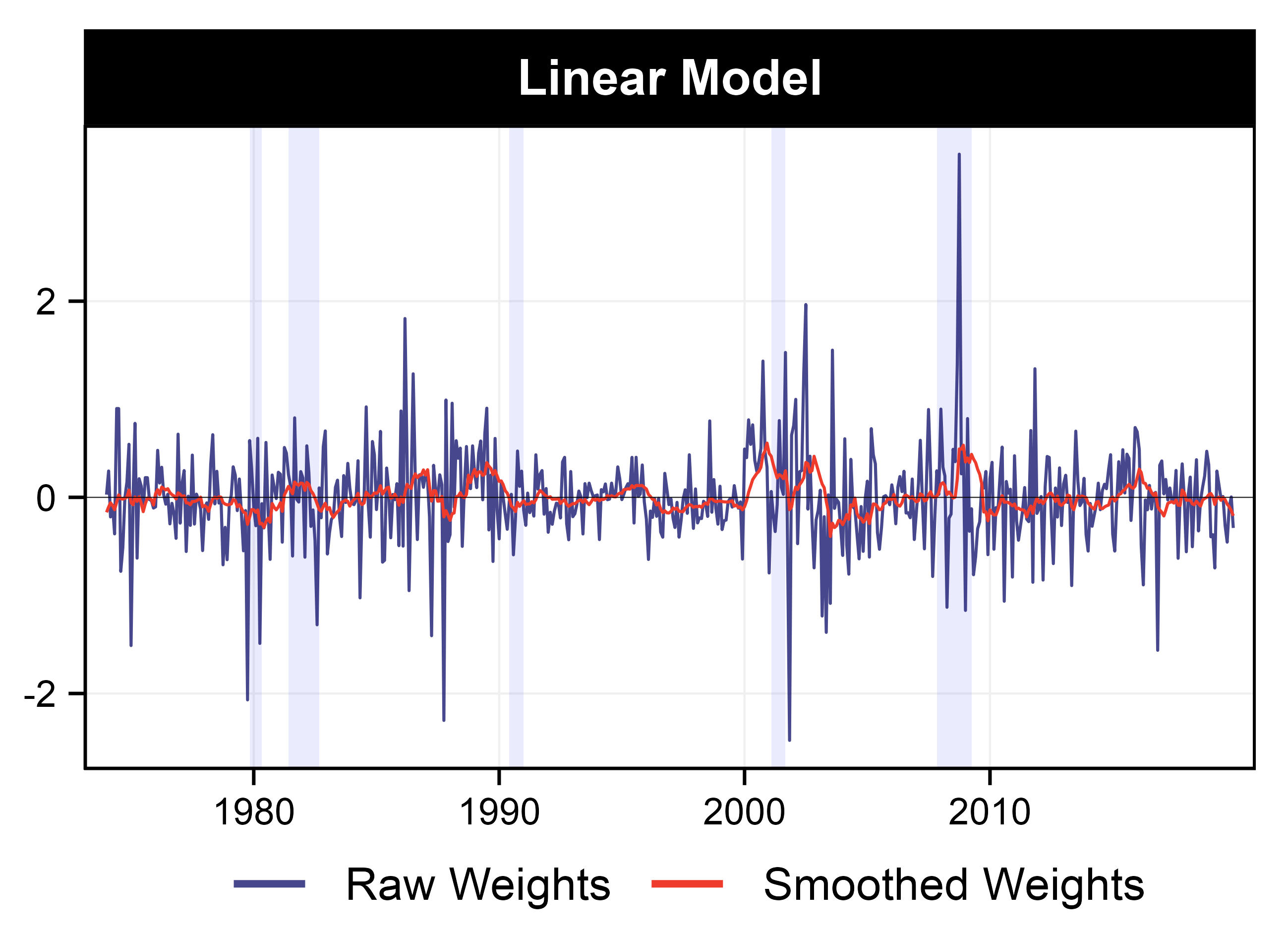}
    \end{subfigure}%
    \begin{subfigure}[t]{0.5\textwidth}
      \centering
      \includegraphics[width=\textwidth, trim = 0mm 0mm 0mm 0mm, clip]{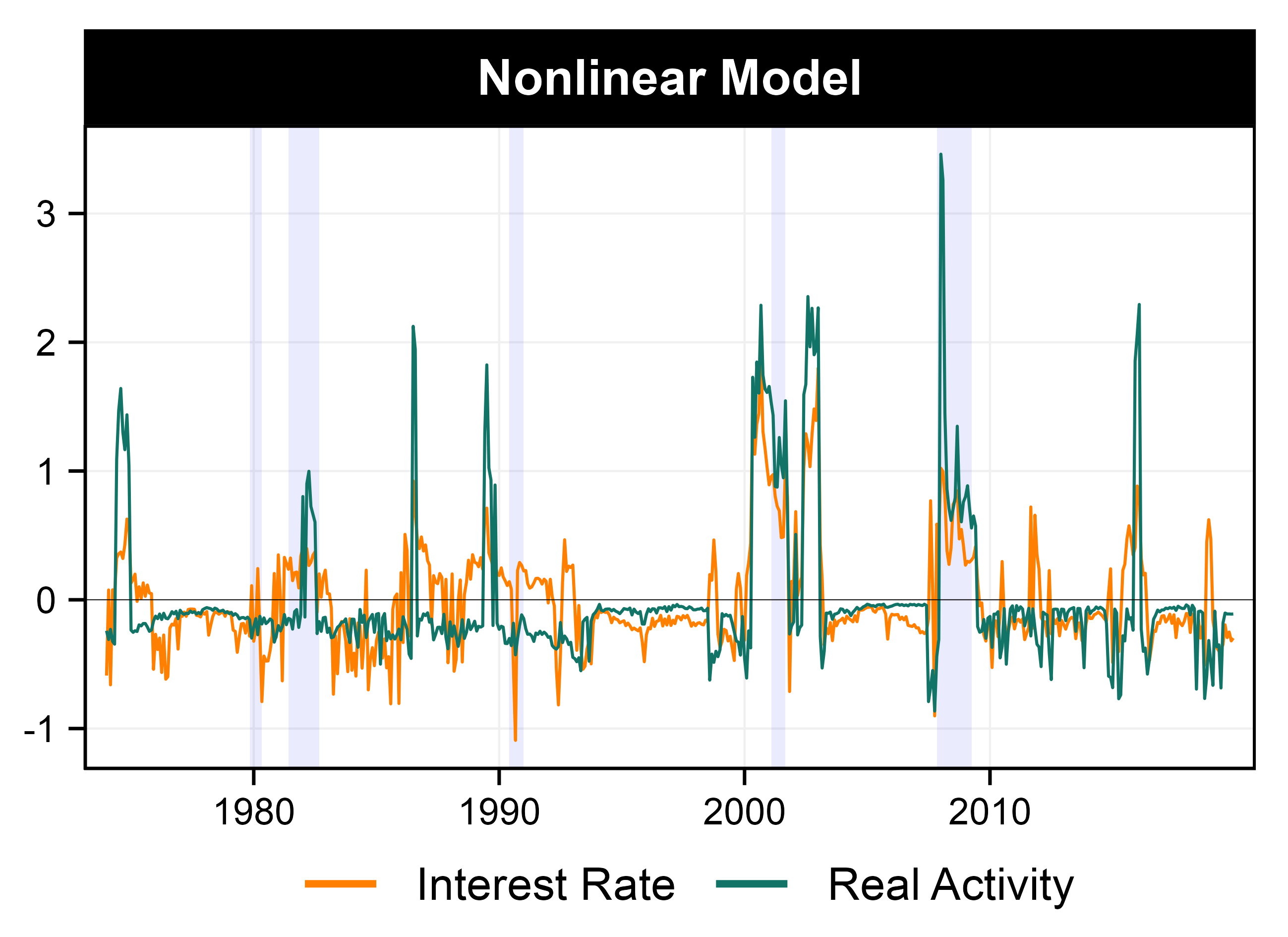}
    \end{subfigure}
    
  \end{center}
        \vspace*{-0.6cm}
    \begin{threeparttable}
    \centering
    \begin{minipage}{\textwidth}
      \begin{tablenotes}[para,flushleft]
    \setlength{\lineskip}{0.2ex}
    \notsotiny 
  {\textit{Notes}: The left panel shows proximity scores from the linear model, with smoothed weights averaged over 12 months. The right panel shows proximity weights from the nonlinear model, which can differ between targets. The estimation sample starts in 1974M1 and ends in 2019M12. Lavender shading corresponds to NBER recessions.}
    \end{tablenotes}
  \end{minipage}
  \end{threeparttable}
\end{figure}

\vspace{0.25em}
{\noindent \sc \textbf{Analysis of Proximity Scores.}} 
The distributions of weights, as shown in Figure \ref{fig:fin_weights}, are particularly telling about the fundamental differences between the nonlinear and linear model. While certain features of the linear model’s weighting structure carry over to the nonlinear framework, we find that the Random Forest offers greater interpretability. Its sparse weighting scheme arguably helps to assess contributions and understand the responses, which can be more challenging in denser or volatile structures, such as those found in linear model. For the latter, the raw weighting structure remains identical across target variables. While smoothing techniques aid in revealing patterns, they do not fundamentally alter the interpretation. In contrast, the nonlinear model offers a more transparent means of comparing distinct financial episodes, making it particularly useful for examining economic shocks beyond standard regression coefficient analysis.

For real activity, the nonlinear model’s weighting structure suggests that the response to large negative financial shocks can be effectively understood in terms of two distinct clusters: 1) \textit{normal financial conditions}, characterized by minor or positive shocks, 2) \textit{financial distress episodes}, marked by significant downturns, with the most extreme event occurring at the onset of the GFC (peaking in 2008M1 for real activity). While this peak is also present in the linear model, the nonlinear approach provides a more precise identification of key episodes, particularly in the 2000s (2000M9 and 2002M8), corresponding to the stock market turbulences from the dot-com bubble, and a few events in the 1980s (1986M8 and 1982M3). Additionally, the nonlinear model uniquely captures certain spikes, such as 2015M12 and 1989M7, which are either absent or less pronounced in the linear model. \textcolor{black}{The 1980s dates coincide with heightened stock market volatility combined with a restrictive monetary policy stance, whereas 2015M12 marks the first rate hike since 2006.} Notably, these spikes appear primarily in the real activity weighting, with no corresponding patterns in interest rate responses. 

While the GFC is key in driving the real activity response, the early 2000s recession has a milder impact, reflecting the milder contraction. The remaining spikes' contribution to the IRF is limited, as they do not coincide with extreme realization in real activity. However, their identification suggests that the model recognizes plausible intervention episodes related to adverse financial shocks.

The weighting structure for interest rate responses in the nonlinear model exhibits greater sparsity than for real activity. A few critical events dominate, particularly the response to the dot-com bubble and the September 2001 terrorist attacks. Nonlinear proximity scores peak in 2000M9 and 2002M11. In contrast, the model assigns less importance to the GFC in driving interest rate responses, likely because monetary policy was constrained by the zero lower bound, limiting conventional rate adjustments to financial shocks. The weighting scheme thus highlights how financial shocks prompted rapid and aggressive monetary easing, whereas the GFC exerts a comparatively weaker influence within this framework.

\vspace{0.25em}
{\noindent \sc \textbf{Uncovering Nonlinearities via Clustering.}} 
Finally, we apply our clustering approach as described in Section \ref{sec:ml}, which helps to uncover additional nonlinearities. As shown in Figure \ref{fig:fin_cluster_rf} (Appendix), we identify two distinct clusters for both variables.\footnote{Note that cluster labels are assigned independently for each variable. Real activity and interest rate clusters are estimated separately, and their numbering does not imply direct correspondence.} In both cases, Cluster 1 exhibits a fast and strong response, reinforcing the patterns observed in the aggregate nonlinear responses. In contrast, Cluster 2 closely resembles the linear model, trending downwards after 2000 and yielding a response of similar magnitude for both variables. A closer look at the corresponding weighting structure in Figure \ref{fig:fin_cluster_weights} (Appendix) reveals the sources of nonlinearities and suggests a structural break for Cluster 1. The latter clearly highlights periods in the 1970s and 1980s—proximities not observed in Cluster 2 and the linear model.

\section{Conclusion and Directions for Future Research}\label{sec:con} 

Local projections are widely used in empirical macroeconomics to flexibly and straightforwardly estimate impulse response functions. While the properties of the estimator itself are now well-known, that of a given \textit{estimate} constructed from a given finite sample is application-dependent and equally important. This paper introduces a decomposition of LP estimates into a sum of historical contributions, where each period’s contribution is the product of a weight and the response variable’s realization. Under least squares, these weights correspond to proximity scores between the projected policy intervention and past interventions in the sample.

This decomposition serves as a diagnostic tool. First, it reveals whether an estimate is informed by a broad range or a narrow subset of historical episodes, with concentration statistics documenting the extent of support. Second, it allows us to assess whether weights and contributions align with historical narratives, helping diagnose empirical puzzles and refine identification strategies.

Beyond linear models, our framework extends naturally to nonlinear machine learning-based impulse responses. Many such algorithms yield responses that, while nonlinear in regressors, ultimately represent a weighted sum of past realizations. Our approach thus provides a means to interpret and explain these more flexible methods with the same level of transparency as traditional LPs.

We illustrate the framework with applications to monetary, fiscal, climate, and financial shocks. The most salient facts are:
\begin{itemize} 
    \item \textbf{Monetary policy:} Cholesky VAR shocks produce a price puzzle due to misinterpretation of 1970s stagflation episodes, while \cite{romerromer2004} shocks rely almost entirely on politically driven monetary loosening in the 1970s. Nonlinear IRFs with Random Forest further refine the set of events used for identification and pin down specific events of political interference with the Fed in the early 1970s.
    \item \textbf{Fiscal policy:} \cite{RZ2018} state-dependent fiscal multipliers in recessions are largely driven by a single historical event---World War II. Other estimates of fiscal policy effects are also quite dependent on single military events, like the Korean War, raising general concerns about external validity.
    \item \textbf{Climate shocks:} The long-run GDP damage from global temperature shocks reported in \cite{bilal2024} appears fragile, being primarily driven by a single 1960s volcanic eruption coupled with sustained post-war economic growth. The medium term effects appear more robust and fairly well supported by various data points. 
    \item \textbf{Financial shocks:} Comparing linear and nonlinear responses highlights which historical events are upweighted to generate commonly observed size and sign-dependent effects. \textcolor{black}{Main differences appear in the treatment of the period before 2000 and the Great Financial Crisis.} We also highlight that the weights retrieved from Random Forest, by the virtue of being much sparser in terms of proximity scores, turn out to be much easier to interpret than that of linear models.  
\end{itemize}

\noindent By shedding light on the historical foundations of LP estimates, this framework enhances transparency and interpretability. Moreover, interpretation via data points instead of predictors allows to bridge the gap between traditional econometric methods and more flexible machine learning approaches.

\vspace{0.25em}

{\noindent \sc \textbf{Some Avenues for Future Research.}} We focused on local projections, which naturally accommodate our proposed decomposition since impulse response coefficients are linear functions of the target variable at each horizon. In contrast, while vector autoregressions are linear models, their impulse response functions are nonlinear in the targets—particularly beyond horizon 1—due to the recursive multiplication of the dynamic impact matrix. As a result, decomposing the structural VAR form is straightforward, but doing so for the vector moving average representation, which delivers the parameters of real interest, is more complex.

In such settings, one could resort to sampling-based methods or more sophisticated, computationally intensive tools such as Shapley values, which can decompose the output of a broad class of nonlinear operators \citep{anatomy2}. For instance, the data-Shapley framework of \citet{ghorbani2019data} could be employed to attribute IRF estimates to specific realizations of all dependent variables in the system. This general approach could also be applied to historically decompose other popular yet opaque objects that are nonlinear combination of data points, such as Cholesky-factorized matrices.

In this paper, we largely abstract from the consequential choices researchers make when selecting controls in local projections. While control selection is typically specified, it is well understood that a manual process underlies these decisions, inevitably affecting the sign and magnitude of estimated causal effects. Here, we adhere to the authors’ chosen specifications when available or follow standard practices in the literature. However, given the significant role of control selection---particularly for subtle economic shocks such as post-1990 monetary policy surprises---it would be valuable to decompose local projections into the sum of contributions from different controls, whether the conditional mean model is estimated via least squares, Lasso, or Random Forests. This would provide greater transparency by clarifying how variable inclusion or exclusion influences impulse response functions.

\clearpage

\setlength\bibsep{5pt}
               
\bibliographystyle{apalike}
 
\setstretch{0.75}

\def\dboxpath{/karinklieber} 
\def\dboxpath{/UQAM} 

\bibliography{ref_pgc_v250209_Gabcopy}

\clearpage
 
\appendix
\newcounter{saveeqn}
\setcounter{saveeqn}{\value{section}}
\renewcommand{\theequation}{\mbox{\Alph{saveeqn}.\arabic{equation}}} \setcounter{saveeqn}{1}
\setcounter{equation}{0}
\setstretch{1.25}
 
\pagebreak
 

\appendix
 
\section{Appendix}


\subsection{Additional Graphs and Tables}\label{sec:addresults}

\begin{figure}[h]
  \caption{\normalsize{Proximity Weights for Price Puzzle Resolutions}} \label{fig:mp_weights_var_var}
    \centering
    \vspace*{-0.2cm}
    \includegraphics[width=\textwidth, trim = 0mm 0mm 0mm 0mm, clip]{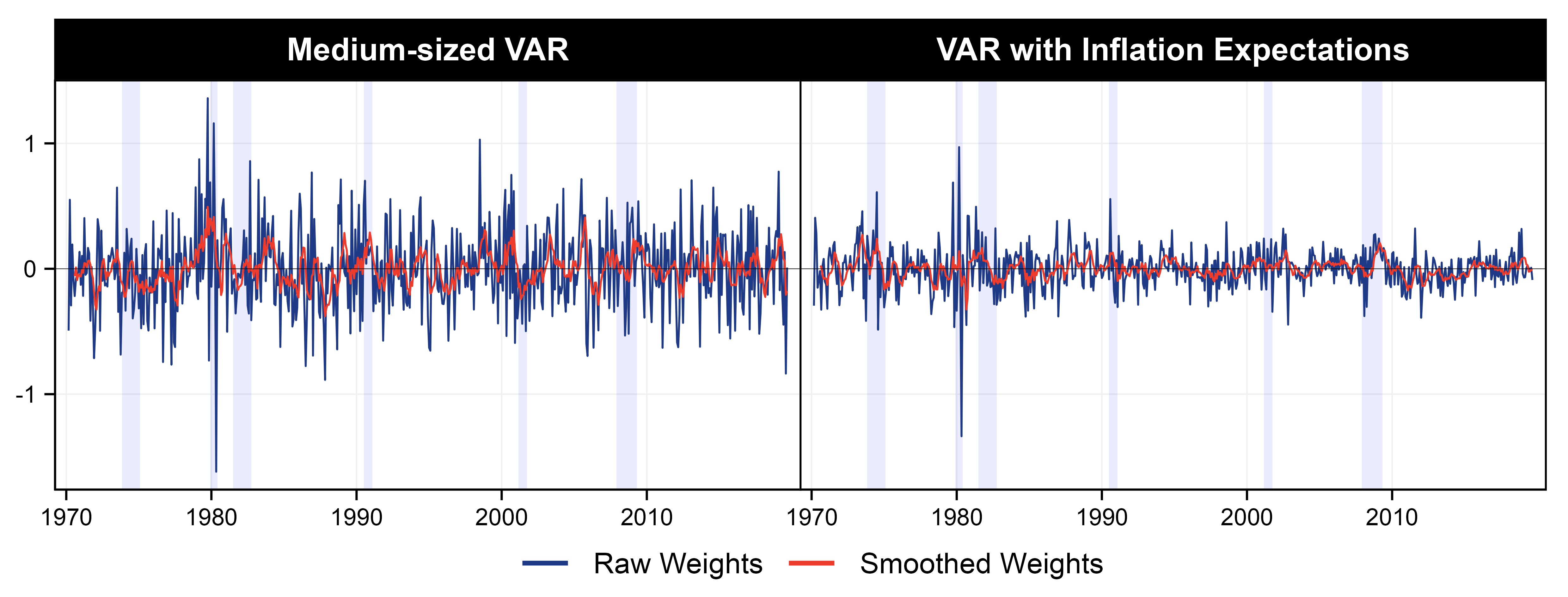}
     
    \begin{threeparttable}
    \centering
    \vspace*{-0.6cm}
    \begin{minipage}{\textwidth}
      \begin{tablenotes}[para,flushleft]
    \setlength{\lineskip}{0.2ex}
    \notsotiny 
  {\textit{Notes}: The plot displays proximity scores between the intended policy intervention and past interventions. Smoothed weights are averaged over six months. The left panel is estimated with a medium-sized VAR with short run restrictions. The right panels use inflation expectations in addition to the 4-variable VAR. In both cases, the estimation sample starts in 1970M3 and ends in 2019M9. Lavender shading corresponds to NBER recessions. }
    \end{tablenotes}
  \end{minipage}
  \end{threeparttable}

\end{figure}

 \begin{figure}[h]
  \caption{\normalsize{Moving Averages of Contributions for Different Specifications for Monetary Policy}}\label{fig:mp_contrib_ma}
    \centering
    \vspace*{-0.2cm}
    \includegraphics[width=\textwidth, trim = 0mm 0mm 0mm 0mm, clip]{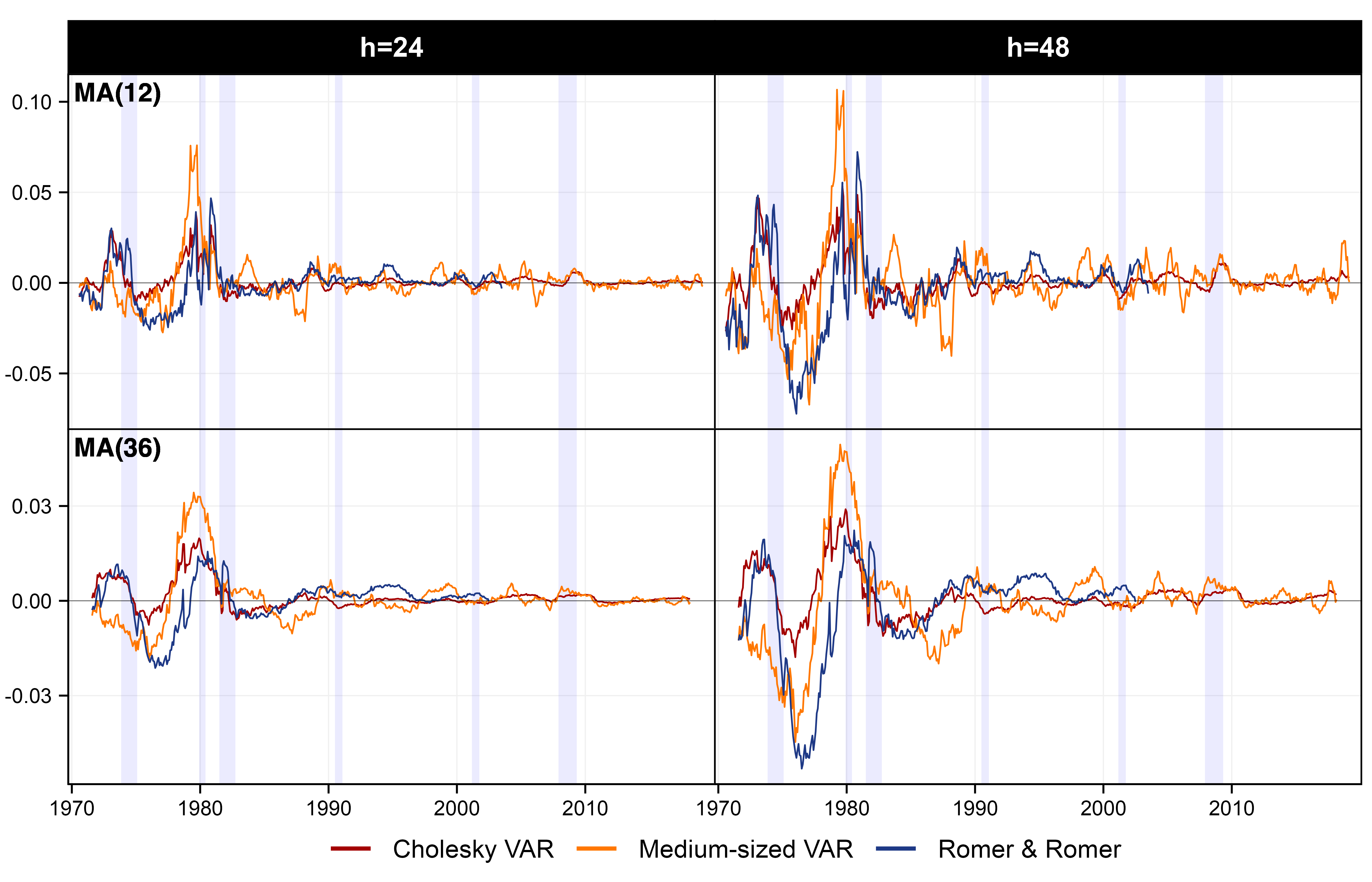}
     
    \begin{threeparttable}
    \centering
    \vspace*{-0.6cm}
    \begin{minipage}{\textwidth}
      \begin{tablenotes}[para,flushleft]
    \setlength{\lineskip}{0.2ex}
    \notsotiny 
  {\textit{Notes}: The figure shows moving averages over 12 months (upper panels) and 36 months (lower panels) for different specifications used to analyze the effects of monetary policy shocks at two horizons ($h=24$ and $h=48$). The Cholesky VAR and the medium-sized VAR span from 1970M3 to 2019M9, while the specification using Romer \& Romer shocks extends from 1970M3 to 2004M1.}
    \end{tablenotes}
  \end{minipage}
  \end{threeparttable}
\end{figure}

\begin{figure}[t]
  \caption{\normalsize{Cosine Similarities for Monetary Policy Shocks}} \label{fig:mp_cosine}
    \centering
    \vspace*{-0.2cm}
    \includegraphics[width=\textwidth, trim = 0mm 0mm 0mm 0mm, clip]{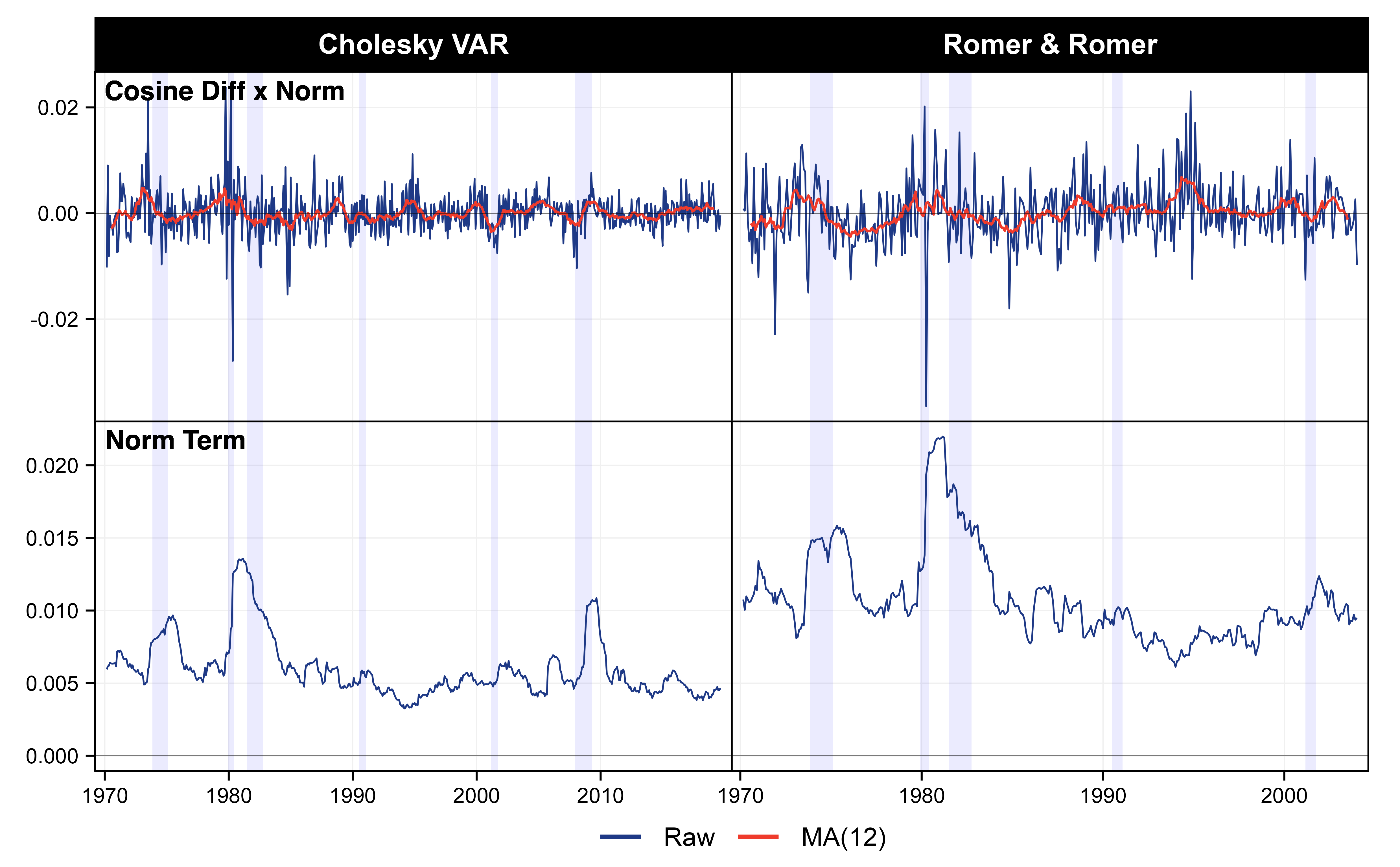}
     
    \begin{threeparttable}
    \centering
    \vspace*{-0.6cm}
    \begin{minipage}{\textwidth}
      \begin{tablenotes}[para,flushleft]
    \setlength{\lineskip}{0.2ex}
    \notsotiny 
  {\textit{Notes}: The plot shows proximity scores $w_t$ expressed as the product of the scale-invariant alignment, i.e., cosine similarities in the upper panel, and the magnitudes, measured by the $l_2$-norm of $F_t$ in the lower panel. For more details see Section \ref{sec:prox}. The Cholesky VAR sample spans from 1970M3 to 2019M9, while the specification using Romer \& Romer shocks extends from 1970M3 to 2004M1}
    \end{tablenotes}
  \end{minipage}
  \end{threeparttable}

\end{figure}

\begin{figure}[h]
  \caption{\normalsize{Proximity Scores for Nonlinear Estimation of Monetary Policy Shocks}} \label{fig:mp_weights_rf}
    \centering
    \vspace*{-0.2cm}
    \includegraphics[width=\textwidth, trim = 0mm 0mm 0mm 0mm, clip]{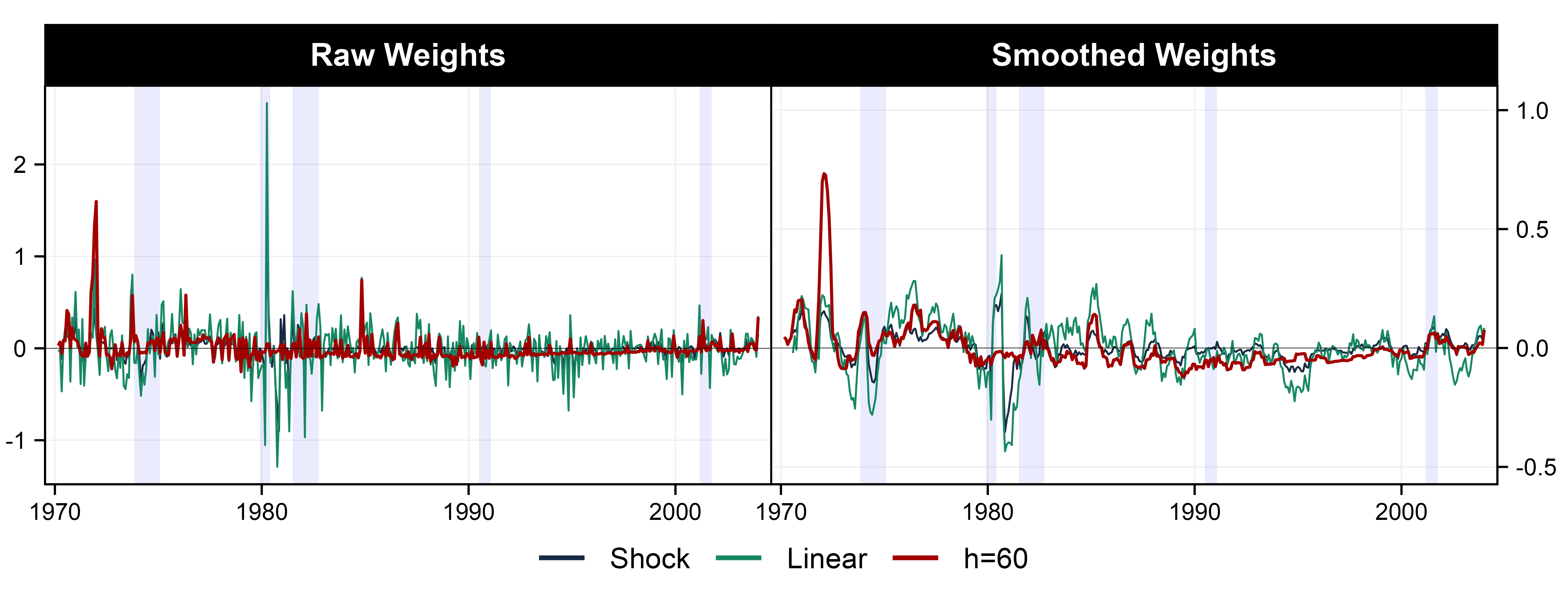}
     
    \begin{threeparttable}
    \centering
    \vspace*{-0.6cm}
    \begin{minipage}{\textwidth}
      \begin{tablenotes}[para,flushleft]
    \setlength{\lineskip}{0.2ex}
    \notsotiny 
  {\textit{Notes}: The plot displays proximity scores between the intended policy intervention and past interventions estimated with a Random Forest. Note that for nonlinear models proximity scores differ for each horizon. Smoothed weights are averaged over six months. The shock series is scaled to match the mean absolute deviation of the weights series. Lavender shading corresponds to NBER recessions.}
    \end{tablenotes}
  \end{minipage}
  \end{threeparttable}
  
  \end{figure}

 \begin{figure}[h]
  \caption{\normalsize{Clusters in the Nonlinear Response of Inflation to Monetary Policy Shocks}}\label{fig:mp_cluster_rf}
    \centering
    \vspace*{-0.2cm}
    \includegraphics[width=\textwidth, trim = 0mm 0mm 0mm 0mm, clip]{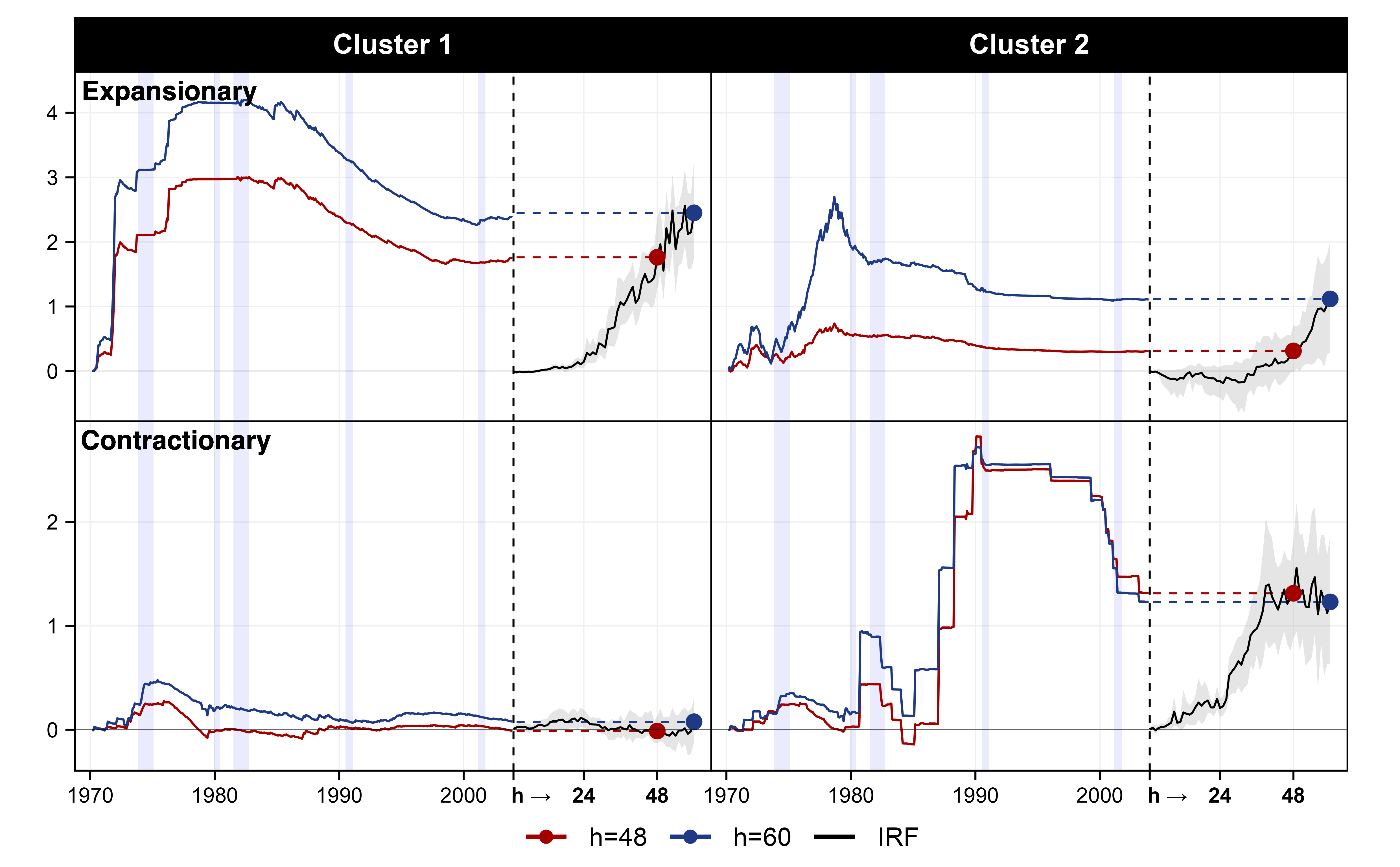}
     
    \begin{threeparttable}
    \centering
    \vspace*{-0.6cm}
    \begin{minipage}{\textwidth}
      \begin{tablenotes}[para,flushleft]
    \setlength{\lineskip}{0.2ex}
    \notsotiny 
  {\textit{Notes}: The plot partitions the main results of nonlinear responses of inflation to monetary policy shocks, as shown in Figure \ref{fig:mp_contrib_rf}, into two clusters using k-means clustering. Cluster 1 for the expansionary monetary policy shock comprises 57\% of the data, while cluster 2 accounts for 43\%. For the contractionary shock, 95\% of data falls into cluster 1, with only 5\% assigned to cluster 2. Note the cluster labels are assigned independently for each shock sign. Clusters for the expansionary and the contractionary shock are estimated separately, and the numbering does not imply correspondence between them. Cumulative contributions $\sum_{t=1}^T c_{th}$ are presented in solid colored lines, summing to the final predicted value shown as dots. Full impulse response functions are shown in black with 84\% confidence bands, defined by the mean and the 16$^{th}$ and 84$^{th}$ percentile of the tree distribution (see Section \ref{sec:RFrev}). Lavender shading corresponds to NBER recessions.}
    \end{tablenotes}
  \end{minipage}
  \end{threeparttable}
\end{figure}

\begin{figure}[h]
  \caption{\normalsize{Proximity Scores for Government Spending Shock}} \label{fig:fiscal_weights_stat}
    \centering
    \vspace*{-0.2cm}
    \includegraphics[width=\textwidth, trim = 0mm 0mm 0mm 0mm, clip]{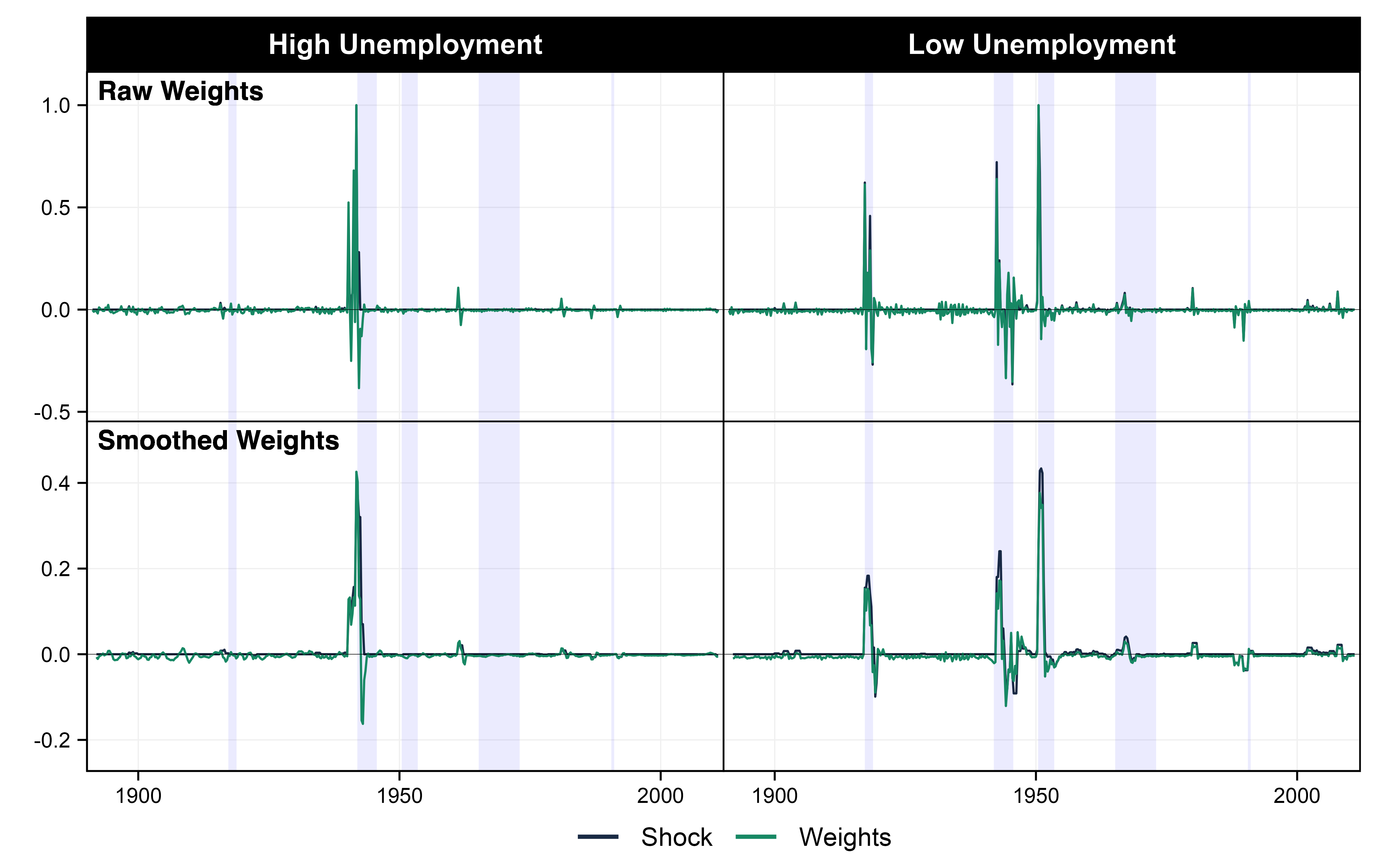}
     
    \begin{threeparttable}
    \centering
    \begin{minipage}{\textwidth}
    \vspace*{-0.2cm}
      \begin{tablenotes}[para,flushleft]
    \setlength{\lineskip}{0.2ex}
    \notsotiny 
    {\textit{Notes}: The plot displays the government spending shock series as identified by \cite{RZ2018}, and the proximity scores between the intended intervention and past interventions. Upper panels show the raw weights and shock series, lower panels present weights and shock series smoothed over four quarters. Both series are scaled relative to the peak values in the raw series. Lavender shading corresponds to WWI, WWII, Korean war, Vietnam war, and Gulf war.}
    \end{tablenotes}
  \end{minipage}
  \end{threeparttable}

\end{figure}

\begin{figure}[h]
  \caption{\normalsize{Responses of Real GDP to Government Spending Shock in Levels}} \label{fig:fiscal_contrib_levels}
    \centering
    \vspace*{-0.2cm}
    \includegraphics[width=\textwidth, trim = 0mm 0mm 0mm 0mm, clip]{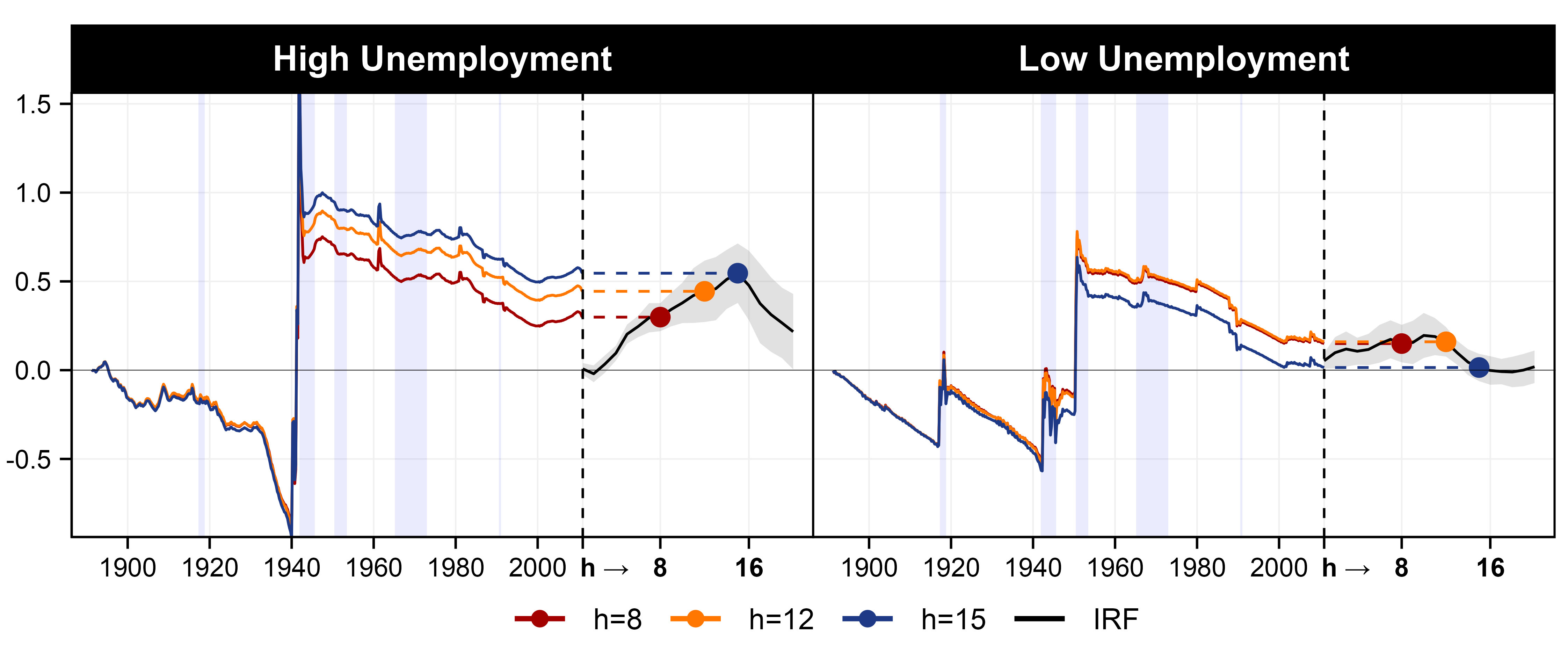}
     
    \begin{threeparttable}
    \centering
    \begin{minipage}{\textwidth}
    \vspace*{-0.2cm}
      \begin{tablenotes}[para,flushleft]
    \setlength{\lineskip}{0.2ex}
    \notsotiny 
    {\textit{Notes}: The plot shows responses of real GDP in levels to a government spending shock as defined in \cite{RZ2018} in periods of high unemployment versus low unemployment (above/below 6.5\%). Data enters estimation in levels. The estimation sample starts in 1891Q1 and ends in 2010Q4. Cumulative contributions $\sum_{t=1}^T c_{th}$ are presented in solid colored lines, summing to the final predicted value shown as dots. Full impulse response functions are shown in black with 95\% confidence bands applying Newey-West standard errors. Lavender shading corresponds to WWI, WWII, Korean war, Vietnam war, and Gulf war.}
    \end{tablenotes}
  \end{minipage}
  \end{threeparttable}

\end{figure}

\begin{figure}[h]
  \caption{\normalsize{Proximity Scores for Government Spending Shock in Levels}} \label{fig:fiscal_weights_levels}
    \centering
    \vspace*{-0.2cm}
    \includegraphics[width=\textwidth, trim = 0mm 0mm 0mm 0mm, clip]{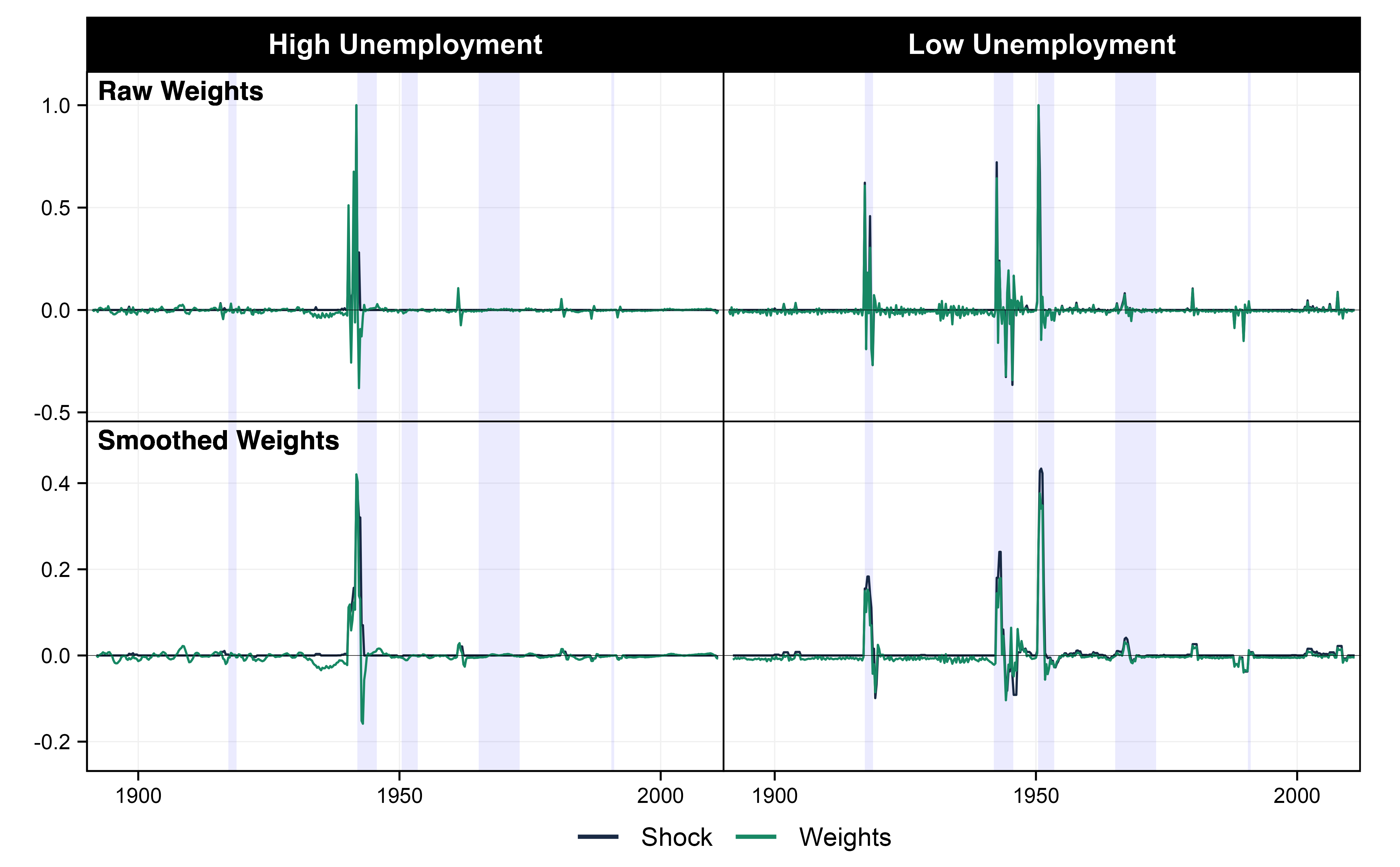}
     
    \begin{threeparttable}
    \centering
    \begin{minipage}{\textwidth}
    \vspace*{-0.2cm}
      \begin{tablenotes}[para,flushleft]
    \setlength{\lineskip}{0.2ex}
    \notsotiny 
    {\textit{Notes}: The plot displays the government spending shock series as identified by \cite{RZ2018}, and the proximity scores between the intended intervention and past interventions when estimating with data in levels. Upper panels show the raw weights and shock series, lower panels present weights and shock series smoothed over four quarters. Both series are scaled relative to the peak values in the raw series. Lavender shading corresponds to WWI, WWII, Korean war, Vietnam war, and Gulf war.}
    \end{tablenotes}
  \end{minipage}
  \end{threeparttable}

\end{figure}

\begin{figure}[t!]
  \caption{\normalsize{Alternative Government Spending Shocks}} \label{fig:fiscal_contrib_benzeev}
    \centering
    \vspace*{-0.2cm}
    \includegraphics[width=\textwidth, trim = 0mm 0mm 0mm 0mm, clip]{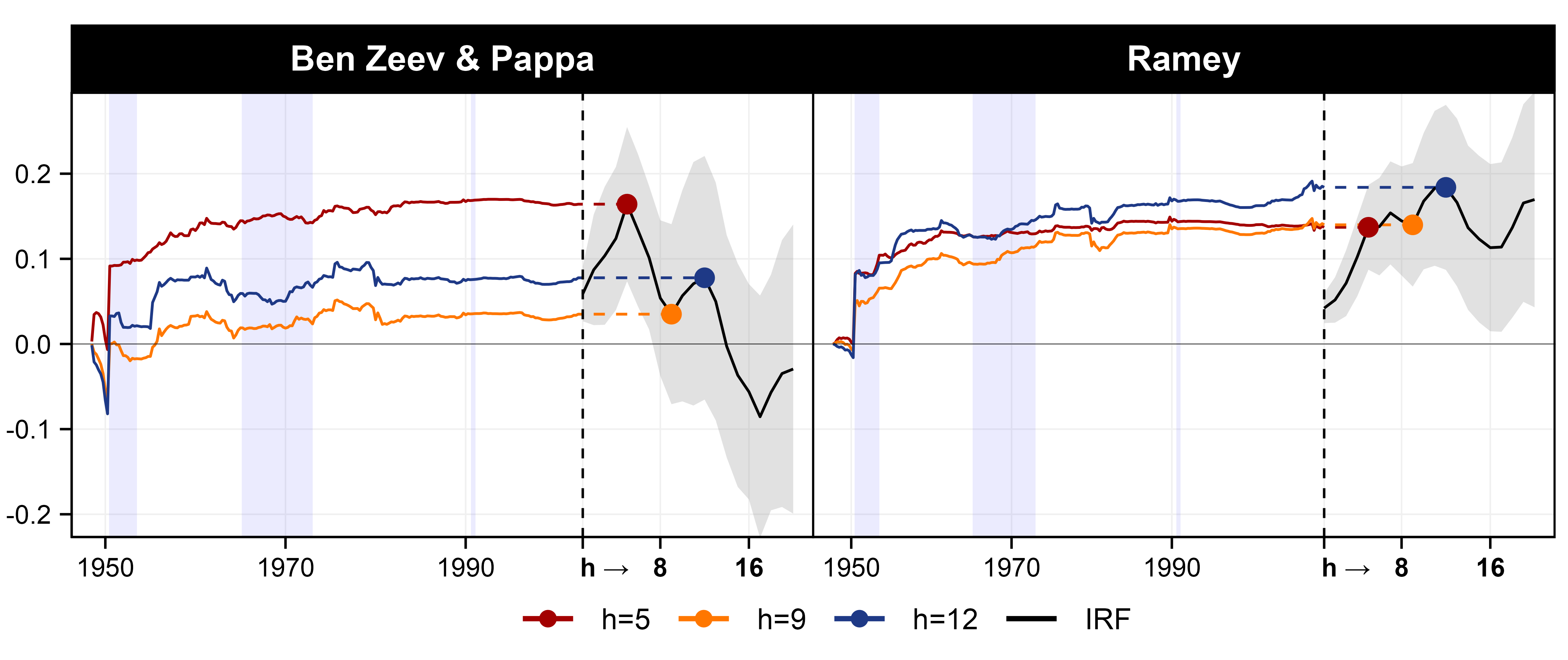}
     
    \begin{threeparttable}
    \centering
    \begin{minipage}{\textwidth}
    \vspace*{-0.2cm}
      \begin{tablenotes}[para,flushleft]
    \setlength{\lineskip}{0.2ex}
    \notsotiny 
    {\textit{Notes}: The plot shows responses of real GDP to a government spending shock as defined in \cite{RZ2018} and \cite{benzeev2017}. The estimation sample starts in 1891Q1 and ends in 2010Q4. Cumulative contributions $\sum_{t=1}^T c_{th}$ are presented in solid colored lines, summing to the final predicted value shown as dots. Full impulse response functions are shown in black with 95\% confidence bands applying Newey-West standard errors. Lavender shading corresponds to WWI, WWII, Korean war, Vietnam war, and Gulf war.}
    \end{tablenotes}
  \end{minipage}
  \end{threeparttable}

\end{figure}

\begin{figure}[h]
  \caption{\normalsize{Proximity Scores for Alternative Government Spending Shocks}} \label{fig:fiscal_weights_benzeev}
    \centering
    \vspace*{-0.2cm}
    \includegraphics[width=\textwidth, trim = 0mm 0mm 0mm 0mm, clip]{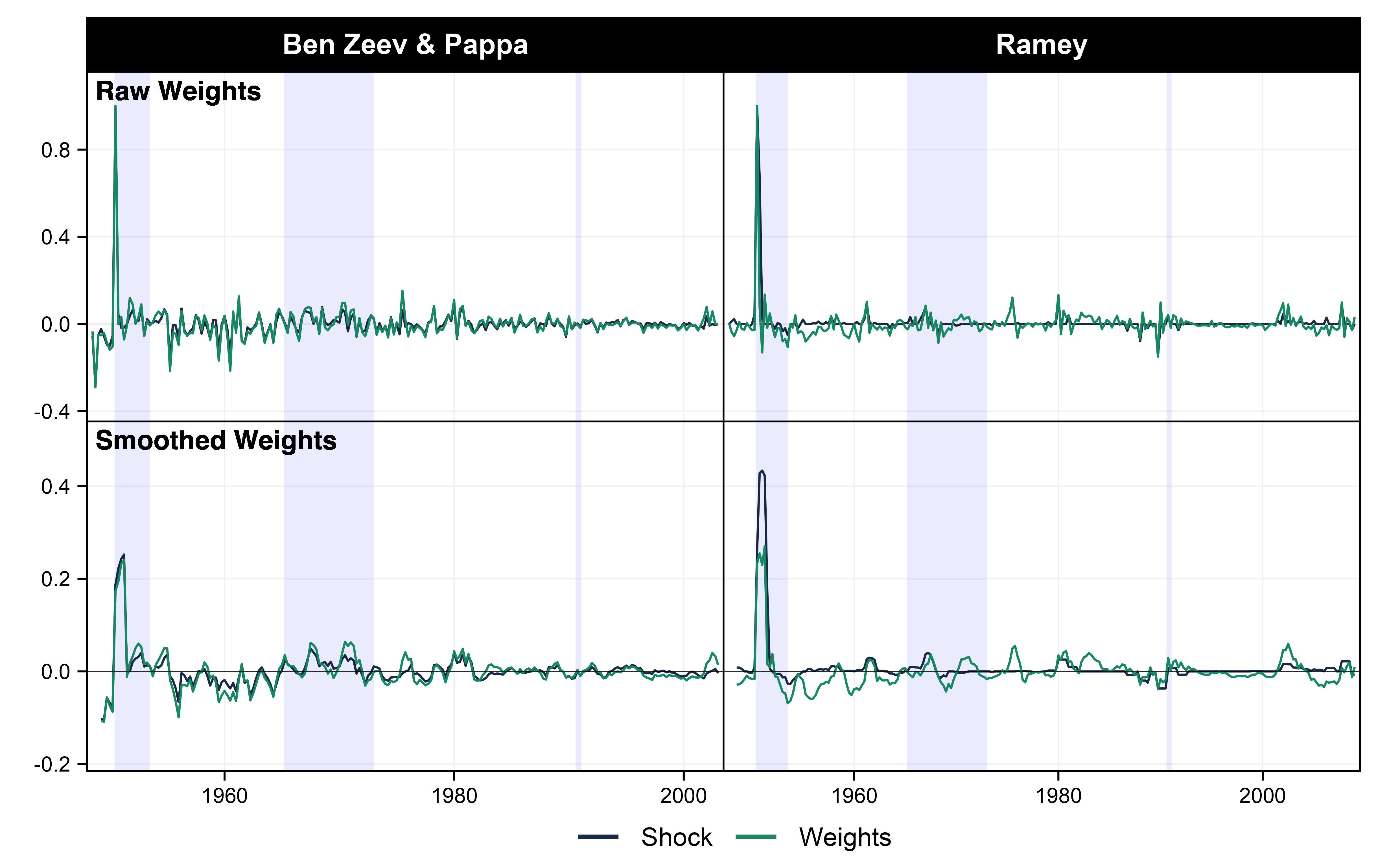}
     
    \begin{threeparttable}
    \centering
    \begin{minipage}{\textwidth}
    \vspace*{-0.2cm}
      \begin{tablenotes}[para,flushleft]
    \setlength{\lineskip}{0.2ex}
    \notsotiny 
    {\textit{Notes}: The plot displays the government spending shock series as identified by \cite{RZ2018} and \cite{benzeev2017}, and the proximity scores between the intended intervention and past interventions. Upper panels show the raw weights and shock series, lower panels present weights and shock series smoothed over four quarters. Lavender shading corresponds to WWI, WWII, Korean war, Vietnam war, and Gulf war.}
    \end{tablenotes}
  \end{minipage}
  \end{threeparttable}

\end{figure}

\begin{figure}[h]
  \caption{\normalsize{Responses of Real GDP to Alternative Government Spending Shocks for Different Subsamples}} \label{fig:fiscal_trim_benzeev}
    \centering
    \vspace*{-0.2cm}
    \includegraphics[width=\textwidth, trim = 0mm 0mm 0mm 0mm, clip]{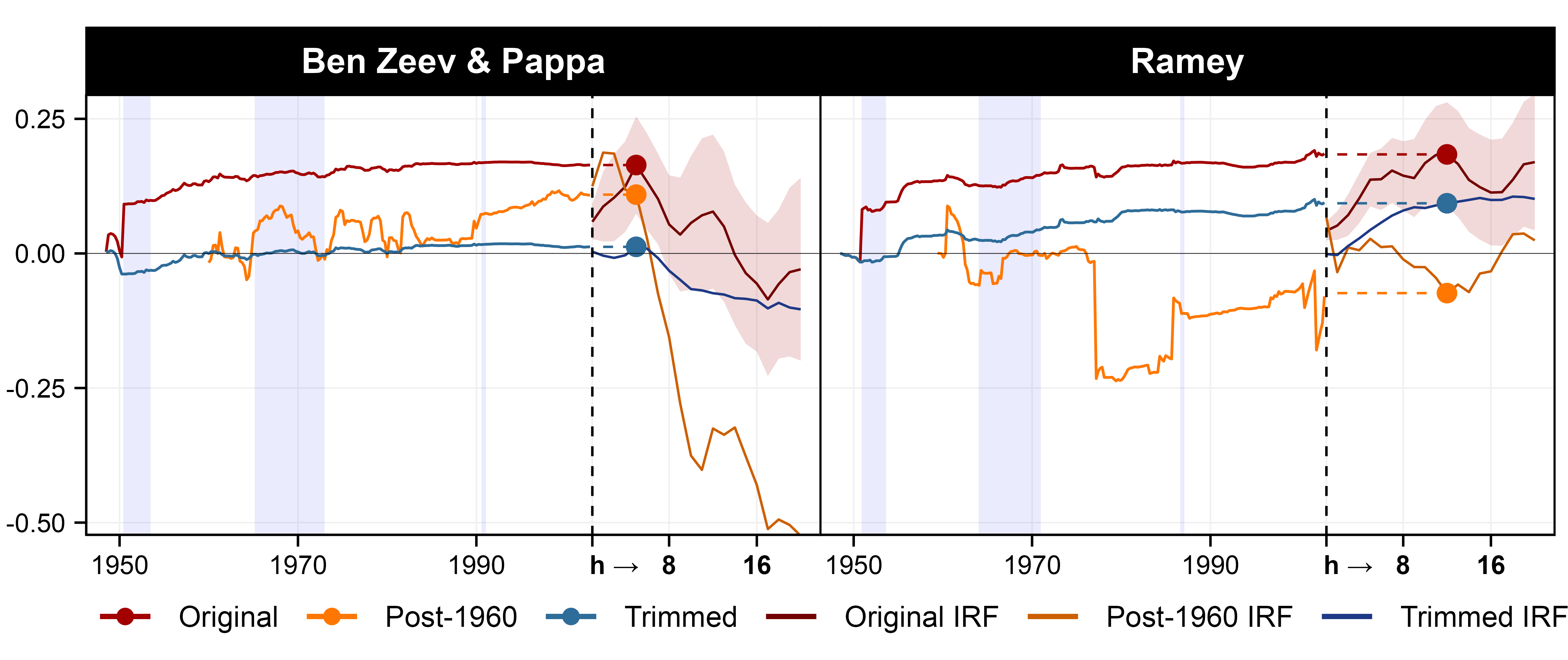}
     
    \begin{threeparttable}
    \centering
    \begin{minipage}{\textwidth}
    \vspace*{-0.2cm}
      \begin{tablenotes}[para,flushleft]
    \setlength{\lineskip}{0.2ex}
    \notsotiny 
    {\textit{Notes}: The plot shows responses of real GDP growth to a government spending shock, as identified in \cite{RZ2018} and \cite{benzeev2017}, for three different estimation samples. \textit{Post-1960 IRF} results from estimating the model starting in 1960Q1 and ending in 2010Q4. For \textit{Trimmed IRF}, we set weights at the 1st and 99th percentiles to zero for each horizon, and re-cumulate the remaining contributions to obtain the IRF. \textit{Original IRF} refers to results obtained using the full sample, as done in the main analysis (Figure \ref{fig:fiscal_contrib_benzeev}). For the evidence curves we focus on the peak horizon of the original IRF. Cumulative contributions $\sum_{t=1}^T c_{th}$ are presented in solid colored lines, summing to the final predicted value shown as dots. Full impulse response functions are shown in black with 95\% confidence bands applying Newey-West standard errors. Lavender shading corresponds to WWI, WWII, Korean war, Vietnam war, and Gulf war.}
    \end{tablenotes}
  \end{minipage}
  \end{threeparttable}

\end{figure}

 \begin{figure}[h]
  \caption{\normalsize{Clusters in the Nonlinear Responses to Financial Shocks}}\label{fig:fin_cluster_rf}
    \centering
    \vspace*{-0.2cm}
    \includegraphics[width=\textwidth, trim = 0mm 0mm 0mm 0mm, clip]{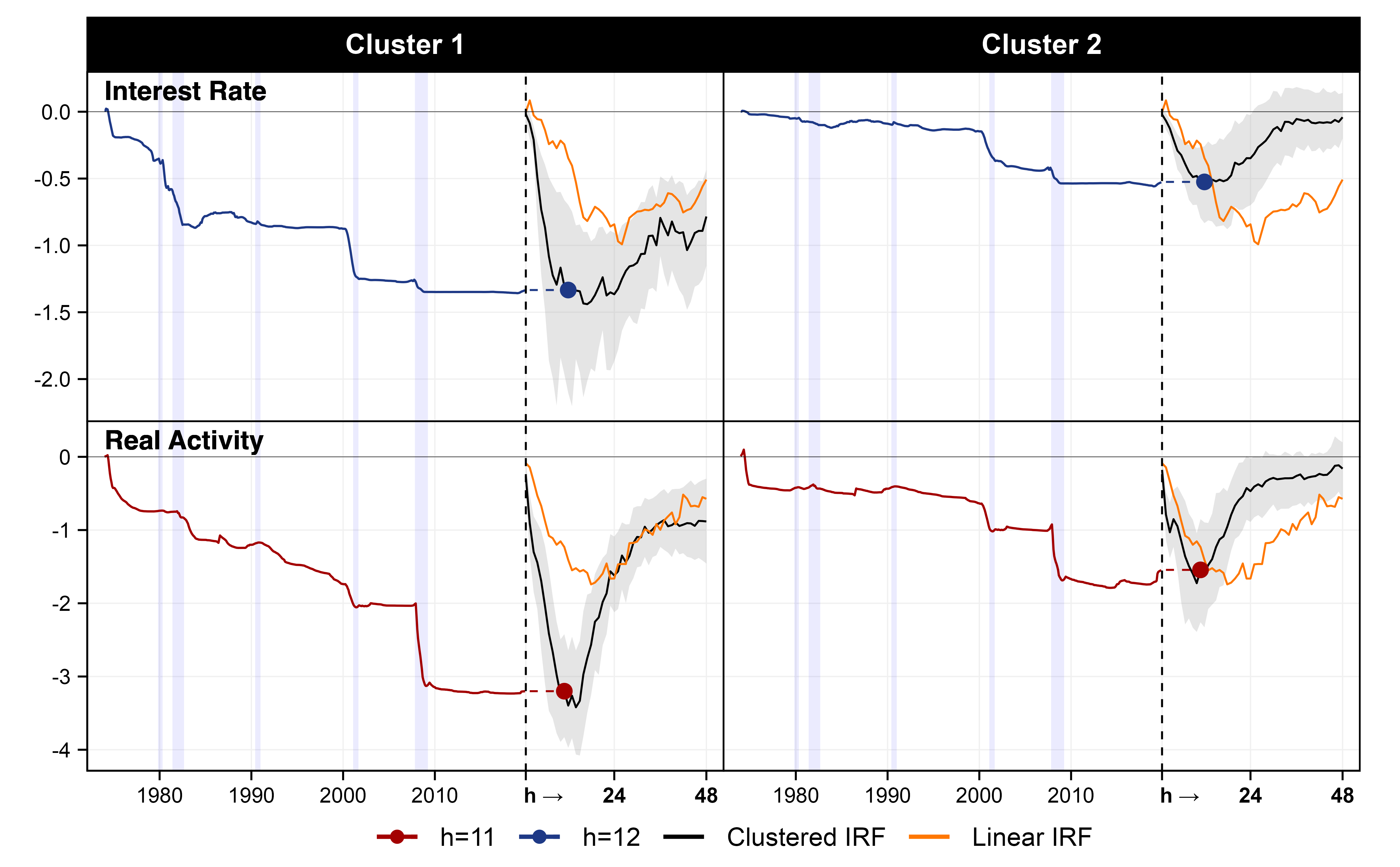}
     
    \begin{threeparttable}
    \centering
    \vspace*{-0.6cm}
    \begin{minipage}{\textwidth}
      \begin{tablenotes}[para,flushleft]
    \setlength{\lineskip}{0.2ex}
    \notsotiny 
  {\textit{Notes}: The plot partitions the main results of nonlinear responses to financial shocks, as shown in Figure \ref{fig:fin_contrib_rf}, into two clusters using k-means clustering. Cluster 1 for the response of real activity comprises 39\% of the data, while cluster 2 accounts for 61\%. For the response of the interest rate, 31\% of data falls into cluster 1, with only 69\% assigned to cluster 2. Note that the cluster labels are assigned independently for each variable. Clusters for real activity and interest rates are estimated separately, and the numbering does not imply correspondence between them. Cumulative contributions $\sum_{t=1}^T c_{th}$ are presented in solid colored lines, summing to the final predicted value shown as dots. Full impulse response functions are shown in black with 84\% confidence bands, defined by the mean and the 16$^{th}$ and 84$^{th}$ percentiles of the tree distribution (see Section \ref{sec:RFrev}). Lavender shading corresponds to NBER recessions.}
    \end{tablenotes}
  \end{minipage}
  \end{threeparttable}
\end{figure}

 \begin{figure}[h]
  \caption{\normalsize{Proximity Scores for Clustered Nonlinear Responses to Financial Shocks}}\label{fig:fin_cluster_weights}
    \centering
    \vspace*{-0.2cm}
    \includegraphics[width=\textwidth, trim = 0mm 0mm 0mm 0mm, clip]{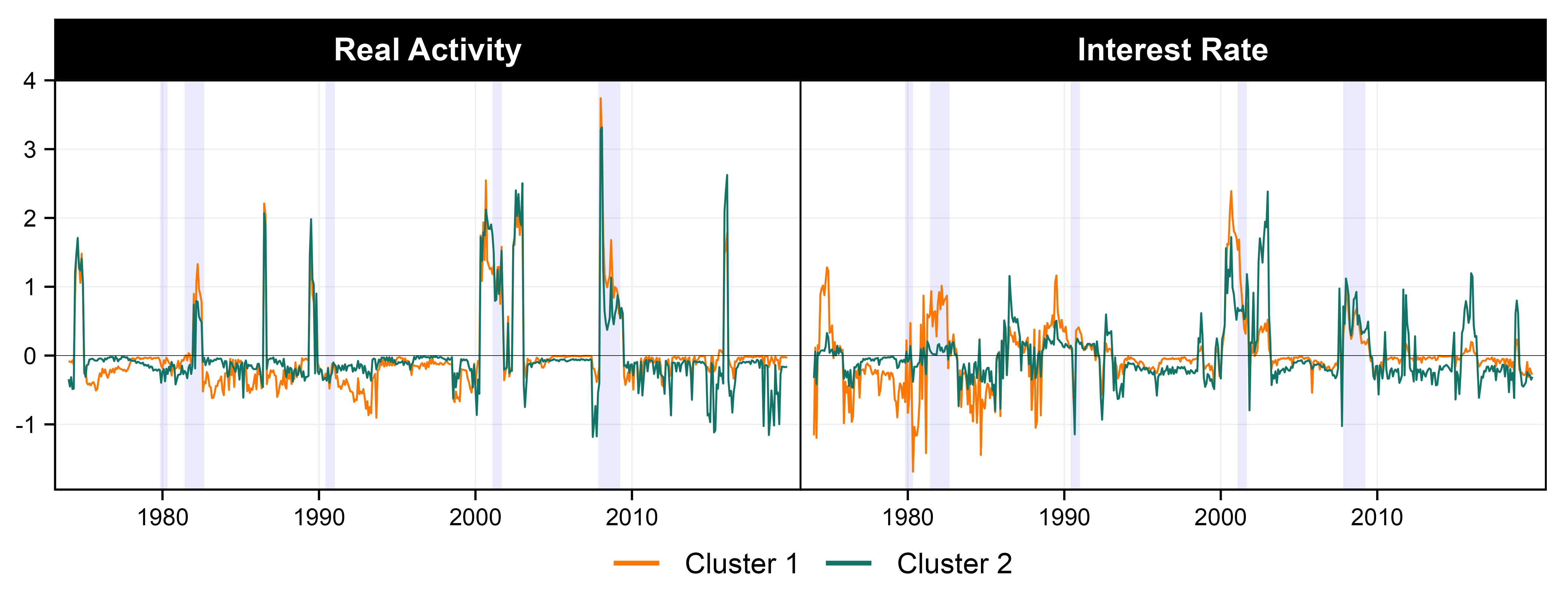}
     
    \begin{threeparttable}
    \centering
    \vspace*{-0.6cm}
    \begin{minipage}{\textwidth}
      \begin{tablenotes}[para,flushleft]
    \setlength{\lineskip}{0.2ex}
    \notsotiny 
  {\textit{Notes}: The plot shows proximity scores when partitioning the main results of nonlinear responses to financial shocks into two clusters using k-means clustering. Cluster 1 for the response of real activity comprises 39\% of the data, while cluster 2 accounts for 61\%. For the response of the interest rate, 31\% of data falls into cluster 1, with only 69\% assigned to cluster 2. Note that the cluster labels are assigned independently for each variable. Clusters for real activity and interest rates are estimated separately, and the numbering does not imply correspondence between them. Lavender shading corresponds to NBER recessions.}
    \end{tablenotes}
  \end{minipage}
  \end{threeparttable}
\end{figure}

\begin{table}[t]
\vspace{1em}
  \footnotesize
  \centering
  \begin{threeparttable}
  \caption{\normalsize {Concentration Measures for Monetary Policy} \label{tab:mp}
    \vspace{-0.3cm}}
    \setlength{\tabcolsep}{0.6em} 
    \setlength\extrarowheight{2.9pt}

     \begin{tabular}{l| rrr|rrr}
    \toprule \toprule
    \addlinespace[2pt]
    \multicolumn{1}{l|}{} & \multicolumn{3}{c|}{\textbf{WC}}  & \multicolumn{3}{c}{\textbf{CC}}  \\ 
    \cmidrule(lr){2-4} \cmidrule(lr){5-7}
    \multicolumn{1}{l|}{} & \multicolumn{1}{c}{$h=24$} & \multicolumn{1}{c}{$h=48$} & \multicolumn{1}{c|}{$h=60$}  & \multicolumn{1}{c}{$h=24$} & \multicolumn{1}{c}{$h=48$} & \multicolumn{1}{c}{$h=60$}  \\ 
    \midrule
  \addlinespace[5pt] 
  \rowcolor{gray!15} 
  \multicolumn{2}{l}{\textbf{Inflation}}   && \multicolumn{2}{l}{} & & \cellcolor{gray!15}  \\ \addlinespace[2pt]
  Romer \& Romer & 0.33 & " " & " " & 0.44 & 0.40 & 0.40  \\ 
  Cholesky VAR & 0.37 & " " & " " & 0.53 & 0.49 & 0.48  \\ 
  Medium-sized VAR & 0.27 & " " & " "& 0.42 & 0.40 & 0.39  \\ 
  VAR with Inflation Expectations & 0.31 & " " & " "& 0.48 & 0.44 & 0.43  \\ 
  \addlinespace[5pt] 
  \rowcolor{gray!15} 
  \multicolumn{2}{l}{\textbf{Unemployment}}   && \multicolumn{2}{l}{} & &  \cellcolor{gray!15}  \\ \addlinespace[2pt]
  Romer \& Romer & 0.33 & " " & " "& 0.49 & 0.40 & 0.41  \\ 
  Cholesky VAR & 0.38 & " " & " "& 0.52 & 0.43 & 0.43  \\ 
  Medium-sized VAR & 0.27 & " " & " "& 0.40 & 0.39 & 0.39  \\ 
  VAR with Inflation Expectations & 0.31 & " " & " "& 0.47 & 0.37 & 0.37  \\ 
  \addlinespace[5pt] 
  \rowcolor{gray!15} 
  \multicolumn{2}{l}{\textbf{Nonlinear Estimation}}   && \multicolumn{2}{l}{} & &  \cellcolor{gray!15}  \\ \addlinespace[2pt]
  Inflation (expansionary) & 0.36 &0.38 & 0.33 & 0.45 & 0.50 & 0.44  \\ 
  Inflation (contractionary) & 0.41 &0.42 & 0.39 & 0.43 & 0.46 & 0.47  \\ 
   \bottomrule \bottomrule 
\end{tabular}
\begin{tablenotes}[para,flushleft]
  \footscript 
    \textit{Notes}: The table summarizes concentration statistics as discussed in Section \ref{sec:derivatives}. We choose $Q=10$.
  \end{tablenotes}
\end{threeparttable}
\end{table}

\begin{table}[t]
\vspace{1em}
  \footnotesize
  \centering
  \begin{threeparttable}
  \caption{\normalsize {Concentration Measures for Government Spending} \label{tab:fiscal}
    \vspace{-0.3cm}}
    \setlength{\tabcolsep}{0.6em} 
    \setlength\extrarowheight{2.9pt}

    \begin{tabular}{l| rrr|rrr}
    \toprule \toprule
    \addlinespace[2pt]
    \multicolumn{1}{l|}{} & \multicolumn{3}{c|}{\textbf{WC}}  & \multicolumn{3}{c}{\textbf{CC}}  \\ 
    \cmidrule(lr){2-4} \cmidrule(lr){5-7}
    \multicolumn{1}{l|}{} & \multicolumn{1}{c}{$h=8$} & \multicolumn{1}{c}{$h=12$} & \multicolumn{1}{c|}{$h=15$}  & \multicolumn{1}{c}{$h=8$} & \multicolumn{1}{c}{$h=12$} & \multicolumn{1}{c}{$h=15$}  \\ 
    \midrule
  \addlinespace[5pt] 
  \rowcolor{gray!15} 
  \multicolumn{2}{l}{\textbf{High Unemployment}}   && \multicolumn{2}{l}{} & & \cellcolor{gray!15}  \\ \addlinespace[2pt]
  Full Sample  & 0.66 & " " & " " & 0.89 & 0.90 & 0.90  \\ 
  Trimmed Sample & 0.36 & " " & " " & 0.55 & 0.54 & 0.56  \\ 
  Post-1960 & 0.52 & " " & " "& 0.62 & 0.59 & 0.56  \\ 
  \addlinespace[5pt] 
  \rowcolor{gray!15} 
  \multicolumn{2}{l}{\textbf{Low Unemployment}}   && \multicolumn{2}{l}{} & & \cellcolor{gray!15}  \\ \addlinespace[2pt]
  Full Sample & 0.66 & " " & " " & 0.81 & 0.81 & 0.77  \\ 
  Trimmed Sample & 0.49 & " " & " " & 0.70 & 0.67 & 0.68  \\ 
  Post-1960 & 0.61 & " " & " "& 0.72 & 0.72 & 0.71  \\ 
  \addlinespace[5pt] 
  \rowcolor{gray!15} 
  \multicolumn{2}{l}{\textbf{Ben Zeev}}   && \multicolumn{2}{l}{} & &  \cellcolor{gray!15}  \\ \addlinespace[2pt]
  Full Sample & 0.40 & " " & " "& 0.57 & 0.52 & 0.58  \\ 
  Trimmed Sample & 0.29 & " " & " "& 0.41 & 0.46 & 0.45  \\ 
  Post-1960 & 0.32 & " " & " "& 0.43 & 0.45 & 0.46  \\ 
   \bottomrule \bottomrule 
\end{tabular}
\begin{tablenotes}[para,flushleft]
  \footscript 
    \textit{Notes}: The table summarizes concentration statistics as discussed in Section \ref{sec:derivatives}. We choose $Q=10$.
  \end{tablenotes}
\end{threeparttable}
\end{table}

\begin{table}[t]
\vspace{1em}
  \footnotesize
  \centering
  \begin{threeparttable}
  \caption{\normalsize {Concentration Measures for Global Temperature Shock} \label{tab:climate}
    \vspace{-0.3cm}}
    \setlength{\tabcolsep}{0.6em} 
    \setlength\extrarowheight{2.9pt}

    \begin{tabular}{l| rrr|rrr}
    \toprule \toprule
    \addlinespace[2pt]
    \multicolumn{1}{l|}{} & \multicolumn{3}{c|}{\textbf{WC}}  & \multicolumn{3}{c}{\textbf{CC}}  \\ 
    \cmidrule(lr){2-4} \cmidrule(lr){5-7}
    \multicolumn{1}{l|}{} & \multicolumn{1}{c}{$h=3$} & \multicolumn{1}{c}{$h=6$} & \multicolumn{1}{c|}{$h=10$}  & \multicolumn{1}{c}{$h=3$} & \multicolumn{1}{c}{$h=6$} & \multicolumn{1}{c}{$h=10$}  \\ 
    \midrule
  \addlinespace[5pt] 
  \rowcolor{gray!15} 
  \multicolumn{2}{l}{\textbf{World GDP}}   && \multicolumn{2}{l}{} & & \cellcolor{gray!15}  \\ \addlinespace[2pt]
  Full Sample  & 0.25 & " " & " " & 0.27 & 0.28 & 0.33  \\ 
  Sample 1965-2019 & 0.22 & " " & " " & 0.22 & 0.22 & 0.27  \\ 
  Sample 1963-1990 & 0.24 & " " & " "& 0.34 & 0.30 & 0.31  \\ 
  Sample 1980-2010 & 0.21 & " " & " "& 0.23 & 0.23 & 0.25  \\ 
  Cubic Trend & 0.22 & " " & " "& 0.25 & 0.26 & 0.30  \\ 
  \addlinespace[5pt] 
  \rowcolor{gray!15} 
  \multicolumn{2}{l}{\textbf{Global Average Temperature}}   && \multicolumn{2}{l}{} & &  \cellcolor{gray!15}  \\ \addlinespace[2pt]
  Full Sample & 0.27 & " " & " "& 0.42 & 0.43 & 0.33  \\ 
   \bottomrule \bottomrule 
\end{tabular}
\begin{tablenotes}[para,flushleft]
  \footscript 
    \textit{Notes}: The table summarizes concentration statistics as discussed in Section \ref{sec:derivatives}. We choose $Q=10$.
  \end{tablenotes}
\end{threeparttable}
\end{table}

\begin{table}[t]
\vspace{1em}
  \footnotesize
  \centering
  \begin{threeparttable}
  \caption{\normalsize {Concentration Measures for Financial Shock} \label{tab:fin}
    \vspace{-0.3cm}}
    \setlength{\tabcolsep}{0.6em} 
    \setlength\extrarowheight{2.9pt}

    \begin{tabular}{l| rrr|rrr}
    \toprule \toprule
    \addlinespace[2pt]
    \multicolumn{1}{l|}{} & \multicolumn{3}{c|}{\textbf{WC}}  & \multicolumn{3}{c}{\textbf{CC}}  \\ 
    \cmidrule(lr){2-4} \cmidrule(lr){5-7}
    \multicolumn{1}{l|}{} & \multicolumn{1}{c}{$h=11$} & \multicolumn{1}{c}{$h=12$} & \multicolumn{1}{c|}{$h=26$}  & \multicolumn{1}{c}{$h=11$} & \multicolumn{1}{c}{$h=12$} & \multicolumn{1}{c}{$h=26$}  \\ 
    \midrule
  \addlinespace[5pt] 
  \rowcolor{gray!15} 
  \multicolumn{2}{l}{\textbf{Interest Rate}}   && \multicolumn{2}{l}{} & & \cellcolor{gray!15}  \\ \addlinespace[2pt]
  Linear Model  & 0.35 & " " & " " & 0.55 & 0.54 & 0.48  \\ 
  Nonlinear Model & 0.35 & 0.33 & 0.29 & 0.57 & 0.56 & 0.48  \\ 
  \addlinespace[5pt] 
  \rowcolor{gray!15} 
  \multicolumn{2}{l}{\textbf{Real Activity}}   && \multicolumn{2}{l}{} & &  \cellcolor{gray!15}  \\ \addlinespace[2pt]
  Linear Model  & 0.35 & " " & " " & 0.47 & 0.45 & 0.41  \\ 
  Nonlinear Model & 0.46 & 0.45 & 0.43 & 0.61 & 0.62 & 0.40  \\ 
   \bottomrule \bottomrule 
\end{tabular}
\begin{tablenotes}[para,flushleft]
  \footscript 
    \textit{Notes}: The table summarizes concentration statistics as discussed in Section \ref{sec:derivatives}. We choose $Q=10$.
  \end{tablenotes}
\end{threeparttable}
\end{table}

\clearpage

\subsection{A Review of Random Forest}\label{sec:RFrev}

Random Forest \citep[RF, ][]{breiman2001} is an ensemble learning method that averages over the output of multiple regression trees. To introduce the algorithm, we first describe how a single regression tree is estimated using a greedy algorithm, and then explain how multiple trees are aggregated to form a forest.

A tree takes the vector of features $X_t$ for a single observation $t$ as input and outputs the corresponding fitted value for the dependent variable $y_{t}$. Formally, this can be written as: 
$$y_{t} = \mathcal{T}(X_t) + \epsilon_t,$$
where $\mathcal{T}$ denotes the tree. 
For a graphical representation of a tree, we consider a simplified example that models inflation $\pi_t$ with two features in $X_t$, the nominal rate of interest $r_t$ and a measure of the output gap $g_t$. A simple decision tree could take the following form: 

\vspace{1em}
\Tree[.{$X_t$} 
[.{$g_t \geq 0 $} 
[.{$r_{t} \geq 4 \%$} {$\phantom{--} \pi_t= 2 +\epsilon_t \phantom{--}$} ]
[.{$r_{t}  < 4 \% $} {$\phantom{--} \pi_t = 5 +\epsilon_t \phantom{--}$} ]
 ]
[.{$g_t  < 0 $} {$\phantom{--} \pi_t = 1 +\epsilon_t \,  .     \phantom{--}$} ]
 ]
 \vspace{1em}

\noindent The estimation of such a tree typically relies on a \textit{greedy} algorithm that \textit{recursively} splits the data, a method pioneered in \cite{breiman1984classification}'s work on Classification and Regression Trees (CART). The splitting is determined  by
\begin{equation}\label{treesplit}
\begin{aligned}
\min\limits_{k\in \mathcal{K}, d \in {\rm I\!R}}\Bigg[ \min\limits_{\mu_{1}} \sum \limits_{\{ t \in \mathcal{L}| X_{t,k} \leq c  \}} \left(y_{t}-\mu_{1}\right)^{2} 
 + \min\limits_{\mu_{2}} \sum \limits_{\{ t \in \mathcal{L} | X_{t,k} > d  \}} \left(y_{t}-\mu_{2}\right)^{2}\Bigg].
\end{aligned}
\end{equation}
In this equation, $\min\limits_{k\in \mathcal{K}, d \in {\rm I\!R}}$ refers to the minimization over all possible splits, with $K$ the number of available features,  and $d$ a real number representing the split point.  $\mathcal{L}$ denotes a leaf, which reflects the subset of features considered by the algorithm to estimate the next split. Initially, $\mathcal{L}$ consists of the entire training sample.  The algorithm  then recursively partitions $\mathcal{L}$ into smaller subsets until a stopping criterion is met, resulting in a set of terminal nodes. Note that in our illustrative example of modeling inflation, the tree has three terminal nodes. The optimized values $ k^*$ and $c^*$ as well as the predicted values $\mu_{1}$ and $\mu_{2}$, presenting within-leaf sample averages, are obtained by minimizing the total within-leaf sum of squared errors.

In a Random Forest, trees serve as base learners, and the final step is to combine them by taking the average of their predictions. For this process, we rely on three steps. First, to minimize bias, trees are typically grown to substantial depth. This implies continuing the splitting process in \eqref{treesplit} until all terminal nodes contain very few observations (usually less than five and set via \texttt{minimal.node.size}). Second, applying the \textit{Bagging} strategy \citep[Bootstrap Aggregation, as in ][]{breiman1996bagging}, we generate $B$ bootstrap samples, selecting $[y_{t} \enskip X_t]$ pairs with replacement. Each tree in the forest is constructed using a single bootstrapped sample $b \in 1,\dots, B$. Each tree is trained on one such sample, introducing variability that is critical for ensemble effectiveness. Third, the tree-growing process incorporates feature selection randomness: at each split, a random subset $\mathcal{K^{-}\subset \mathcal{K}}$ of predictors is considered, the relative size of which is controlled via the \texttt{mtry} parameter. This further decorrelates the trees, which is essential for reducing variance when predictions are averaged across the forest. The final RF prediction is the simple average over all $B$ tree predictions, resulting in $\hat{y}_{t+h} = \frac{1}{B} \sum_{b=1}^B \mathcal{T}_{bh}(X_t)$.


\subsubsection{Implementation Details}

In our applications, we implement Random Forests using the \texttt{ranger} package in \texttt{R}. All argument names referenced below correspond to those in that package. Specifically, we set \texttt{min.node.size = 5} and use a relatively low \texttt{mtry} value, equal to $\sfrac{1}{15}$ of the total number of predictors, to encourage tree diversity and mitigate overfitting. To ensure that key features—namely, lags of the target variable and the shock series—remain central throughout the model, we invoke the \texttt{always.split.variables} option. This forces these variables into the candidate set $\mathcal{K}^-$ at every split, thereby anchoring the model on features that are of primary interest for our impulse response analysis.

This combination of low \texttt{mtry} and forced splitting on shock-related variables introduces a form of selective regularization: it nudges the ensemble toward repeatedly using the variable of interest (e.g., the policy shock), thus increasing its visibility in the forest’s predictive structure. Without this adjustment, most shock series considered in our applications would be effectively ignored by RF, as their standalone predictive power is low relative to other available predictors. In this sense, the approach aligns with a Bayesian philosophy, where the treatment variable is regularized less aggressively than other features—not through explicit priors, but through design choices that prioritize its inclusion. This avoids the complexity and data inefficiency of more formal alternatives such as double machine learning or honest forests, which typically require splitting the sample into auxiliary subsamples—an approach less amenable to historical macroeconomic analysis. 

Empirically, this strategy proves reasonable: in both the monetary policy and financial shock applications, the Random Forest impulse responses are broadly consistent with, and often comparable in magnitude to, those from the linear model. Moreover, we tested the sensitivity of our results to the choice of \texttt{min.node.size = 5}, which can affect in-sample overfitting, and found the results to be robust. Thus, our use of imbalanced regularization in favor of the shock variable appears both principled and effective. Nonetheless, integrating our interpretability framework with more sophisticated causal forest methods remains a promising avenue for future work.
 
\end{document}